\documentclass[aps,prx,twocolumn,superscriptaddress,longbibliography,10pt]{revtex4-2}

\usepackage{lmodern} %improves the font rendering in the pdf
\usepackage{microtype}

\usepackage[colorlinks=true, allcolors=blue]{hyperref}

\usepackage{amsmath}
\usepackage{amsfonts}
\usepackage{amssymb}
\usepackage{amstext}
\usepackage[sort&compress]{natbib}
\usepackage{amsthm}
\usepackage{graphicx}
\usepackage{textcomp}
\usepackage{color}
\usepackage{mathtools}
\usepackage{here}
\usepackage{multirow}
\usepackage[capitalize]{cleveref}
\usepackage{paralist}
\usepackage{multirow}
\usepackage[normalem]{ulem} %for \sout

\usepackage[dvipsnames]{xcolor}
\usepackage{tikz}
\usetikzlibrary{patterns} % Load the patterns library

\usepackage{thmtools,thm-restate}
\usepackage{regexpatch}
\makeatletter
\xpatchcmd\thmt@restatable{%
\csname #2\@xa\endcsname\ifx\@nx#1\@nx\else[{#1}]\fi
}{%
\ifthmt@thisistheone
\csname #2\@xa\endcsname\ifx\@nx#1\@nx\else[{#1}]\fi
\else
\csname #2\@xa\endcsname[{restated}]
\fi}{}{}
\makeatother
\usepackage{comment}

\newtheorem{theorem}{Theorem}

\newtheorem{corollary}{Corollary}

\theoremstyle{definition}
\newtheorem{definition}{Definition}
\newtheorem{problem}{Problem}

\newtheorem*{Proof}{Proof}

\newcommand{\ket}[1]{|#1\rangle} %ket
\newcommand{\bra}[1]{\langle#1|} %bra
\newcommand{\ketbra}[2]{|#1\rangle\langle#2|} %ketbra
\newcommand{\braket}[2]{\langle #1 \vert #2 \rangle}

\newcommand{\ctrl}{\mathrm{ctrl}}
\newcommand{\hp}{\mathchar`-}

\DeclarePairedDelimiterX\Set[1]\{\}{%
  
  #1
}

\newcommand{\func}[1]{{\ensuremath{\mathsf{#1}}}}
\newcommand{\class}[1]{{\ensuremath{\mathsf{#1}}}}
\newcommand{\lang}[1]{{\ensuremath{\mathsf{#1}}}}

\newcommand{\poly}{\func{poly}}

\newcommand{\expf}{\func{exp}}

\newcommand{\StDes}{\operatorname{\lang{StDes}}}
\newcommand{\UniDes}{\operatorname{\lang{UniDes}}}
\newcommand{\compSFP}{\operatorname{\lang{compSFP}}}
\newcommand{\compUFP}{\operatorname{\lang{compUFP}}}
\newcommand{\SFP}{\operatorname{\lang{SFP}}}
\newcommand{\UFP}{\operatorname{\lang{UFP}}}

\newcommand{\BPP}{\class{BPP}}
\newcommand{\BQP}{\class{BQP}}

\newcommand{\PP}{\class{PP}}
\newcommand{\PQP}{\class{PQP}}
\newcommand{\FP}{\class{FP}}
\newcommand{\sharpP}{{\#\class{P}}}

\usepackage{etoolbox}
\DeclarePairedDelimiterX{\abs}[1]{\lvert}{\rvert}{%
  \ifblank{#1}{\,\cdot\,}{#1}
}   % absolute value

\DeclarePairedDelimiterX\norm[1]\lVert\rVert{%
  \ifblank{#1}{\,\cdot\,}{#1}
}   % norm

\newcommand{\CC}{\mathbb{C}}% complex numbers
\newcommand{\ZZ}{\mathbb{Z}}% integers
\newcommand{\cA}{{\mathcal{A}}}

\newcommand{\cE}{{\mathcal{E}}}

\newcommand{\cG}{{\mathcal{G}}}
\newcommand{\cH}{{\mathcal{H}}}

\newcommand{\cO}{{\mathcal{O}}}

\newcommand{\cT}{{\mathcal{T}}}

\newcommand{\cX}{{\mathcal{X}}}

\newcommand{\sfU}{\textsf{U}}
\newcommand{\sfV}{\textsf{V}}
\newcommand{\sfS}{\textsf{S}}
\newcommand{\sfD}{\textsf{D}}
\newcommand{\Haar}{\textsf{H}}

\newcommand{\tr}[1]{\mathrm{Tr}\bigl[ #1 \bigr]}

\def\lsim{\mathrel{\rlap{\lower4pt\hbox{\hskip1pt$\sim$}}
		\raise1pt\hbox{$<$}}}                % less than or approx. symbol
\def\gsim{\mathrel{\rlap{\lower4pt\hbox{\hskip1pt$\sim$}}
		\raise1pt\hbox{$>$}}}                % greater than or approx. symbol

\newcommand{\argdot}{{\,\cdot\,}} % for a dot as an argument

\definecolor{martin}{rgb}{0,.4,1}

\begin{document}

\begin{titlepage}
\thispagestyle{empty}

\begin{flushright}
YITP-24-141
\end{flushright}

\title{On Computational Complexity of Unitary and State Design Properties}
	
\author{Yoshifumi Nakata}
\affiliation{Yukawa Institute for Theoretical Physics, Kyoto university, Kitashirakawa Oiwakecho, Sakyo-ku, Kyoto, 606-8502, Japan}
\author{Yuki Takeuchi}
\thanks{The current affiliation is Information Technology R\&D Center, Mitsubishi Electric Corporation.}
\affiliation{NTT Communication Science Laboratories, NTT Corporation, 3-1 Morinosato Wakamiya, Atsugi, Kanagawa 243-0198, Japan}
\affiliation{NTT Research Center for Theoretical Quantum Information, NTT Corporation, 3-1 Morinosato Wakamiya, Atsugi, Kanagawa 243-0198, Japan}
\author{Martin Kliesch}
\affiliation{Institute for Quantum Inspired and Quantum Optimization, Hamburg University of Technology, Germany}
\author{Andrew Darmawan}
\affiliation{Yukawa Institute for Theoretical Physics, Kyoto university, Kitashirakawa Oiwakecho, Sakyo-ku, Kyoto, 606-8502, Japan}

%\date{\today}
	
\begin{abstract}
We investigate unitary and state $t$-designs from a computational complexity perspective.
First, we address the problems of computing frame potentials that characterize (approximate) $t$-designs. 
We present a quantum algorithm for computing frame potentials and establish the following: (1) exact computation can be achieved by a single query to a $\sharpP$-oracle and is $\sharpP$-hard; (2) for state vectors, deciding whether the frame potential is larger than or smaller than certain values is $\BQP$-complete, provided the promise gap between the two values is inverse-polynomial in the number of qubits; and (3) for both state vectors and unitaries, this promise problem is $\PP$-complete if the promise gap is exponentially small. 
Second, we address the promise problem of deciding whether or not a given set is a good approximation to a design. Given a certain promise gap that could be constant, we show that this problem is $\PP$-hard, highlighting the inherent computational difficulty of determining properties of unitary and state designs.
We further identify implications of our results, including variational methods for constructing designs, diagnosing quantum chaos, and exploring emergent designs in Hamiltonian systems.
\end{abstract}

\maketitle
\end{titlepage}
%%%========== PDF meta data =====================
 \hypersetup{
	     pdfsubject = {Quantum complexity theory},
	     pdfkeywords = {random quantum circuits, unitary, projective, spherical, designs, frame potential, quantum, randomness, OTOC, complexity, class, PP, BQP, sharpP, PQP, random state, random unitary, quantum pseudorandomness},
	    }
%%%=============================================

%%% ==============================================
\section{Introduction}
%%% ==============================================

Quantum randomness is a fundamental resource in quantum information processing. It is commonly formulated by a \emph{Haar random unitary}, a unitary drawn uniformly at random with respect to the uniform probability measure on the unitary group, i.e., the normalized Haar measure.
Applications of quantum randomness range from quantum cryptography~\cite{AS2004,HLSW2004,kretschmer:LIPIcs.TQC.2021.2, 10.1007/978-3-031-15802-5_10,10.1007/978-3-031-15802-5_8}, algorithms~\cite{S2005,BH2013,B2018, G2019,BFNV2019}, and sensing~\cite{KRT2014,KL15,KZD2016,OAGKAL2016}, to quantum communication~\cite{D2005,DW2004,GPW2005,ADHW2009,DBWR2010,SDTR2013,HOW2005,HOW2007,NWY2021,WN2023}. 
In theoretical physics, quantum randomness provides powerful tools to explore far-from-equilibrium dynamics, yielding novel insights into thermalization~\cite{PSW2006,LPSW2009,dRHRW2016,PhysRevA.101.042126,IH2022,BDP2023}, the holographic principle~\cite{HP2007,NWK2023,NMK2025}, and scrambling~\cite{SS2008,LSHOH2013,MSS2016,RY2017,PRXQuantum.4.010311,PhysRevX.14.041051,PhysRevX.14.041059}. Experimentally, it plays a key role in benchmarking noisy quantum devices~\cite{EAZ2005,KLRetc2008,MGE2011,MGE2012, PhysRevLett.112.240504, PhysRevA.93.012301, garion2020experimental, OWE2019,HROWE2022,Heinrich22GeneralGuarantees} and studying complex quantum many-body dynamics~\cite{Choi2023,MCSEC2023}.

A Haar random unitary is, however, computationally expensive as most unitaries require exponentially many quantum gates to implement~\cite{NC2010}.
This fact necessitates approximations of a Haar random unitary~\cite{10.1007/978-3-319-96878-0_5,MPSY2024}, among which a \emph{unitary $t$-design} is a commonly used formulation. A unitary $t$-design is a random unitary that has the same $t$-th moment as a Haar random one.
Similarly, a \emph{state $t$-design} is defined as a random state vector that has the same $t$-th moment as a \emph{Haar random state}, i.e., a state vector obtained by applying a Haar random unitary to a fixed state vector.
In many applications, unitary and state $t$-designs can be used instead of Haar random ones. 

Given their significance, implementations of designs have long been studied. While it has been known in combinatorial mathematics~\cite{DGS1975,DGS1977} that unitary and state $t$-designs exist for any $t$, their explicit constructions have been found only recently~\cite{BNOZ2022,NZOBSTHYZTTN2021}. 
In quantum information, approximate implementations have attracted substantial attention. A standard approach is to use quantum circuits on $n$ qubits~\cite{HL2009TPE,BHH2016,NHKW2017,OnoBueKli17,H2022,Harrow2023,Haferkamp2023,metger2024simpleconstructionslineardepthtdesigns,chen2024efficientunitarydesignspseudorandom,laracuente2024approximateunitarykdesignsshallow,schuster2024randomunitariesextremelylow}, and it has recently been shown that $\cO(t \log n/\delta)$ depth circuits suffice for generating the $\delta$-approximation of a unitary $t$-design~\cite{schuster2024randomunitariesextremelylow,laracuente2024approximateunitarykdesignsshallow}.
Another approach leverages the \emph{frame potential}, which achieves its minimum for $t$-designs, enabling variational constructions of designs~\cite{GAE2007}. Additionally, implementations of designs that exploit quantum many-body dynamics offer a more exotic method~\cite{PRXQuantum.4.010311,Choi2023,Chan_2024,mok2024optimalconversionclassicalquantum,HC2022, IH2023,PhysRevLett.131.250401},  further introducing novel frameworks to study complex quantum many-body systems, such as \emph{deep thermalization}~\cite{IH2022,BDP2023} and \emph{complete Hilbert space ergodicity}~\cite{PhysRevX.14.041051,PhysRevX.14.041059}.

In recent years, computation-theoretic properties of designs have also been studied since designs and related concepts~\cite{10.1007/978-3-319-96878-0_5, Choi2023} have proven to be extremely useful for, e.g., cryptography uses~\cite{kretschmer:LIPIcs.TQC.2021.2, 10.1007/978-3-031-15802-5_10,10.1007/978-3-031-15802-5_8}.
A fundamental question in this context is: \emph{Can one efficiently determine whether a given distribution constitutes a design?}
Since designs can be characterized by the frame potential, this question can be reframed as: \emph{Can one efficiently compute the frame potential?} Given the numerous applications of designs, these questions are not only foundational but also of practical importance.

In this paper, we formulate these questions as computational problems and initiate a computational complexity theoretic analysis of unitary and state design properties. 
We begin by providing a quantum algorithm for estimating the frame potential and show that, while the algorithm achieves $1/\poly(n)$ accuracy for state vectors in polynomial time, where $n$ is the number of qubits, achieving exponential accuracy or estimating the frame potential of unitaries takes exponential time. 

To check if more efficient algorithms exist, we introduce and investigate computational problems about the frame potential. First, we show that \emph{exactly} computing the frame potential can be achieved by a single query to a $\sharpP$ oracle and is $\sharpP$-hard. 
We then consider a problem of \emph{approximately} computing frame potentials. We formulate this as a promise problem to decide whether the frame potential of a given set is larger or smaller than certain values, with the promise that either one of the two is the case. For state vectors, this problem is shown to be $\BQP$-complete and $\PP$-complete if the promise gap between the two values is $\Theta(1/\poly(n))$ and $\Theta(2^{-\poly(n)})$, respectively. For unitaries, the problem is $\PP$-complete if the promise gap is $\Theta(2^{-\poly(n)})$.
These results imply that general algorithms, whether quantum or classical, for computing the frame potential with exponential accuracy in polynomial time are unlikely to exist, indicating that our quantum algorithm cannot be drastically improved.

We also propose and investigate the problems more directly related to designs, i.e., promise problems to decide whether a given set is a $\delta$-approximate $t$-design or not a $\delta'$-approximate $t'$-design, where $t\leq t'$, under the promise that only one of the two is the case. 
We show that this promise problem is in $\PP$ for $t=t'$ if the gap between $\delta'$ and $\delta$ is not too small. We also show that, for any $t \leq t'$, this problem is $\PP$-hard if $\delta' -\delta \in \Theta(2^{c(t'-t)n} \delta)$ for $t<t'$ and $\delta' -\delta \in \Theta(2^{-c n} \delta)$ for $t=t'$, where $c=1$ for a set of state vectors and $c=2$ for a set of unitaries. 
This result is particularly remarkable when $t'=t+1$. By setting $\delta$ as $\Theta(2^{-n})$ for state vectors and $\Theta(2^{-2n})$ for unitaries, our result implies that it is $\PP$-hard to decide if a given set is an exponentially-accurate approximation to a $t$-design or a worse-than-constant approximation to a $(t+1)$-design that is not an exponentially-accurate approximate $t$-design. This large gap in the approximation accuracy, in particular, demonstrates the computationally hard nature of checking unitary and state design properties.

Due to the broad generality of our complexity-theory-based approach and the universal significance of unitary and state designs, our results have the following immediate implications.
\begin{enumerate}
	\item \emph{Variational constructions of designs.} A promising application of the frame potentials is a variational construction of designs~\cite{GAE2007}. Our results suggest that variational methods with a general ansatz do not scale due to the absence of efficient algorithms for computing frame potentials, highlighting the crucial importance of choosing a good ansatz.
	\item \emph{Quantum chaos through OTOCs.} The frame potential of a set of unitaries is closely related to the out-of-time-ordered correlators (OTOCs)~\cite{RY2017}. Therefore, our hardness results for computing frame potentials directly apply to OTOCs. As OTOCs are used to diagnose quantum chaos~\cite{MSS2016}, this reveals the complex nature of quantum chaos from a computational viewpoint, establishing a connection between quantum many-body physics and complexity theory.
	\item \emph{Designs in quantum many-body systems.} Several lines of evidence suggest that sufficiently complex quantum many-body Hamiltonian dynamics are highly useful for generating a unitary or state design~\cite{PRXQuantum.4.010311,Choi2023,Chan_2024,mok2024optimalconversionclassicalquantum,HC2022, IH2023,PhysRevLett.131.250401}. Except a few solvable cases in the thermodynamic limit~\cite{HC2022, IH2023}, most studies rely on numerical simulations. Our results indicate that direct numerical simulations do not scale unless they are specifically tailored to the Hamiltonian dynamics being investigated.
\end{enumerate}

This paper is organized as follows.
We begin with preliminaries in~\cref{S:Preliminaries}, where we explain our notation and overview basic properties of unitary and state designs. Our results and their implications are all summarized in~\cref{S:MainResults}. Proofs are provided in~\cref{S:CompFPs,S:promiseFP,S:promiseDES}. 
After the conclusion in~\cref{S:conclusion}, we provide supplementary proofs in Appendices.

%%% ==============================================
\section{Preliminaries} \label{S:Preliminaries}
%%% ==============================================

We introduce our notation in~\cref{SS:Notation}. A  brief overview of unitary and state designs is provided in~\cref{SS:UnitaryandStateDesigns}. The computational complexity classes that we use are summarized in~\cref{SS:CpomplexityClasses}.

\subsection{Notation} \label{SS:Notation}

We use the Schatten $p$-norms for linear operators, which are defined by $\| X \|_p \coloneqq ( \tr{|X|^{p}} )^{1/p}$ ($p \in [1, \infty]$) with $|X| = \sqrt{X^{\dagger}X}$. 
We particularly use the trace ($p=1$), the Hilbert-Schmidt ($p=2$), and the spectral ($p = \infty$) norms. They satisfy
\begin{multline}
    \| O\|_{\infty} \leq  \| O\|_2 \leq \| O\|_1 \\
    \leq \sqrt{\textrm{rank}(O)} \| O\|_2  \leq \textrm{rank}(O) \|O\|_{\infty}. \label{Eq:Schatten}
\end{multline}
For superoperators, we use the diamond norm $\| \argdot \|_{\diamond}$, which is defined by $\| \cT \|_{\diamond} = \max_{n \geq 1} \| \cT \otimes {\rm id}_n \|_{1 \rightarrow 1}$, where ${\rm id}_n$ is the identity map on the $n$-dimensional Hilbert space, and $\| \cE\|_{1 \rightarrow 1} = \max_{X} \|\cE(X)\|_1/ \|X\|_1$ with the maximization over all non-zero operators $X$. 

Throughout the paper, we consider probability measures over multisets of state vectors or unitaries. An average is written as $\mathbb{E}_{\ket{\psi} \sim \mu}[\argdot]$ for the probability measure $\mu$ of state vectors, and as $\mathbb{E}_{U \sim \nu}[\argdot]$ for the probability measure $\nu$ of unitaries.
When a finite multiset of state vectors or unitaries is given, the uniform distribution over the multiset is naturally defined. Hence, with a slight abuse of notation, we use the same symbol for the multiset and the uniform distribution over the multiset. For instance, for a finite multiset of state vectors $\sfS = \{ \ket{\psi_j} \}_{j=1, \dots, K}$, we express its average such as $\mathbb{E}_{\ket{\psi} \sim \sfS}[\argdot] = 1/K \sum_{j=1}^K (\argdot)$. Hereafter, we refer to a multiset as a set unless it is necessary to clearly distinguish them. 

An important probability measure on the unitary group $\mathbb{U}(d)$ is the (normalized) \emph{Haar} measure $\Haar$. A unitary $U$ drawn uniformly at random with respect to the Haar measure $\Haar$ is called a Haar random unitary and is conventionally denoted by $U \sim \Haar$.

By applying a Haar random unitary to a fixed state vector $\ket{\varphi_0}$, a unitarily invariant probability measure on the set of all state vectors can be defined, which we call a \emph{Haar random state}. Due to the unitarily invariant property of a Haar random unitary, a Haar random state is independent of the choice of the fixed vector $\ket{\varphi_0}$. Similarly to a Haar random unitary, we use the notation $\ket{\varphi} \sim \Haar$ for a Haar random state.

\subsection{Unitary and state designs} \label{SS:UnitaryandStateDesigns}

We provide a brief overview of definitions and important properties of state and unitary designs.

\subsubsection{State designs}
In the original definition from design theory~\cite{DGS1975,DGS1977}, a state $t$-design ($t \in \ZZ^+$) is defined as the uniform distribution over a finite set (not a multiset) of state vectors that has the same $t$-th moment as a Haar random state. 
However, in quantum information, it is common to define a design as the uniform distribution over a finite multiset, which more closely aligns with the concept of a \emph{weighted} $t$-design in design theory. We follow the convention in quantum information and adopt the following definition of a state design.

\begin{definition}[$\delta$-approximate state $t$-design] \label{Def:statedesign}
    When the uniform probability measure over a finite multiset $\sfS$ of $d$-dimensional state vectors satisfies 
    \begin{equation}
        \Bigl\| \mathbb{E}_{\ket{\psi} \sim \sfS} \bigl[\ketbra{\psi}{\psi}^{\otimes t}\bigr] - \mathbb{E}_{\ket{\psi} \sim \Haar} \bigl[\ketbra{\psi}{\psi}^{\otimes t}\bigr] \Bigr\|_{\infty} \leq \frac{\delta}{d_t}, \label{Eq:DefStateDesign}
    \end{equation}
    it is called a $\delta$-approximate state $t$-design. Here, $d_t = \binom{d + t -1}{t}$ is the dimension of the symmetric subspace in $(\CC^d)^{\otimes t}$. 
    When $\delta = 0$, it is called an \emph{exact} state $t$-design.    
\end{definition}

The expectation of the second term in~\cref{Eq:DefStateDesign} can be explicitly spelt out using Schur-Weyl duality. Denoting by $\Pi_{\rm sym}$ the projection onto the symmetric subspace in $(\CC^d)^{\otimes t}$, we have
\begin{equation}
    \mathbb{E}_{\ket{\psi} \sim \Haar} \bigl[\ketbra{\psi}{\psi}^{\otimes t}\bigr]
    =
    \frac{\Pi_{\rm sym}}{d_t}. \label{Eq:symproj}
\end{equation}

In~\cref{Def:statedesign}, we measure the approximation precision by using the spectral norm following one of the original proposals~\cite{AE2007}. In recent years, the trace norm has also been used instead, to define a $\delta$-approximate state $t$-design by
\begin{equation}
	\Bigl\| \mathbb{E}_{\ket{\psi} \sim \sfS} \bigl[\ketbra{\psi}{\psi}^{\otimes t}\bigr] - \mathbb{E}_{\ket{\psi} \sim \Haar} \bigl[\ketbra{\psi}{\psi}^{\otimes t}\bigr] \Bigr\|_{1} \leq \delta. \label{Eq:DefStateDesign1norm}
\end{equation}
This definition is weaker than~\cref{Def:statedesign} in the sense that~\cref{Eq:DefStateDesign} implies~\cref{Eq:DefStateDesign1norm}, which is due to~\cref{Eq:Schatten} and the fact that the support of the operators is the symmetric subspace in $(\mathbb{C}^d)^{\otimes t}$. Hence, we use~\cref{Def:statedesign} in this paper.

A useful quantity that characterizes a state design is a \emph{state frame potential}.

\begin{definition}[State frame potential]
    Let $\mu$ be a probability measure on a multiset of state vectors. The state frame potential of degree $t$, $F_t(\mu)$, is defined as
    \begin{align}
        F_t(\mu) &\coloneqq \Bigl\| \mathbb{E}_{\ket{\psi} \sim  \mu} \bigl[ \ketbra{\psi}{\psi}^{\otimes t} \bigr] \Bigr\|_2^2\label{Eq:StFrame2norm}\\
        &= \mathbb{E}_{\ket{\psi}, \ket{\phi} \sim \mu} \Bigl[ \bigl| \braket{\psi}{\phi} \bigr|^{2t} \Bigr].
    \end{align}
\end{definition}

Important properties of the state frame potential are summarized in~\cref{Prop:StateFPBasicProperties}.
These properties are well-known, but for completeness, we provide a proof in~\cref{App:FramePotential}.
Note that the assumption $t \leq d$ in~\cref{Prop:StateFPBasicProperties} is for the sake of convenience. Even if it is not satisfied, similar statements hold.

\begin{restatable}{proposition}{StateFPBasicProperties} \label{Prop:StateFPBasicProperties}
    Let $d, t \in \ZZ^+$ be such that $t \leq d$, and let $d_t = \binom{d+t-1}{t}$. The following hold for the state frame potential of degree $t$:
    \begin{enumerate}
        \item $F_t(\Haar) = 1/d_t$. \label{item:propertyStateHaar}
        \item \label{item:propertyFHleqFS}
        For any probability measure $\mu$ on a non-empty set of state vectors, 
        \begin{equation}
            F_t(\mu) \in \bigl[1/d_t,\, 1\bigr],    
        \end{equation}
        where equality of the lower bound holds if and only if it is an exact state $t$-design. For the uniform measure on a finite set $\sfS$ with cardinality $K$, 
        \begin{equation}
            F_t(\sfS) \in \bigl[\max\bigl\{  1/d_t,\,  1/K \bigr\},\, 1\bigr].
        \end{equation}
        \item \label{item:property:FSbound}
        If a finite set $\sfS$ of state vectors is a $\delta$-approximate state $t$-design, then
        \begin{equation}
            F_t(\sfS) \leq \frac{1}{d_t} + \frac{\delta^2}{d_t}.
        \end{equation}
        \item \label{item:property:approxDesign}
        If a finite set $\sfS$ of state vectors satisfies 
        \begin{equation}
            F_t(\sfS) \leq \frac{1}{d_t} + \frac{\delta^2}{d_t^2},    
        \end{equation}
        then $\sfS$ is a $\delta$-approximate state $t$-design.
        \item \label{item:property:K}
        For a finite set $\sfS$ of state vectors to be a $\delta$-approximate state $t$-design, it has to 
        contain at least $K_{\rm st}(d, t, \delta)$ state vectors, where
        \begin{equation}
            K_{\rm st}(d, t, \delta) \coloneqq \frac{d_t}{1 + \delta^2}.
        \end{equation}
    \end{enumerate}
\end{restatable}

From the property~\ref{item:propertyFHleqFS} in~\cref{Prop:StateFPBasicProperties}, one observes that the frame potential is indeed a potential in the sense that a state $t$-design can be obtained by minimizing the frame potential of degree $t$.
This is extended to approximate state designs by the properties~\ref{item:property:FSbound} and~\ref{item:property:approxDesign}. 
However, if the frame potential $F_t(\sfS)$ takes the value such as
\begin{equation}
    \frac{1}{d_t} + \frac{\delta^2}{d_t^2} < F_t(\sfS) < \frac{1}{d_t} + \frac{\delta^2}{d_t},
\end{equation}
we cannot conclude whether or not the set is a $\delta$-approximate unitary $t$-design.
The last property~\ref{item:property:K} implies that, if a set has small cardinality, it cannot be a state design. This is naturally expected as state designs should be able to mimic the uniform distribution over the set of all state vectors.

\subsubsection{Unitary designs}

To define a $\delta$-approximate unitary $t$-design, it is convenient to use the \emph{moment operator of degree $t$} of a probability measure of unitaries. 
So, let $\mu$ be a probability measure on the unitary group $\mathbb{U}(d)$. The moment operator is given by
\begin{equation}
    M^{(t)}_{\mu} \coloneqq \mathbb{E}_{U \sim \mu}[ U^{\otimes t} \otimes \bar{U}^{\otimes t}],
\end{equation}
where $\bar{\argdot}$ denotes the complex conjugate in a fixed orthonormal basis.

Based on the moment operator, a unitary $t$-design can be defined as follows.

\begin{definition}[$\delta$-approximate unitary $t$-design] \label{Def:AppUniDes}
     When the uniform probability measure over a finite multiset $\sfU = \{U_j\}_{j=1, \dots, K}$ satisfies
    \begin{equation}
        \bigl\| M^{(t)}_\sfU-M^{(t)}_\Haar \bigr\|_1 \leq \delta, \label{Eq:DefUnitaryDesign}
    \end{equation}
    it is called a \emph{$\delta$-approximate unitary $t$-design}. 
    When $\delta = 0$, it is called an \emph{exact} unitary $t$-design.    
\end{definition}

Similarly to state designs, Schur-Weyl duality allows us to identify the expectation in~\cref{Eq:DefUnitaryDesign} over the Haar measure. Let $\pi_t: U \in \mathbb{U}(d) \mapsto U^{\otimes t} \otimes \bar{U}^{\otimes t}$ be a representation of the unitary group $\mathbb{U}(d)$. Then, taking the expectation yields a projection onto trivial irreducible representations of $\pi_t$.

Approximate unitary designs can also be defined in terms of different norms to measure the approximation. One of the standard definitions is based on the diamond norm of superoperators. For a given probability measure $\mu$ on the unitary group $\mathbb{U}(d)$, let $\cG^{(t)}_{\mu}$ be the CPTP map given by 
\begin{equation}
	\cG^{(t)}_{\mu}(X) =  \mathbb{E}_{U \sim \mu}\bigl[ U^{\otimes t} X U^{\dagger \otimes t} \bigr], \label{Eq:DesSuperOp}
\end{equation}
where $X$ is an operator on a $d^t$-dimensional Hilbert space. A $\delta$-approximate unitary $t$-design in terms of the diamond norm is the uniform probability measure over a finite multiset $\textsf{U}$ that satisfies 
\begin{equation}
    \bigl\| \cG^{(t)}_{\textsf{U}} - \cG^{(t)}_{\Haar} \bigr\|_{\diamond} \leq \delta. \label{Eq:DiamondNorm}    
\end{equation}

While these two definitions are not exactly the same, using either one generally does not yield significant differences in applications of unitary designs. Indeed, if $\sfU$ is a $\delta$-approximate unitary $t$-design in one of the two definitions, then $\sfU$ is a $(\poly(d^t) \delta)$-approximate unitary $t$-design in the other definition. This is guaranteed by 
\begin{multline}
    d^{-2t}\bigl\| M^{(t)}_\sfU-M^{(t)}_\Haar \bigr\|_1 \leq \bigl\| \cG^{(t)}_{\sfU} - \cG^{(t)}_{\Haar}  \bigr\|_{\diamond} \\
    \leq  d^t \bigl\| M^{(t)}_\sfU-M^{(t)}_\Haar \bigr\|_1. \label{Eq:DifferencesDesignDef}
\end{multline}
For the derivation, see the comment after~\cref{Prop:UnitaryFPBasicProperties}.
As most circuit constructions of $\delta$-approximate unitary $t$-designs on $n$ qubits use $\poly(n, t, \log 1/\delta)$ gates~\cite{HL2009TPE,BHH2016,NHKW2017,OnoBueKli17,H2022,Harrow2023,Haferkamp2023,metger2024simpleconstructionslineardepthtdesigns,chen2024efficientunitarydesignspseudorandom,laracuente2024approximateunitarykdesignsshallow, schuster2024randomunitariesextremelylow}, the overhead of the number of gates due to the factor $\poly(d^t) = \poly(2^{tn})$ in the approximation accuracy $\delta$ is only linear in $t n$, which is not dominant in most cases.

It is also important that one can, in general, amplify the approximation accuracy simply by repeating the unitaries. That is, for any $\delta$-approximate unitary $t$-design $\mathsf{U} = \{U_j\}_j$ in terms of the moment operators (\cref{Def:AppUniDes}), $\mathsf{U}^m = \{U_{j_m} \dots U_{j_1}\}_{j_1, \dots, j_m}$ is a $(d^t \delta^m)$-approximate or even better unitary $t$-design in terms of the diamond norm. Hence, if one has an accurate approximate unitary $t$-design in terms of the moment operators, such as $\delta = d^{-1}$, it can be converted to an accurate approximate $t$-design in the diamond norm by a constant number of repetitions.

For these reasons, we mostly focus on the definition based on~\cref{Def:AppUniDes} in this work.

Unitary designs are also characterized by the unitary frame potential. With a slight abuse of notation, we use the same symbol $F_t$ as the state frame potential.

\begin{definition}[Unitary frame potential]
    Let $\mu$ be a probability measure of unitaries. The unitary frame potential of degree $t$ is defined as
    \begin{equation}
        F_t(\mu) \coloneqq \| M^{(t)}_{\mu} \|^2_2 = \mathbb{E}_{U,V \sim \mu}\Bigl[\bigl| \tr{U^{\dagger}V} \bigr|^{2t}\Bigr].
    \end{equation}    
\end{definition}

The unitary frame potential has similar properties to the state frame potential, which are summarized in~\cref{Prop:UnitaryFPBasicProperties}. The assumption $t \leq d$ in the proposition is for the sake of simplicity.
A proof is provided in~\cref{App:FramePotential} to be self-contained.

\begin{restatable}{proposition}{UnitaryFPBasicProperties} \label{Prop:UnitaryFPBasicProperties}
    Let $d, t \in \ZZ^+$ be such that $t \leq d$. The following hold for the unitary frame potential.
    \begin{enumerate}
        \item $F_t(\Haar) =  t!$. \label{item:uniFH}
        \item For any probability measure $\mu$ on a non-empty set in $\mathbb{U}(d)$, 
        \begin{equation}
            F_t(\mu) \in \bigl[t!, \, d^{2t}\bigr],    
        \end{equation}
        where equality of the lower bound holds if and only if it is an exact unitary $t$-design. For the uniform measure on a finite set $\sfU$ with cardinality $K$, 
        \begin{equation}
            F_t(\sfU) \in \bigl[\max\bigl\{  t!,\,  d^{2t}/K \bigr\},\,  d^{2t}\bigr].
        \end{equation}\label{item:uniFleq}
        \item If a finite set $\sfU$ of unitaries is a $\delta$-approximate unitary $t$-design, then
        \begin{equation}
            F_t(\sfU) \leq t! + \delta^2.
        \end{equation} \label{item:uniLB}
        \item If a finite set $\sfU$ of unitaries satisfies 
        \begin{equation}
            F_t(\sfU) \leq t! + \frac{\delta^2}{d^{2t}}, 
        \end{equation}
        then $\sfU$ is a $\delta$-approximate unitary $t$-design.\label{item:uniUB}
        \item For a finite set $\sfU$ of unitaries to be a $\delta$-approximate unitary $t$-design, it has to 
        contain at least $K_{\rm uni}(d, t, \delta)$ unitaries, where
        \begin{equation}
            K_{\rm uni}(d, t, \delta) \coloneqq \frac{d^{2t}}{t! + \delta^2}.
        \end{equation} \label{item:uniK}
        \item
        Let $\cG^{(t)}_{\mu}$ be the superoperator defined by~\cref{Eq:DesSuperOp} for a probability measure $\mu$ on $\mathbb{U}(d)$. For a finite set $\sfU$ of unitaries, it satisfies
        \begin{equation}
        \begin{aligned}
             d^{-t} \sqrt{F_t(\sfU) - F_t (\Haar)}  &\leq \bigl\| \cG^{(t)}_{\sfU} - \cG^{(t)}_{\Haar}  \bigr\|_{\diamond} 
             \\ &
             \leq d^t \sqrt{F_t(\sfU) - F_t (\Haar)}. 
        \end{aligned} 
        \end{equation}
        \label{item:DiamondFramePotential}
    \end{enumerate}
\end{restatable}

The last property~\ref{item:DiamondFramePotential} implies~\cref{Eq:DifferencesDesignDef} by using the fact that
\begin{equation}
    \bigl\| M^{(t)}_\sfU-M^{(t)}_\Haar \bigr\|_2 = \sqrt{F_t(\sfU) - F_t (\Haar)},
\end{equation}
and~\cref{Eq:Schatten}.

\subsection{Complexity classes} \label{SS:CpomplexityClasses}
The standard idea for characterizing the difficulty of a computational problem is to identify which complexity class the problem belongs to and to show that it is among the most difficult problems in that class. 
We briefly review the complexity classes $\sharpP$, $\PP$, $\PQP$, $\BQP$, and $\FP^\sharpP$ which are important for this work. 

The class $\sharpP$ is a set of all problems of counting the number of accepting paths of a non-deterministic polynomial-time Turing machine. 
By definition, the output of a $\sharpP$-problem is a non-negative integer. 

We take the common view in quantum information theory that the classes $\PP$, $\PQP$, and $\BQP$ are \emph{promise problems} \cite{Goldreich06OnPromiseProblems,Watrous2009}. 
For such a problem the answer is either true or false but has to be given (correctly) only for certain problem instances, which are specified by a promise. That is, each problem comes along with a potential promise and only has to be solved when it is fulfilled.

The class $\PP$ is the set of promise problems for each of which some classical algorithm with polynomially bounded runtime
(more formally, a probabilistic polynomial-time Turing machine) 
can output their correct answer with probability strictly larger than $1/2$. 
Since the probability may be exponentially close to $1/2$, $\PP$ contains the problems that are unlikely to be solvable by classical algorithms in polynomial time.
The class $\PQP$ is a quantum analogue of $\PP$, in which the probabilistic polynomial time Turing machine is replaced with a polynomial-time quantum algorithm. 
The class $\BQP$ is defined similarly to $\PQP$ but with a higher success probability of $\geq 2/3$. 

Obviously, $\BQP \subseteq \PQP$ and $\PP \subseteq \PQP$. It is also known that $\PQP \subseteq \PP$~\cite{Watrous2009}. Hence, $\BQP \subseteq \PQP=\PP$. 
It is expected  that $\BQP\neq\PQP$, i.e., $\PQP = \PP$ is expected to contain problems that are intractable even for scalable universal quantum computers.

We will rely on the concepts of hard and complete problems. 
Informally, a problem is \emph{hard} for some class if it is at least as difficult as the most difficult problem of that class in the sense that a solver for that problem can be used to solve any other problem of the class with some ``small overhead''. 
Technically, hardness is always defined with respect to a specific type of reduction, which depends on the class in question. 
If a problem is hard for some class and, at the same time, a member of that class, then it is \emph{complete} for that class.

Particularly important complete problems in this work are $\lang{\# SAT}$ and $\textsf{MAJ-SAT}$, which are $\sharpP$-complete and $\PP$-complete problems, respectively.

\begin{definition}[$\lang{\# SAT}$]
    Given the input as a classically efficiently computable function $f:\{0,1\}^N\rightarrow\{0,1\}$ specified by a Boolean circuit, output the number of $x\in\{0,1\}^N$ satisfying $f(x)=1$.    
\end{definition}
Note that this problem $\lang{\# SAT}$ is equivalent to calculating the number of satisfying assignments of a Boolean formula, which is the origin of its name.

\begin{definition}[\textsf{MAJ-SAT}~\cite{Gill1974}]
    Let $f: \{0,1\}^m \rightarrow \{0,1\}$ be a Boolean function that is computable in classical polynomial time, and let $s(f)$ be $\sum_{x\in\{0,1\}^{m}}f(x)$.
    Decide whether $0\leq s(f)<2^{m-1}$ or $2^{m-1} \leq s(f) \leq 2^{m}$.
\end{definition}

\begin{theorem}[See, e.g.,~\cite{cox2018overviewsemanticsyntacticcomplexity}] \label{Thm:MAJSAT}
    $\lang{\# SAT}$ is $\sharpP$-complete, and \textsf{MAJ-SAT} is $\PP$-complete.
\end{theorem}

Finally, $\FP^\sharpP$ is the class of functions that can be computed by a deterministic algorithm in polynomial time with additional query access to a solver for $\lang{\# SAT}$.
That is, once a part of the algorithm has specified an instance of $\lang{\# SAT}$, its solution can be obtained in a single computation step.

%%% ==============================================
\section{Main results} \label{S:MainResults}
%%% ==============================================

In this section, we provide our main results about complexities of various computational problems. 
We begin with a condition that must be satisfied by the sets of states or unitaries for our complexity  results to hold, and then summarize our results.
A summary of our complexity results is also given in~\cref{Tab:SummaryComplexity}.
The proofs will be provided in later sections. As usual, if necessary, complex numbers are represented by floating point numbers with finite precision controlled by the number of digits.

\begin{table*}[tb!]
    \centering
    \caption{A summary of the problems that we introduce in this paper and our results about their computational complexity. 
    As our main focus is on state and unitary designs, we have assumed here that the cardinality of the set is at least $K_{\rm st}(2^n, t, \delta)$ for a set of state vectors and $K_{\rm uni}(2^n, t, \delta)$ for a set of unitaries. See the respective
    theorems for the cases that do not satisfy these assumptions.
    } \label{Tab:SummaryComplexity}
    \begin{tabular}{ |c|c||c|c|c|c|  }
    \hline
     \multicolumn{2}{|c||}{} & Compute FP & \multicolumn{2}{|c|}{Decide: FP $\geq \alpha$ vs.\ FP $\leq \alpha-\epsilon$} & \begin{tabular}{c}Decide: a $\delta$-approx. $t$-design \\
     or not a $\delta'$-approx. $t'$-design.\end{tabular}\\
    \hline \hline
    \begin{tabular}{c} State \\ vectors\end{tabular} & Problem & \begin{tabular}{c}$\compSFP(t)$ \\ (\cref{Prob:compSFP}) \end{tabular}  & \multicolumn{2}{|c|}{\begin{tabular}{c}$\SFP(t,\epsilon)$ \\ (\cref{Prob:SFP})\end{tabular}}  & \begin{tabular}{c}$\StDes(t, t', \delta, \delta')$ \\ (\cref{Prob:StDes})\end{tabular} \\ \cline{2-6} 
    & Complexity  & \begin{tabular}{c}$\FP^{\sharpP} \cap \sharpP$-hard \\ (\cref{Thm:compSFPin,Thm:compSFPhard}) \end{tabular}& \begin{tabular}{c} $\BQP$-complete for \\ $\epsilon \in \Omega(1/\poly(n))$ \\ (\cref{Thm:SFPBQPin,Thm:SFPBQPhard})\end{tabular} &  \begin{tabular}{c} $\PP$-complete for\\ $\epsilon \in \Theta(2^{-\poly(n)})$ \\ (\cref{Thm:approxSFPexpin,Thm:approxSFPexphard}) \end{tabular} & \begin{tabular}{c}   $\PP$ for $t=t'$ and suitable $\delta$ and $\delta'$.
    \\ $\PP$-hard if\\
    $\delta' - \delta \in \begin{cases} \Theta(2^{(t'-t)n}\delta) & \text{for $t<t'$,}\\ \Theta(2^{-n}\delta)  & \text{for $t=t'$}. \end{cases}$ \\
    (\cref{Thm:StDESin,Thm:StDEShard})\end{tabular}\\ \hline
    Unitaries & Problem  &  \begin{tabular}{c}$\compUFP(t)$ \\ (\cref{Prob:compUFP}) \end{tabular}  & \multicolumn{2}{|c|}{\begin{tabular}{c}$\UFP(t,\epsilon)$ \\ (\cref{Prob:UFP}) \end{tabular}}  & \begin{tabular}{c}$\UniDes(t, t',\delta, \delta')$ \\ (\cref{Prob:UniDes}) \end{tabular}\\ \cline{2-6}
    &  Complexity & \begin{tabular}{c}$\FP^{\sharpP} \cap \sharpP$-hard \\ (\cref{Thm:compUFPin,Thm:compUFPhard}) \end{tabular} & \multicolumn{2}{|c|}{\begin{tabular}{c}$\PP$-complete for $\epsilon \in \Theta(2^{-\poly(n)})$ \\ (\cref{Thm:approxUFPin,Thm:approxUFPhard}) \end{tabular} }  &  \begin{tabular}{c}  
    $\PP$ for $t=t'$ and suitable $\delta$ and $\delta'$.
    \\ $\PP$-hard if\\
    $\delta' - \delta \in \begin{cases} \Theta(2^{2(t'-t)n}\delta) & \text{for $t<t'$,}\\ \Theta(2^{-2n}\delta)  & \text{for $t=t'$}. \end{cases}$ \\ (\cref{Thm:UniDESin,Thm:UniDEShard})\end{tabular} \\
    \hline
    \end{tabular}
\end{table*}

\subsection{Data format for sets of states and unitaries}
When we consider computational problems related to unitary or state designs, it is natural to suppose that the input of the problem is a set of unitaries acting on $\CC^d$ or states in $\CC^d$.
If a set consists of $\poly(d)$ elements, such problems may be tractable in polynomial time with respect to $d$. However, in a quantum setting, it is standard to start with quantum circuits on $n=\log_2 d$ qubits consisting of $\poly(n)$ gates and to ask about the efficiency in terms of $n$. Throughout the paper, we focus on such situations.

When concerned with designs on $n$ qubits, we need to deal with the set consisting of $\Omega(\exp(n))$ elements. In fact, the property~\ref{item:property:K} in~\cref{Prop:StateFPBasicProperties} and the property~\ref{item:uniK} in~\cref{Prop:UnitaryFPBasicProperties} imply that a set $\sfS$ of $n$-qubit state vectors with cardinality $|\sfS| < K_{\rm st}(2^n, t, \delta)$ is not a $\delta$-approximate state $t$-design, and that a set $\sfU$ of $n$-qubit unitaries with cardinality 
$|\sfU| < K_{\rm uni}(2^n, t, \delta)$ 
is not a $\delta$-approximate unitary $t$-design. Note that both $K_{\rm st}(2^n, t, \delta)$ and $K_{\rm uni}(2^n, t, \delta)$ are exponentially large in $n$.

To efficiently handle a set with an exponentially large cardinality, the set should be well structured. We formalize this by introducing \emph{a set of unitaries with a computationally efficient description}.

\begin{definition}[Computationally efficient descriptions]
    A multiset $\mathsf U = \Set{U_j}_{j=1}^K \subseteq \mathbb U(2^n)$ of $n$-qubit unitaries, where $K \in \cO(2^{\poly(n)})$, is said to have a \emph{computationally efficient description}, if it is specified by a partial function $f$ defined on $\Set{0,1}^m$ with $m\coloneqq\lceil\log_2 K\rceil$, which is computable in classical polynomial time, that outputs a classical description $f(j)$ ($j \in \Set{0,1}^{m}$) of a $\poly(n)$-sized quantum circuit of $U_j$. 
\end{definition}

Trivial instances of a set of unitaries with a computationally efficient description are a set of $n$-qubit Pauli operators and that of $n$-qubit Clifford unitaries. 
The former can be specified by $2n$ bits $x \in \{0, 1\}^{2n}$, and the description $f(x)$ specifies a tensor product of single-qubit Pauli unitaries, which is a quantum circuit with depth $1$.
The latter can be similarly specified by $\cO(\poly(n))$ bits with a suitable efficiently computable function $f$ that identifies a Clifford circuit of circuit size $\cO(n^2/\log(n))$, which is sufficient to represent any Clifford unitary \cite{Aaronson2004ImprovedSimulationOf}. 

A set of of $n$-qubit unitaries appearing in \emph{1-dimensional local random quantum circuit (1D LRC)} with $\cO(\poly(n))$ depth is also an important instance of a set with a computationally efficient description. A 1D LRC repeats two layers of two-qubit unitaries organized in a brickwork fashion. 
In the first layer, $2$-qubit gates, each randomly and independently drawn from a fixed gate set $\cG$, are applied to odd-even qubits in parallel. 
Similarly, in the second layer, random and independent $2$-qubit gates from $\cG$ are applied to even-odd qubits.
To specify each quantum circuit generated in this way, we need $\cO( \{\mathrm{\#\  of\  gates}\} \log_2 |\cG|)$ bits, where $|\cG|$ is the cardinality of the gate set $\cG$.
The number of gates is $\cO(\poly(n))$ whenever the depth is $\cO(\poly(n))$. 
In a standard case, $|\cG|$ is constant, but even if $|\cG| \in \cO(\expf(n))$, each quantum circuit generated in this way can be specified by $\poly(n)$ bits (in the latter case, we need to assume that each gate in $\cG$ has a constant-size description). 
This is also the case for local random quantum circuits on any graph in which the graph describes the two qubits to which a $2$-qubit gate is applied.

Other important sets of unitaries are those generated by Hamiltonian dynamics. When a Hamiltonian $H$ satisfies certain conditions, it is known that the corresponding time evolution $e^{-itH}$ can be simulated efficiently in terms of the time $t$ by quantum circuits with small error (see e.g.~\cite{PhysRevX.11.011020}).
Hence, by sampling discrete times $t_j \in [0, \poly(n)]$ ($j=1 \dots, K$ with $K \in \Theta(2^{\poly(n)})$) in an efficiently computable manner, we can define a set $\{ e^{- i H t_j}\}_{j=1}^K$ of unitaries with computationally efficient descriptions. 
This type of set of unitaries may be of particular interest in numerical or experimental approaches to the emergent designs in quantum many-body systems~\cite{PRXQuantum.4.010311,Choi2023,Chan_2024,mok2024optimalconversionclassicalquantum,HC2022, IH2023,PhysRevLett.131.250401}. Note that, in numerical or experimental settings, continuous degrees of freedom typically must be discretized and, hence, it is natural to consider discrete choices of time. Instead of discrete times, one may also consider a set of Hamiltonians $\{H_j\}_{j=1}^K$ and may define $\{ e^{- i H_j t}\}_{j=1}^K$ at a fixed time $t \in \cO(\poly(n))$. This set also has a computationally efficient description as far as the Hamiltonian dynamics can be simulated efficiently by quantum circuits.

We also introduce a \emph{set of state vectors with a computationally efficient description}. It is a set of state vectors generated by applying a set of unitaries with a computationally efficient description to a fixed initial state vector that can be prepared in a quantum polynomial time, such as a computational-basis state.

\subsection{Quantum algorithm for computing frame potentials} \label{SS:QAFP}

\cref{Prop:StateFPBasicProperties,Prop:UnitaryFPBasicProperties} imply that we can check if a given set of unitaries or states is a unitary or state design by computing their frame potentials.
This motivates us to first provide quantum algorithms for estimating the unitary and state frame potentials.
The idea behind the quantum algorithm comes from the fact that the frame potentials are defined by a sum of inner products between state vectors or unitaries. By `encoding' all the inner products into a single quantum state, we can compute the frame potential. 

\begin{figure}[tb!] 
  \centering
  \includegraphics[width=.4\textwidth,clip]{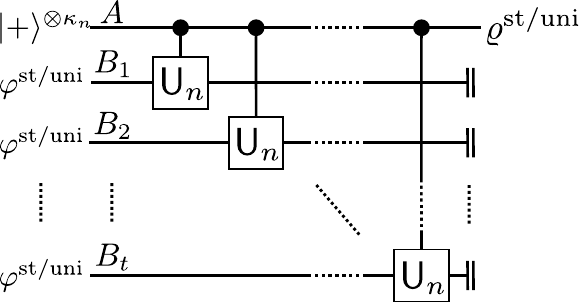}
  \caption{A key quantum circuit for estimating the state or unitary frame potentials of degree $t$. The system $A$ consists of $\kappa_n = \log K_n$ qubits, initially prepared in $\ket{+}^{\otimes \kappa_n}$. Each $B_m$ ($m=1, \dots, t$) is an $n$-qubit system and is initially prepared in $\varphi^{\rm st} \coloneqq \ketbra 0{0}^{\otimes n}$
  and $\varphi^{\rm uni} \coloneqq I/2^n$ for computing the state and unitary frame potentials, respectively. Each multi-qubit-controlled unitary applies $U_j$ onto $B_m$ when $A$ is in $\ket{j}$. The value of the state or unitary frame potential can be obtained from the purity of the output state $\varrho^{\rm st/uni}$ on $A$. 
  }
  \label{Fig:QC}
\end{figure}

In~\cref{Fig:QC}, we provide a key quantum circuit of our quantum algorithm, which is based on a multi-qubit controlled unitary $\ctrl \hp \sfU_n$ constructed from the set $\sfU_n$ of unitaries to be investigated. In the case of state vectors, $\sfU_n$ is a set of unitaries that generates the state vectors. 
Let us denote $\sfU_n=\{ U_{j} \}_{j=1,\dots,K_n}$, and let $A$ be a $\kappa_n$-qubit system ($\kappa_n \coloneqq \log K_n$), and $B_m$ be an $n$-qubit system ($m = 1,\dots, t$).  
We define the controlled unitary $\ctrl^A \hp \sfU_n^{B_m}$ as
\begin{equation}
    \ctrl^A\hp\sfU_n^{B_m} \coloneqq \sum_{j=1}^{K_n} \ketbra{j}{j}^A \otimes U_j^{B_m}.
\end{equation}
Note that when $\sfU_n$ has a computationally efficient description, the controlled unitary $\ctrl \hp \sfU_n$ can be efficiently implemented by quantum circuits obtained from classically controlling the Boolean function $f$ that classically specifies the $\poly(n)$-sized quantum circuit.
Denoting $B_1 \dots B_t$ simply by $B$, which is in total $(tn)$-qubit system, we define
\begin{align}
    \mathrm{ctrl}^A\mathchar`-\sfU_n^{B} &\coloneqq \prod_{m=1}^t \mathrm{ctrl}^A\mathchar`-\sfU_n^{B_m}. \label{Eq:ctrlU}
\end{align}

Denoting $\ket{+}^{\otimes \kappa_n}$ by $\ket{+_{\kappa_n}}$, $\ket{0}^{\otimes n}$ by $\ket{0_n}$, and a maximally entangled state between $B_m$ and $R_m$ by $\ket{\Phi_n}^{B_mR_m}$, we introduce
\begin{align}
    &\ket{\varrho_{\rm st}}^{AB} 
    = \mathrm{ctrl}^A\mathchar`-\sfU_n^{B} \left(\ket{+_{\kappa_n}}^{A} \bigotimes_{m=1}^t \ket{0_n}^{B_m} \right), \label{Eq:defvarrhost}\\
    &\ket{\varrho_{\rm uni}}^{ABR} = 
    \mathrm{ctrl}^A\mathchar`-\sfU^{B}_n \left(\ket{+_{\kappa_n}}^{A} \bigotimes_{m=1}^t \ket{\Phi_n}^{B_mR_m} \right). \label{Eq:StateUFPEncoded}
\end{align}
Note that the former is the output state of the circuit in~\cref{Fig:QC} when the initial states in $B_m$ are $\varphi^{\textrm{st}}$, and the latter is a purification of the output state of the circuit in~\cref{Fig:QC}, when the initial states in $B_m$ are $\varphi^{\textrm{uni}}$, where $R = R_1 \dots R_t$ is the purifying system.

We then have the following lemma about the frame potentials.

\begin{restatable}{lemma}{FPencodedState} \label{Lemma:FPencodedState}        
    Let $\varrho_{\rm st}^A$ and $\varrho_{\rm uni}^A$ be the reduced density matrix in $A$ of $\ket{\varrho_{\rm st}}^{AB}$ and $\ket{\varrho_{\rm uni}}^{ABR}$, respectively. Then, the state frame potential of $\sfS_n = \{ U_j \ket{0_n} \}_{j=1, \dots, K_n}$ and the unitary frame potential of $\sfU_n$ are given by
    \begin{align}
        &F_t(\sfS_n) = {\rm Tr} \bigl[ ( \varrho_{\rm st}^A)^2 \bigr], 
        &F_t(\sfU_n) = 2^{2nt} {\rm Tr} \bigl[ ( \varrho_{\rm uni}^A)^2 \bigr], \label{Eq:ar;ogknaerge3ar}
    \end{align}
    respectively.
\end{restatable}

\cref{Lemma:FPencodedState} implies that the frame potentials can be evaluated by the purities of the corresponding quantum states, which can be achieved by, e.g., the swap test or an amplitude estimation algorithm. This leads to the following theorem.

\begin{restatable}{theorem}{QAforFP} \label{Thm:QA_FP}
    Let $\sfU_n$ be a set of $n$-qubit unitaries with a computationally efficient description, the cardinality of which is $K_n$, and let $\sfS_n$ be a set of $n$-qubit state vectors with a computationally efficient description that is generated by applying the $\sfU_n$ to a computational zero state $\ket{0_n}$. The following hold:
    \begin{itemize}
        \item For $\eta \in (\max\{ d_t^{-1}, K_n^{-1} \} , 1]$ and $\epsilon \in (0, \eta- K_n^{-1})$, it takes  $tM_{\rm st}$ queries to $\mathrm{ctrl}\mathchar`-\sfU_n$ and its inverse to determine whether $F_t(\sfS_n) \geq \eta $ or $F_t(\sfS_n) \leq \eta - \epsilon$ with success probability at least $2/3$, where 
        \begin{equation}
            M_{\rm st} \coloneqq \cO\biggl(\frac{1}{\epsilon} \min \bigl\{ \sqrt{\eta K_n - 1}, 1 \bigr\}  \biggr).
        \end{equation}
        \item For $\eta \in (\max\{ t!, 2^{2nt} K_n^{-1} \} , 2^{2nt}]$ and $\epsilon \in (0, \eta-2^{2nt} K_n^{-1})$, it takes  $tM_{\rm uni}$ queries to $\mathrm{ctrl}\mathchar`-\sfU_n$ and its inverse to determine whether $F_t(\sfU_n) \geq \eta $ or $F_t(\sfU_n) \leq \eta - \epsilon$ with success probability at least $2/3$, where 
        \begin{equation}
            M_{\rm uni} \coloneqq \cO\biggl(\frac{2^{nt}}{\epsilon} \min \bigl\{ \sqrt{\eta K_n - 2^{2nt}}, 2^{nt} \bigr\}  \biggr). \label{Eq:Muni}
        \end{equation}
    \end{itemize}
\end{restatable}

The details of~\cref{Lemma:FPencodedState} and~\cref{Thm:QA_FP} are provided in~\cref{S:QAforFP}.

For a set of state vectors, the complexity of this quantum algorithm is determined merely by the required accuracy $\epsilon$. This is because the algorithm uses $\kappa_n = \log K_n$ ancillary qubits and $t M_{\rm st}$ queries to $\ctrl$-\sfU, both of which are efficient if a set of state vectors has a computationally efficient description. Also, $M_{\rm st}$ is at most $\cO(\epsilon^{-1})$. Thus, one can estimate the state frame potential with $1/\poly(n)$ accuracy in $\poly(n)$ time for sets of $n$-qubit state vectors with computationally efficient descriptions. 
On the other hand, the algorithm fails to estimate the state frame potential with $1/\exp(n)$ accuracy in polynomial time. We will see below that improving this algorithm is unlikely to be possible.

In contrast, this algorithm fails to compute the unitary frame potential in polynomial time even for, e.g., constant accuracy. This is merely due to the factor $2^{nt}$ in~\cref{Eq:Muni}, which can be further traced back to the fact that $\tr{(\varrho^{\rm uni})^2}$ should be evaluated with at least exponential accuracy for estimating $F_t(\sfU)$ with constant accuracy (see~\cref{Eq:ar;ogknaerge3ar}).

\subsection{Complexity of computing frame potentials}

To address the question of whether the quantum algorithm can be improved, we formulate and investigate the problems of computing the frame potentials in a complexity-theoretic manner.
We first formulate the problem of computing the state frame potential as $\compSFP(t)$.

\begin{problem}[$\compSFP(t)$] \label{Prob:compSFP}
    \hfill\\ 
    \underline{Input:} 
    \begin{compactitem}
        \item a set $\sfS_n$ of state vectors of $n$ qubits with a computationally efficient description
        \item $t \in \ZZ^+$ (degree)
    \end{compactitem}
    \underline{Output:} the state frame potential $F_t( \sfS_n )$ of degree $t$.
\end{problem}

In the definition of $\compSFP(t)$, we do not put any assumption on the structure of $\sfS_n$. However, since our primary concern is a state $t$-design, we mainly deal with candidate sets of designs and do not consider any sets that are apparently not a state $t$-design. For instance, if the cardinality of a given set after removing the multiplicity is substantially smaller than $K_{\rm st}(2^n, t, \delta)$, the set is clearly not a $\delta$-approximate state $t$-design due to the property~\ref{item:property:K} in~\cref{Prop:StateFPBasicProperties}. Such sets are out of our interest.

In the definition of $\compSFP(t)$, we do not include these assumptions on $\sfS_n$ simply to keep the problem as minimal as possible. When necessary, we explicitly mention that we consider only candidate sets of designs. This remark applies to all computational problems that we introduce below.

We can show that, for any constant $t \in \ZZ^+$, $\compSFP(t)$ can be solved by using a single query to a $\sharpP$ oracle and is $\sharpP$-hard. See~\cref{SS:CompSFP} for the proofs.

\begin{restatable}{theorem}{compSFPin}\label{Thm:compSFPin}
    For any constant $t \in \ZZ^+$, $\compSFP(t)\in\FP^\sharpP$.
\end{restatable}

\begin{restatable}{theorem}{compSFPhard}\label{Thm:compSFPhard}
    For any constant $t \in \ZZ^+$, $\compSFP(t)$ is $\sharpP$-hard. This is the case even if the cardinality $K_n$ of the set satisfies $K_n \geq K_{\rm st}(2^n, t, 0)$.
\end{restatable}

From~\cref{Thm:compSFPin,Thm:compSFPhard}, we observe a gap between the inclusion and the hardness of $\compSFP(t)$. 
This is merely due to the fact that the answer to $\compSFP(t)$ is a non-negative \emph{real number}, but answers to the problems in $\sharpP$ are non-negative \emph{integers}.
This gap can be closed by restricting the class of state vectors and by considering a normalized frame potential. The corresponding computational problem, defined similarly to $\compSFP(t)$, can be shown to be in $\sharpP$ for any constant $t \in \ZZ^+$. See~\cref{Thm:compSFPin2} in~\cref{S:CompFPs} for the complete statement.\\

We next consider the problem of computing the unitary frame potential $\compUFP(t)$.
When one is concerned with unitary designs, one may additionally put assumptions on the set.

\begin{problem}[{$\compUFP(t)$}] \label{Prob:compUFP}
\hfill\\ 
\underline{Input:} 
    \begin{compactitem}
        \item a set $\sfU_n$ of unitaries on $n$ qubits with a computationally efficient description
        \item $t \in \ZZ^+$ (degree)
    \end{compactitem}
\underline{Output:} the unitary frame potential $F_t( \sfU_n )$.     
\end{problem}

We obtain the results similar to those for $\compSFP(t)$. The proofs are given in~\cref{SS:CompUFP}.

\begin{restatable}{theorem}{compUFPin} \label{Thm:compUFPin}
    For any constant $t \in \ZZ^+$, $\compUFP(t) \in\FP^\sharpP$.
\end{restatable}

\begin{restatable}{theorem}{compUFPhard}
    \label{Thm:compUFPhard}
    For any constant $t \in \ZZ^+, \compUFP(t)$ is $\sharpP$-hard. This is the case even if the cardinality $K_n$ of the set satisfies $K_n \geq K_{\rm uni}(2^n, t, 0)$.
\end{restatable}

\subsection{Promise problems on frame potentials}

From the results in the previous section, it follows that an efficient general algorithm, whether classical or quantum, for exactly computing frame potentials, is unlikely to exist. It is then natural to consider approximate computation. One approach to address this question is to formulate it as a promise problem.

We introduce the state frame potential problem $\SFP(t, \epsilon)$ as follows.

\begin{problem}[$\SFP(t,\epsilon)$]  \label{Prob:SFP}
\hfill\\
    \underline{Input:} 
    \begin{compactitem}
        \item a set $\sfS_n$ of $K_n$ $n$-qubit state vectors with a computationally efficient description
        \item $t \in \ZZ^+$ (degree)
        \item $\alpha > \beta \geq \max\{ 1/d_t,\, 1/K_n\}$ such that $\alpha-\beta>\epsilon>0$
    \end{compactitem}
    \underline{Promise:} the state frame potential $F_t(\sfS_n)$ is either $\geq \alpha$ or $\leq \beta$.\\
    \underline{Output:} decide which is the case.  
\end{problem}

From the definition of $\SFP(t, \epsilon)$ as a promise problem, the frame potential $F_t(\sfS_n)$ is guaranteed to either be $\geq \alpha$ or $\leq \beta$, where $\alpha > \beta$. The difference between $\alpha$ and $\beta$ is characterized by $\epsilon$, which is called a \emph{promise gap}.
Note that we have assumed $\beta \geq \max\{ 1/d_t,\, 1/K_n\}$ simply to avoid a trivial situation. See the property~\ref{item:propertyFHleqFS} in~\cref{Prop:StateFPBasicProperties}.

An algorithm for solving $\SFP(t,\epsilon)$ would allow us to approximately compute the frame potential by, e.g., a binary search with logarithmic overhead. 

In~\cref{SS:QAFP}, we have already provided a polynomial-time quantum algorithm for computing the state frame potential within the inverse-polynomial accuracy (see~\cref{Thm:QA_FP}). 
In terms of the computational complexity, this implies the following.

\begin{restatable}{theorem}{SFPBQPin} \label{Thm:SFPBQPin}
    For any constant $t \in \ZZ^+$, $\SFP(t, \epsilon) \in \BQP$ if $\epsilon \in \Omega(1/\poly(n))$.
\end{restatable}

We can further show that $\SFP(t, \epsilon)$ is $\BQP$-hard even if $\epsilon$ is constant. See~\cref{SS:SFPpoly} for the proof.

\begin{restatable}{theorem}{SFPBQPhard} \label{Thm:SFPBQPhard}
    For any constant $t\in\ZZ^+$, there exists a constant $\epsilon$ such that $\SFP(t,\epsilon)$ is $\BQP$-hard. This is the case even if the cardinality $K_n$ of the set satisfies $K_n \geq K_{\rm st}(2^n, t, 0)$.
\end{restatable}

Combining these two, we conclude that, for any constant $t\in\ZZ^+$, $\SFP(t,\Omega(1/\poly(n)))$ is $\BQP$-complete.

The inverse-polynomial accuracy of the state frame potential is, however, not sufficient for checking whether or not a given set is a state design. In fact, properties~\ref{item:property:FSbound} and~\ref{item:property:approxDesign} in~\cref{Prop:StateFPBasicProperties} imply that an inverse-exponential accuracy is needed to decide which is the case.
For this reason, we investigate $\SFP(t,\epsilon)$ for exponentially small $\epsilon$ and show the following. See~\cref{SS:SFPexp} for the proof.

\begin{restatable}{theorem}{approxSFPexpin}\label{Thm:approxSFPexpin}
    For any constant $t \in \ZZ^+$, $\SFP(t,\epsilon) \in \PP$ if $\epsilon \in \Omega\bigl(2^{-\poly(n)}\bigr)$.
\end{restatable}

\begin{restatable}{theorem}{approxSFPexphard}\label{Thm:approxSFPexphard}
    For any constant $t \in \ZZ^+$, $\SFP(t,\epsilon)$ is \PP-hard if $\epsilon\in \cO(1/(K_n^2 2^n))$, where $K_n$ is the cardinality of the set. This is the case even if $K_n \geq K_{\rm st}(2^n, t, 0)$.
\end{restatable}

Note that $K_n \in \cO(2^{\poly(n)})$ for a set of state vectors with a computationally efficient description. Hence, \cref{Thm:approxSFPexpin,Thm:approxSFPexphard} imply that $\SFP \bigl(t,\Theta(2^{-\poly(n)})\bigr)$ is \PP-complete for any constant $t \in \ZZ^+$.\\

We can similarly formulate the unitary frame potential problem $\UFP(t,\epsilon)$.

\begin{problem}[$\UFP(t,\epsilon)$] \label{Prob:UFP}
\hfill\\
    \underline{Input:} 
    \begin{compactitem}
        \item a set $\sfU_n$ of $K_n$ $n$-qubit unitaries with a computationally efficient description
        \item $t \in \ZZ^+$ (degree)
        \item $\alpha > \beta \geq \max\{t!,\, 2^{2nt}/K_n\}$ such that $\alpha-\beta>\epsilon>0$
    \end{compactitem}
    \underline{Promise:} the unitary frame potential $F_t(\sfU_n)$ is either $\geq \alpha$ or $\leq \beta$.\\
    \underline{Output:} decide which is the case.  
\end{problem}

The lower bound on $\beta$ is for avoiding trivial situations (see the property~\ref{item:uniFleq} in~\cref{Prop:UnitaryFPBasicProperties}).

We obtain the following results about the complexity class of $\UFP(t,\epsilon)$. See~\cref{SS:UFP} for the proofs.

\begin{restatable}{theorem}{approxUFPin}\label{Thm:approxUFPin}
    For any constant $t \in \ZZ^+$, $\UFP(t,\epsilon) \in \PP$ if $\epsilon \in \Omega\bigl(2^{-\poly(n)}\bigr)$.
\end{restatable}

\begin{restatable}{theorem}{approxUFPhard}\label{Thm:approxUFPhard}
    For any constant $t \in \ZZ^+$, $\UFP(t, \cO(2^{2(t-1)n}/K_n^2))$ is \PP-hard, where $K_n$ is the cardinality of the set. This is the case even if $K_n \geq K_{\rm uni}(2^n, t, 0)$.
\end{restatable}

In~\cref{Thm:approxUFPhard}, the required promise gap is $\cO(2^{2(t-1)n}/K_n^2)$, which can be large if the cardinality $K_n$ is small. However, when one is concerned with a set of unitaries that is a candidate of a unitary $t$-design, its cardinality should be at least $K_{\rm uni} \approx 2^{2tn}$. In this case, the required promise gap is $\cO(2^{-\poly(n)})$. That is, it is, in general, hard to compute the unitary frame potential for a candidate set of unitary designs with exponential accuracy.\\

These results imply that it is unlikely to drastically improve the quantum algorithm in the previous section. A general algorithm, whether quantum or classical, that computes the state frame potential with exponential accuracy in polynomial time is highly unlikely to exist.

\subsection{Promise problems about designs}

\begin{figure}[tb!] 
  \centering
  \includegraphics[width=0.4\textwidth,clip]{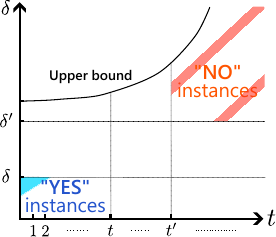}
  \caption{
  A conceptual visualization of $\StDes(t,t',\delta,\delta')$ and $\UniDes(t,t',\delta,\delta')$. The ``yes'' instances, namely, $\delta$-approximate $t$-designs, are indicated by a blue box, and the ``no'' instances, i.e., the sets that are not $\delta'$-approximate $t'$-design, are indicated by a red box.
  They form boxes because, e.g., a $\delta$-approximate $t$-design is a $\delta$-approximate $\tilde{t}$-design for any $\tilde{t} \leq t$. Note that the ``yes'' and ``no'' instances are not exclusive in general, but in $\StDes(t,t',\delta,\delta')$ and $\UniDes(t,t',\delta,\delta')$, it is promised that a given instance belongs to only one of the ``yes'' or ``no'' instances. See the main text. 
  The upper bound in the figure is given by $\approx d_t$ for $\StDes$ and $\approx 2^{2nt}$ for $\UniDes$.
  In particular, we show that $\StDes(t,t',\delta, \delta')$ and $\UniDes(t,t',\delta, \delta')$ with $t'=t+1$, $\delta\in\Theta(2^{-cn})$, and $\delta'\in\Theta(1)$ is $\PP$-hard, where $c=1$ for $\StDes$ and $c=2$ for $\UniDes$. 
  }
  \label{Fig:DesProblemDiagram}
\end{figure}

So far, we have considered problems related to frame potentials. We can also consider a problem of deciding whether a given set is a good approximation to a design.  This problem directly captures the main computational complexity of designs as it is natural to expect that more fine-grained problems related to quantum designs are at least as difficult.

To this end, we introduce the state design problem $\StDes(t,t', \delta, \delta')$ for a set of state vectors as follows.

\begin{problem}[$\StDes(t,t',\delta,\delta')$]  \label{Prob:StDes}
\hfill\\
    \underline{Input:} 
    \begin{compactitem}
        \item a set $\sfS_n$ of $K_n$ $n$-qubit state vectors with a computationally efficient description
        \item $t \leq t' \in \ZZ^+$ (degree)
        \item $0 \leq \delta < \delta'$
    \end{compactitem}
    \underline{Promise:} $\sfS_n$ is either a $\delta$-approximate state $t$-design or not a $\delta'$-approximate state $t'$-design. Only one of the two is the case.\\
    \underline{Output:} decide which is the case.  
\end{problem}

The problem $\StDes(t, t', \delta,\delta')$ has four parameters $(t, t', \delta, \delta')$, which can be split to two different types. One is a pair $(t,t')$, corresponding to the degree of designs, and the other is $(\delta,\delta')$, corresponding to the approximation accuracy. We propose the problem with four parameters, $t$, $t'$, $\delta$, and $\delta'$ so as to deal with these different notions in a unified manner. See~\cref{Fig:DesProblemDiagram} for the conceptual visualization of the problem.

Due to the mixture of two conceptually different types of parameters in $\StDes(t, t', \delta,\delta')$, we need to explicitly state that the promise is \emph{exclusive}. Without this exclusive promise, both could be the case as a set of state vectors could be a good approximate $t$-design \emph{and} a bad approximate $t'$-design at the same time, making the problem ill-defined.\\

When $t=t'$, we can use the computational complexity of $\SFP(t,\epsilon)$ for showing that $\StDes(t, t, \delta,\delta')$ is in $\PP$ if the promise gap is not too small.

\begin{restatable}{theorem}{StDESin} \label{Thm:StDESin}
    For any constant $t \in \ZZ^+$ and $\delta'^2 - d_t \delta^2 \in  \Omega\bigl(2^{-\poly(n)}\bigr)$, $\StDes(t, t, \delta, \delta') \in \PP$.
\end{restatable}

In contrast, we can obtain the hardness result for any $t \leq t'$.

\begin{restatable}{theorem}{StDEShard}\label{Thm:StDEShard}
    For any constant $t \leq t' \in \ZZ^+$, $\delta \in \Omega(2^{-\poly(n)})$, and $\delta'$ that satisfies
    \begin{equation}
        \delta' - \delta \in
        \begin{cases}
            \Theta\left(2^{(t'-t)n} \delta \right) & \text{for $t < t'$,}\\
            \Theta\left(2^{-n} \delta \right) & \text{for $t=t'$},
        \end{cases} \label{Eq:StDESGap}
    \end{equation}
    $\StDes(t,t',\delta,\delta')$ is $\PP$-hard under polynomial-time mapping reductions. 
     This is the case even if the cardinality $K_n$ of the set satisfies $K_n \geq K_{\rm st}(2^n, t', 0)$.
\end{restatable}

Note that~\cref{Eq:StDESGap} implies that $\delta' \in \Theta(2^{(t'-t)n} \delta )$ for any $t \leq t'$.
See~\cref{SS:StDes} for the proofs of~\cref{Thm:StDESin,Thm:StDEShard}.

As a direct corollary of~\cref{Thm:StDEShard}, it follows that, for any $t \in \mathbb{Z}^+$ and $\delta \in \Omega(2^{-\poly(n)})$, it is $\PP$-hard to determine whether a given set of state vectors is a $\delta$-approximate state $t$-design or not a $\Theta(2^n \delta)$-approximate state $(t+1)$-design. If the latter is the case, the set is not a $\delta$-approximate state $t$-design due to the promise.
Hence, an efficient, general algorithm is unlikely to exist to decide if a given set is an exponentially-accurate approximation to a state $t$-design or a worse-than-constant approximation to a state $(t+1)$-design that fails to form an accurate $t$-design.

This inherent difficulty of distinguishing good designs from bad designs highlights the computational hardness of general verification methods of state designs. Hence, in practical applications of state designs, developing a verification technique specifically tailored to the method that generated the set of state vectors is of crucial importance.
Our result may also find applications in adversarial scenarios such as quantum cryptography because it implies that an adversary can easily cheat by preparing a bad $(t+1)$-design instead of a good $t$-design. This may potentially offer a new cryptographic primitive, but we leave a concrete application as a problem for future research. \\

We similarly formulate the unitary design problem $\UniDes(t,t',\delta,\delta')$ for a set of unitaries and show similar results. See~\cref{SS:UniDes} for the proofs.

\begin{problem}[$\UniDes(t,t',\delta,\delta')$]  \label{Prob:UniDes}
\hfill\\
    \underline{Input:} 
    \begin{compactitem}
        \item a set $\sfU_n$ of $K_n$ $n$-qubit unitaries with a computationally efficient description
        \item $t \leq t' \in \ZZ^+$ (degree)
        \item $0 \leq \delta < \delta'$
    \end{compactitem}
    \underline{Promise:} $\sfU_n$ is either a $\delta$-approximate unitary $t$-design or not a $\delta'$-approximate unitary $t'$-design. Only one of the two is the case.\\
    \underline{Output:} decide which is the case.  
\end{problem}

The same remarks as those for $\StDes(t,t',\delta,\delta')$ apply to $\UniDes(t,t',\delta,\delta')$. Specifically, the problem has two types of parameter $(t,t')$ and $(\delta, \delta')$, and the \emph{exclusive} promise is also important.

\begin{restatable}{theorem}{UniDESin} \label{Thm:UniDESin}
    For any constant $t \in \ZZ^+$ and $\delta'^2 - 2^{2nt} \delta^2 \in  \Omega(2^{-\poly(n)})$, $\UniDes(t, t, \delta, \delta') \in \PP$.
\end{restatable}

\begin{restatable}{theorem}{UniDEShard} \label{Thm:UniDEShard}
    For any constant $t \leq t' \in \ZZ^+$, $\delta \in \Omega(2^{-\poly(n)})  \cap o(2^{2(t-1)n})$, and $\delta'$ that satisfies
    \begin{equation}
        \delta' - \delta\in
        \begin{cases}
            \Theta\left(2^{2(t'-t)n} \delta \right) & \text{for $t < t'$,}\\
            \Theta\left(2^{-2n} \delta \right) & \text{for $t=t'$},
        \end{cases} \label{Eq:UniDESGap}
    \end{equation}
    $\UniDes(t,t',\delta,\delta')$ is $\PP$-hard under polynomial-time mapping reductions.
    This is the case even if the cardinality $K_n$ of the set satisfies $K_n \geq K_{\rm uni}(2^n, t', 0)$.
\end{restatable}

By setting $t' = t+1$,~\cref{Thm:UniDEShard} implies that it is computationally hard to decide whether a given set of unitaries is a $\delta$-approximate unitary $t$-design or not a $\Theta(2^{2n}\delta)$-approximate unitary $(t+1)$-design. The exclusive promise further guarantees that, if the latter is the case, the set is not a $\delta$-approximate unitary $t$-design. Thus, similarly to a state design, an efficient algorithm is unlikely to exist for distinguishing a good approximate unitary $t$-design from a bad approximate unitary $(t+1)$-design that fails to form a good approximate $t$-design.
This also has implications to verifying unitary designs in practical applications and to future applications in adversarial situations.

\subsection{Implications of our results}

Our approach is based on computational complexity theory. Due to its broad applicability, our results have immediate implications for the topics related to unitary or state designs. Below, we outline three such implications along with a brief overview of the literature.

\subsubsection{Variational approach to constructing designs}

Generating $\delta$-approximate unitary $t$-designs is of fundamental importance. In~\cite{HL2009TPE}, the first efficient circuit construction has been proposed, which uses $O(t n^2(tn + \log 1/\delta))$ gates if $t \in \cO(n/\log n)$ and $t \leq 2^{n/50}$~\cite{BH2008}. 
Later, the number of gates has been improved to $\cO(t^{4 + o(1)} n (t n + \log 1/\delta))$ for any $t$  by local random circuits~\cite{H2022} and to $(t+1)n^2+ n\log 1/\delta + \cO(n)$ for $t \in o(\sqrt{n})$ by phase-Hadamard-cocktail random circuits~\cite{NHKW2017}, where the term `phase-Hadamard-cocktail' has been proposed in~\cite{KQST2023}. 
Recently, structured random circuits with $\cO(t n (\log n + \log 1/\delta))$ gates have been shown to suffice~\cite{schuster2024randomunitariesextremelylow,laracuente2024approximateunitarykdesignsshallow}. Due to the substantial importance of designs and related concepts, more and more circuit constructions have been proposed~\cite{Harrow2023,Haferkamp2023,metger2024simpleconstructionslineardepthtdesigns,chen2024efficientunitarydesignspseudorandom,laracuente2024approximateunitarykdesignsshallow}.

These results have been obtained first by proposing specific quantum circuits, such as local random circuits or phase-Hadamard-cocktail random circuits, and then by analysing their design properties. 

An alternative approach is a variational method, explored in~\cite{GAE2007} to construct a design with small cardinality.
The variational method is based on a variational ansatz set $\textsf{A}$ with computationally efficient descriptions and an update protocol, such as gradient descent, and may proceed as follows:

\begin{compactenum}
    \item Initialize an ansatz set $\textsf{A}$.
    \item Compute the frame potential of $\textsf{A}$ and its derivative.
    \item Update the ansatz set $\textsf{A}$ to reduce its frame potential.
    \item Repeat steps 2 and 3 until the frame potential approaches its minimum sufficiently closely.
\end{compactenum}

Such a variational method performs well in small systems~\cite{GAE2007}. Moreover, \cref{Thm:SFPBQPin} indicates that, for state vectors, this method may work well on quantum computers until the frame potential approaches its minimum with an accuracy of $\Omega(1/ \poly(n))$. 
However, achieving higher accuracy for state vectors, such as $\cO(2^{- \poly(n)})$, and dealing with a set of unitaries requires careful consideration. 
In fact,~\cref{Thm:approxSFPexpin,Thm:approxSFPexphard,Thm:approxUFPin,Thm:approxUFPhard} immediately imply the following.

\begin{corollary}
    If the ansatz set $\textsf{A}$ can represent arbitrary sets with computationally efficient description, there is no variational method that outputs an approximate $t$-design in quantum and classical polynomial-time, assuming that $\PP \neq \BQP$ and $\PP \neq \BPP$, respectively.
\end{corollary}

Our results, therefore, highlight that it is essential to restrict the representability of the ansatz set $\textsf{A}$ for the variational approach to scale up to a large system.

\subsubsection{Computational complexity of chaos via OTOCs}

The unitary frame potential is closely related to OTOCs, which are widely used as diagnostics for quantum chaotic dynamics in many-body systems.
A $2t$-point OTOC is defined for an $n$-qubit state $\rho$, an $n$-qubit unitary $U$, and $2t$ $n$-qubit observables $\{O_j\}_{j=1, \dots, 2t}$ as follows:
\begin{multline}
    {\rm OTOC}_{2t}(\rho, U, \{O_j\}_j) \\
    = 
    {\rm Tr}\Bigl[ \Bigl( O_{2t} O_{2t-1}(U) \dots O_{2} O_{1}(U) \Bigr) \rho \Bigr],
\end{multline}
where $O_j(U) = U^{\dagger} O_j U$. Note that the unitary is applied only to operators with odd indexes.
The OTOCs are important quantities for characterizing chaotic behaviour in quantum many-body systems, where $\rho$ is a thermal state of a given Hamiltonian $H$, $U= e^{-iHT}$ with $T$ being time, and observables are commonly chosen to be single-qubit Pauli operators.
Under these settings, the OTOCs typically decrease as time $T$ increases, and the speed of the decrease characterizes quantum chaotic dynamics in terms of the Lyapunov exponent~\cite{MSS2016}.

It has been shown in Ref.~\cite{RY2017} that the OTOCs are closely related to the frame potential. To explain this, let us set $\rho$ to be the completely mixed state $\pi$, corresponding to the infinite temperature, and let us take the average of the OTOCs over a given set $\sfU_n$ of $n$-qubit unitaries. By summing each observable $O_j$ over an operator basis $\textsf{B}_n$, such as a set of all $n$-qubit Pauli operators, we can define an average $2t$-point OTOC at infinite temperature as
\begin{multline}
    {\rm OTOC}_{2t}(\sfU) \\
    =  \frac{1}{2^{4nt}}\sum_{\{O_j\}_j \in \textsf{B}_n^{2t}} \mathbb{E}_{U \sim \sfU_n} \bigl| {\rm OTOC}_{2t}(\pi, U, \{O_j\}_j) \bigr|^2.
\end{multline}
This quantity is shown to be equivalent to the unitary frame potential of $\sfU_n$ of degree $t$, namely,
\begin{equation}
    {\rm OTOC}_{2t}(\sfU_n) = \frac{1}{2^{2(t+1)n}}F_t(\sfU_n). \label{Eq:OTOCFramePotential}
\end{equation}

From~\cref{Eq:OTOCFramePotential,Thm:compUFPin,Thm:compUFPhard}, it directly follows that computing the average $2t$-point OTOCs at infinite temperature is hard for any constant $t \in \ZZ^+$.

\begin{corollary}
    For any constant $t \in \ZZ^+$, computing ${\rm OTOC}_{2t}(\sfU_n)$ for a set $\sfU_n$ of unitaries with a computationally efficient description is in $\FP^\sharpP$ and $\sharpP$-hard. 
\end{corollary}

This indicates that verifying quantum chaotic dynamics via OTOCs is, in general, computationally hard in large systems, which characterizes the complex nature of chaotic dynamics from the computational complexity perspective.

\subsubsection{Verification of emergent state designs}

The concepts of unitary and state designs have also been applied to investigate higher-order properties of thermalization and ergodicity in quantum many-body systems.

Let $H$ be a Hamiltonian on a composite system $AB$ of $n + m$ qubits, where $n$ and $m$ are the number of qubits in $A$ and $B$, and let $\ket{\Psi(T)} = e^{- i H T} \ket{0}^{\otimes (n +m)}$ be the state vector in $AB$ at time $T$. By measuring the subsystem $B$ in the computational basis, the following ensemble $\textsf{P}$ of pure states is realized in $A$:
\begin{equation}
    \textsf{P} = \bigl\{ p_j, \ket{\Psi_j(T)} \bigr\}_{j=1,\dots,2^{m}},
\end{equation}
where $\ket{\Psi_j(T)} := (I^A \otimes \bra{j}^B)\ket{\Psi(T)} /\sqrt{p_j}$ with $p_j = |(I^A \otimes \bra{j}^B)\ket{\Psi(T)}|^2$. This ensemble is called a \emph{projected state ensemble}.
It is conjectured, with supporting numerical evidence, that if the Hamiltonian $H$ is chaotic and lacks conservation laws, the projected state ensemble obtained by measuring a sufficiently large subsystem $B$ after long time evolution forms an approximate state $t$-design~\cite{PRXQuantum.4.010311,Choi2023}. 
By leveraging the connection between state designs, thermalization, and ergodicity, this insight has been extensively applied to the study of complex quantum many-body systems~\cite{Chan_2024,mok2024optimalconversionclassicalquantum,HC2022, IH2023,IH2022,BDP2023,PhysRevX.14.041051,PhysRevX.14.041059,PhysRevLett.131.250401}. Novel concepts in quantum many-body physics through the framework of designs, such as \emph{deep thermalization}~\cite{IH2022,BDP2023} and \emph{complete Hilbert space ergodicity}~\cite{PhysRevX.14.041051,PhysRevX.14.041059}, have also emerged.

The conjecture has been analytically addressed in a few solvable models in the thermodynamic limit~\cite{HC2022, IH2023}, but numerical approaches are crucial for the further studies. 
While general numerical simulations of a projected state ensemble are difficult, as they have to compute the measurement probability distributions, there may exist a class of Hamiltonians for which this is also computationally tractable. Even in such cases, our results highlight the importance of developing numerical methods that account for the specific dynamics governed by the Hamiltonian $H$ and the measurement. In particular, the following corollary is obtained from~\cref{Thm:StDEShard}.

\begin{corollary} \label{Cor:EmergentStateDesign}
    Let $\textsf{P}$ be a projected state ensemble of $n$ qubits obtained by measuring the $(n+m)$-qubit state generated by Hamiltonian dynamics with Hamiltonian $H$ after time $T$, which has a computationally efficient description. Assuming that $\PP \neq \BQP$ for the quantum case and that $\PP \neq \BPP$ for the classical case, there is no general quantum or classical polynomial-time algorithm  that can check whether $\textsf{P}$ is a $\delta$-approximate state $t$-design or not a $\delta'$-approximate state $t'$-design if $\delta' -\delta \in \Theta(2^{(t'-t)n}\delta)$ for $t<t'$ and $\delta'  -\delta \in \Theta(2^{-n}\delta)$ for $t=t'$.
\end{corollary}

\cref{Cor:EmergentStateDesign} particularly implies that, toward the proof of the conjecture in general situations, it is of crucial importance to develop an algorithm that is specifically tailored to exploit problem-specific properties, such as the property of Hamiltonian dynamics by $H$, or additional constraints beyond the general case.

\subsection{Proof sketch}

Below, we provide a brief overview of our proof methods. We first focus on the inclusion results and then explain the methods that we use for proving the hardness results.

\subsubsection{Methods for the inclusion results}

Our strategy to obtain the inclusion results is to begin with quantum algorithms, which are not necessarily efficient, and to dequantize them. 
The inclusion results for $\compSFP$ and $\compUFP$ are obtained by dequantizing the quantum algorithm based on~\cref{Lemma:FPencodedState}, which is achieved by a similar technique in Ref.~\cite{SWVC2007}. For $\SFP$ and $\UFP$, we propose variants of the quantum algorithm. In this case, the dequantization step follows from the equivalence that $\PQP = \PP$~\cite{Watrous2009}.

From these inclusion results of the computational problems related to frame potentials, the inclusion results of $\StDes$ and $\UniDes$ can be immediately obtained as frame potentials characterize designs.

\subsubsection{Methods for the hardness results}

Proving hardness results is less straightforward. To establish those results, we employ a technique that we call ``hide-a-hard-instance-in-a-set." This consists of three steps. First, we choose a computationally hard instance relevant to the desired proof. Second, we embed the hard instance in state vectors or unitaries and, finally, we hide them in a set of state vectors or unitaries with known properties.

The specific hard instance depends on the statement to be shown. For the $\sharpP$-hardness of $\compSFP$ and $\compUFP$, we use a $\sharpP$-complete problem $\lang{\# SAT}$. For the $\PP$-hardness of $\SFP$ and $\UFP$ with suitable parameters, a $\PP$-complete problem $\textsf{MAJ-SAT}$ is used. For the $\BQP$-hardness of $\SFP(t, \Omega(1/\poly(n)))$, we use a quantum circuit that solves a $\BQP$-complete problem.

These problems are embedded in a pair of state vectors or unitaries with a computationally efficient description. This embedding is such that the solution of the problems is given by the inner product of the pair. 
The pairs are then ``hidden'' in a set of state vectors or unitaries that we know the details. 
In the case of the computational problems related to frame potentials, this is enough for obtaining the hardness result: as the frame potential is a function of inner products, the frame potential of the resulting set should contain the information about the solution of the hard instance.
That is, if one can evaluate the frame potential of the resulting set, one can use the same method to obtain the solution of the hard instance. This provides a polynomial-time reduction from a complete problem to the corresponding computational problem.

This ``hide-a-hard-instance-in-a-set'' technique is also useful for proving the hardness results of $\StDes$ and $\UniDes$, in which the main concern is to evaluate the approximation accuracy of a given set to a design.
In the proofs, we similarly use the $\textsf{MAJ-SAT}$ problem and embed it into a pair of state vectors or unitaries. We then hide the pair in an approximate design. Although these steps are similar to those in the proof of the problems related to frame potentials, the proofs are more intricate due to the following two challenges.

First, unlike the frame potential, it is not immediately obvious if the approximation accuracy of the set is directly related to the inner product. Note that it is the inner product of the hidden pair that contains the solution of $\textsf{MAJ-SAT}$.
Through careful evaluations of matrix norms, we explicitly compute the approximation accuracy of the resulting set and show that it is given as a function of the inner product of the hidden pair. This evaluation is tight, which is also crucial as the tightness is the key factor to guarantee that any Boolean function can be embedded into  $\StDes$ and $\UniDes$ while keeping their exclusive promise.

The second difficulty is that, in general, hiding even a single pair of state vectors or unitaries in approximate designs significantly degrades their approximation accuracy.
This imposes a substantial restriction on the parameter $\delta$ in $\StDes(t,t',\delta,\delta')$ and $\UniDes(t,t',\delta,\delta')$, preventing us from dealing with accurate designs in the problems. That is, we cannot choose, e.g, $\delta  \in \cO(2^{-\poly(n)})$.
We circumvent this limitation by using the multiplicity degree of freedom of approximate designs. Specifically, we use the fact that the multiplicity of approximate designs can be arbitrarily increased without changing the approximation accuracy. By hiding hard instances into a \emph{sufficiently large} set, we can control the restriction on the approximation of the set after hiding and can remove the restriction on $\delta$.

%%% --------------------------------
\section{Analysis of  complexity of \texorpdfstring{$\compSFP$ and $\compUFP$}{compSFP and compUFP}} \label{S:CompFPs}
%%% --------------------------------

In this section, we investigate the computational complexity of $\compSFP$ and $\compUFP$, and provide proofs for~\cref{Thm:compSFPin,Thm:compSFPhard,Thm:compUFPin,Thm:compUFPhard}.
We address $\compSFP$ in~\cref{SS:CompSFP} and $\compUFP$ in~\cref{SS:CompUFP}.

\subsection{Complexity of \texorpdfstring{$\compSFP$}{compSFP}} \label{SS:CompSFP}

We prove~\cref{Thm:compSFPin} by explicitly constructing a classical algorithm to compute the state frame potential that uses a single query to a $\sharpP$ oracle. The algorithm is based on the quantum algorithm in~\cref{Lemma:FPencodedState}.

\compSFPin* 

\begin{Proof}[\cref{Thm:compSFPin}]
    From~\cref{Lemma:FPencodedState}, we have $F_t(\sfS_n) = \tr{(\varrho^A_{\rm st})^2}$, where $\varrho^A_{\rm st}$ is the marginal state of $\ket{\varrho_{\rm st}}^{AB}$ defined in~\cref{Eq:defvarrhost}. Let $A'$ and $B'$ be isomorphic to $A$ and $B$, respectively, and denote by $\mathbb{F}^{AA'} = \sum_{i,j}\ketbra{ij}{ji}^{AA'}$ the swap operator between $A$ and $A'$, where $\{ \ket{j} \}_j$ is an orthonormal basis. The state frame potential can be rephrased as 
    \begin{multline}
        F_t(\sfS_n) \\
        = \bigl(\bra{\varrho_{\rm st}}^{AB} \otimes \bra{\varrho_{\rm st}}^{A'B'} \bigr)(\mathbb{F}^{AA'} \otimes I^{BB'})\bigl(\ket{\varrho_{\rm st}}^{AB} \otimes \ket{\varrho_{\rm st}}^{A'B'} \bigr). \label{Eq:expectationSFP}
    \end{multline}
    Below, we simply denote the right-hand side by $\bra{\varrho_{\rm st}}^{\otimes 2} (\mathbb{F}^{AA'} \otimes I^{BB'})\ket{\varrho_{\rm st}}^{\otimes 2}$.
    The statement of~\cref{Thm:compSFPin} follows from the fact that this expectation value is obtained by counting the number of satisfying instances of an appropriate Boolean function that is computable in classical polynomial time.
    We show this by using a technique similar to the one used in Ref.~\cite{SWVC2007}.

    We first map the state $\ket{\varrho_{\rm st}}^{\otimes 2}$ to the one that is generated by a quantum circuit consisting only of the Hadamard and Toffoli gates. To this end, let $\sum_{\alpha} (x_{\alpha} + i y_{\alpha}) \ket{\alpha}$ $(x_{\alpha}, y_{\alpha} \in \mathbb{R})$ be an expansion of $\ket{\varrho_{\rm st}}^{\otimes 2}$ in the computational basis. Using this notation and~\cref{Eq:expectationSFP}, the frame potential is given as
    \begin{align}
        F_t(\sfS_n) 
        &=\bra{\varrho_{\rm st}}^{\otimes 2} (\mathbb{F}^{AA'} \otimes I^{BB'})\ket{\varrho_{\rm st}}^{\otimes 2}\\
        &=
        \sum_{\alpha, \beta} (x_{\beta} -i y_{\beta})(x_{\alpha} + i y_{\alpha}) h(\alpha, \beta)\\
        &=
        \sum_{\alpha, \beta} (x_{\alpha} x_{\beta} + y_{\alpha} y_{\beta})h(\alpha, \beta), \label{Eq:FPreal}
    \end{align}
    where $h(\alpha, \beta) = \bra{\beta} (\mathbb{F} \otimes I ) \ket{\alpha}$. In the last line, we used the fact that $h(\alpha, \beta) = h(\beta, \alpha)$, which is either $0$ or $1$.
    
    We now introduce one more ancillary qubit and define $\ket{\bar{\varrho}^{\otimes 2}_{\rm st}}$ by
    \begin{equation}
        \ket{\bar{\varrho}^{\otimes 2}_{\rm st}} = \sum_{\alpha} \ket{\alpha} \otimes (x_{\alpha} \ket{0} + y_{\alpha} \ket{1}).
    \end{equation}
    This is a state on $2(\kappa_n + nt) +1$ qubits, where $\kappa_n = \log K_n$, and $K_n$ is the cardinality of $\sfS_n$.
    When $\ket{\varrho_{\rm st}}^{\otimes 2}$ can be efficiently generated by a quantum circuit, which is the case in our problem, it is known that $\ket{\bar{\varrho}^{\otimes 2}_{\rm st}}$ can also be efficiently generated from the computational zero state only by using the Hadamard and Toffoli gates~\footnote{Strictly speaking, the state generated by the quantum circuit composed only of the Hadamard and Toffoli gates is exponentially-close to the original state. We ignore this approximation error as it can be made small arbitrarily within the order of $\Theta(1/\exp(n))$.}.
    Furthermore, it holds that
    \begin{align}
        \bra{\bar{\varrho}^{\otimes 2}_{\rm st}}(\mathbb{F}^{AA'} \otimes I^{BB'} \otimes I )\ket{\bar{\varrho}^{\otimes 2}_{\rm st}}
        &=
        \sum_{\alpha, \beta} (x_{\alpha} x_{\beta} + y_{\alpha} y_{\beta})h(\alpha, \beta),
    \end{align}
    which is equal to the frame potential $F_t(\sfS_n)$ due to~\cref{Eq:FPreal}.
    Hence, the state frame potential can be obtained as an expectation value by the state $\ket{\bar{\varrho}_{\rm st}^{\otimes 2}}$ that is generated by the quantum circuit consisting only of the Hadamard and Toffoli gates. 
    Below, we denote by $M$ the total number of gates in the quantum circuit generating $\ket{\bar{\varrho}_{\rm st}^{\otimes 2}}$ and by $h$ the number of Hadamard gates in the circuit. The number of Toffoli gates is $M-h$.
    
    Let $\tilde{H}$ be a `rescaled' Hadamard gate $\sqrt{2} H$, where $H$ is the Hadamard gate. Denoting by $G_m$ the unitary on $(2 (\kappa_n + nt)+1)$ qubits corresponding to a single $\tilde{H}$ or a Toffoli gate in the quantum circuit, we can rewrite $\ket{\bar{\varrho}_{\rm st}^{\otimes 2}}$ as follows:
    \begin{equation}
        \ket{\bar{\varrho}_{\rm st}^{\otimes 2}} = 2^{-h/2} G_M G_{M-1} \dots G_1 \ket{0}^{\otimes 2(\kappa_n+nt)+1}.
    \end{equation}
    By substituting the identity $I = \sum_{\alpha} \ketbra{\alpha}{\alpha}$ after each $G_m$, where $\ket{\alpha}$ are the computational basis states, we obtain
    \begin{equation}
        \ket{\bar{\varrho}_{\rm st}^{\otimes 2}} =
        2^{-h/2}\sum_{\alpha_1, \dots, \alpha_M} \prod_{m=1}^M g_m(\alpha_m, \alpha_{m-1}) \ket{\alpha_M}, \label{Eq:Expansion}
    \end{equation}
    where $g_m(\alpha_m, \alpha_{m-1}) =\bra{\alpha_m}G_m\ket{\alpha_{m-1}}$, and $\ket{\alpha_0} = \ket{0}^{\otimes 2(\kappa_n+nt)+1}$.
    Since $\ket{\bar{\varrho}_{\rm st}^{\otimes 2}}$ is a state on $2(\kappa_n + nt) +1$ qubits, $\alpha_m \in \{ 0,1\}^{2(\kappa_n + nt) +1}$.
    Below, we use the notation $g(\vec{\alpha}) = \prod_{m=1}^M g_m(\alpha_m, \alpha_{m-1})$ for $\vec{\alpha} = (\alpha_1, \dots, \alpha_M)$. With this notation, we have
    \begin{equation}
        \ket{\bar{\varrho}_{\rm st}^{\otimes 2}} =
        2^{-h/2}\sum_{\vec{\alpha}} g(\vec{\alpha}) \ket{\alpha_M},
    \end{equation}
    and
    \begin{equation}
        F_t(\sfS_n) 
        =
        2^{-h}\sum_{\vec{\alpha}, \vec{\beta}}
       g(\vec{\alpha}, \vec{\beta}) \bar{h}(\alpha_M, \beta_M), \label{Eq:SFPdecomp}
    \end{equation}
    where $g(\vec{\alpha}, \vec{\beta})=g(\vec{\alpha}) g(\vec{\beta})$, and $\bar{h}(\alpha_M, \beta_M)=\bra{\beta_M} \bigl(\mathbb{F}^{AA'} \otimes I^{BB'} \otimes I\bigr) \ket{\alpha_M} \in \{0,1\}$. 
    
    Since $G_m$ is a unitary corresponding to either single $\tilde{H}$ or a single Toffoli gate, $g_m(\alpha_m, \alpha_{m-1})$ is computable in classical polynomial time. This further implies that $g(\vec{\alpha})$ is also computable in classical polynomial time, and so is $g(\vec{\alpha}, \vec{\beta})$.
    Clearly, $\bar{h}(\alpha_M, \beta_M)$ is also computable in classical polynomial time.

    We now define a Boolean function $f: \{0, 1\}^{L+1} \rightarrow \{0,1\}$, where $L = 2M(2 (\kappa_n + nt) + 1 )$, by
    \begin{equation}
        f(\vec{\alpha},\vec{\beta}, b)
        =
        \begin{cases}
            1 & \text{if $g(\vec{\alpha},\vec{\beta}) \bar{h}(\alpha_M, \beta_M) \geq b$,}\\
            0 & \text{otherwise.}
        \end{cases}
    \end{equation}
    As both $g(\vec{\alpha}, \vec{\beta})$ and $\bar{h}(\alpha_M, \beta_M)$ are computable in classical polynomial time, so is $f(\vec{\alpha},\vec{\beta}, b)$.
    
    We claim that 
    \begin{equation}
        F_t(\sfS_n) 
        =
        2^{-h}\bigl( s(f) - 2^L), \label{Eq:Easy}
    \end{equation}
    where $s(f) = \bigl| \{ (\vec{\alpha}, \vec{\beta}, b) \in \{0,1\}^{L+1}: f(\vec{\alpha}, \vec{\beta}, b) = 1 \} \bigr|$.
    This claim implies that the state frame potential can be computed by counting the number $s(f)$ of satisfying instances of an efficiently computable Boolean function $f$. As counting $s(f)$ is in $\sharpP$, one can compute the state frame potential by a single access to a $\sharpP$ oracle.
    
    To show~\cref{Eq:Easy}, let $s_{\pm}$ and $s_0$ be the number of $(\vec{\alpha}, \vec{\beta})$ such that
    \begin{equation}
        g(\vec{\alpha}, \vec{\beta}) \bar{h}(\alpha_M, \beta_M) = \pm 1, 0,
    \end{equation}
    respectively. As the left-hand side is either $\pm 1$ or $0$ for any $\vec{\alpha}$ and $\vec{\beta}$, $s_- + s_0 + s_+ = 2^L$.
    It also follows from~\cref{Eq:SFPdecomp} that $F_t(\sfS_n) = 2^{-h}(s_+ - s_-)$.
    
    By definition, the number of $(\vec{\alpha}, \vec{\beta})$ such that $f(\vec{\alpha}, \vec{\beta}, 0) =1$ and the number of those satisfying $f(\vec{\alpha}, \vec{\beta}, 1) =1$ are $s_+ + s_0$ and $s_+$, respectively. This implies that $s(f) = 2 s_+ + s_0$. As $s_- + s_0 + s_+ = 2^L$, we have $s(f) = 2^L + s_+ - s_-$. This implies~\cref{Eq:Easy}. $\hfill \qed$
\end{Proof}

We next show the hardness of $\compSFP$, i.e.,~\cref{Thm:compSFPhard}. To this end, we provide a polynomial-time reduction from $\lang{\# SAT}$ to $\compSFP$.

\compSFPhard*

\begin{Proof}[\cref{Thm:compSFPhard}]
    For a Boolean function $f$ that is computable in classical polynomial time, consider a set of quantum circuits on $n$ qubits that generates the set $\sfS_n = \{\ket{f}, \ket{p}\} \cup \{ \ket{\psi_j}\}_{j=1}^{K_n-2}$, where
    \begin{align}
        &\ket{f} =2^{-(n-2)/2} \sum_{x \in \{0,1\}^{n-2}} \ket{x}  \ket{f(x)} \ket{0}, \label{Eq:S1}\\
        &\ket{p} =\ket{+}^{\otimes n-2}  \ket{1} \ket{0}, \label{Eq:S2}\\
        &\ket{\psi_j} =\ket{\varphi_{j}} \ket{1}. \label{Eq:S3}
        \end{align}
    Here, $\ket{+} = (\ket{0}+\ket{1})/\sqrt{2}$, and $\sfS_{\varphi}\coloneqq \{\ket{\varphi_j}\}_{j=1}^{K_n-2}$ is any set of $(n-1)$-qubit state vectors with a computationally efficient description such that $F_t(\sfS_{\varphi})$ can be computed in classical polynomial time.  

    The set $\sfS_n$ constructed in this way also has a computationally efficient description. In fact, one can obtain the description of $\sfS_n$ by adding one bit to the computationally efficient description of $\sfS_{\varphi}$ and by using the added degree of freedom for conditioning if we generate either $\ket{f}$ or $\ket{p}$, or some state in $\sfS_{\varphi}$.
    
    Since $\braket{f}{p} = 2^{-(n-2)}s(f)$ and $\braket{f}{\psi_j} =\braket{p}{\psi_j} = 0$ for all $j = 1, \dots, K_n-2$, the state frame potential $F_t(\sfS_n)$ is directly computed as
    \begin{equation}
        F_t(\sfS_n)
        =
        \biggl( 1 - \frac{2}{K_n} \biggr)^2 F_t\bigl(\sfS_{\varphi} \bigr) +
        \frac{2}{K_n^2} \biggl[ 1 + \biggl(\frac{s(f)}{2^{n-2}}\biggr)^{2t}\biggr]. \label{Eq:FtSnLaterUse}
    \end{equation}
    Since $F_t\bigl(\sfS_{\varphi} \bigr)$ is polynomial-time computable by assumption, this equation directly yields a polynomial-time reduction from a $\sharpP$-complete problem, $\lang{\# SAT}$, to $\compSFP$.
    $\hfill \qed$
\end{Proof}

In the proof of~\cref{Thm:compSFPhard}, we have used a set $\sfS_{\varphi}$ of $(n-1)$-qubit state vectors with a computationally efficient description, for which we can compute the state frame potential in polynomial time. For completeness, we below provide three instances of such a set.

The first one works when $K_n-2 \le 2^{n-1}$. In this case, we can trivially use any orthonormal basis $\{ \ket{e_j} \}_{j=1,\dots, K_n-2}$. The frame potential $F_t(\sfS_\varphi)$ is simply $1/(K_n-2)$.

A more non-trivial choice of $\sfS_{\varphi}$ is 
\begin{equation} \label{set1}
    |\varphi_j\rangle=\cfrac{|0\rangle+e^{2ij\pi/(K_n-2)}|1\rangle}{\sqrt{2}}\otimes |0^{n-2}\rangle.
\end{equation}
Let $K\coloneqq K_n-2$ and assume that $K>t$.
The frame potential $F_t(\sfS_\varphi)$ can be computed as
\begin{align}
    F_t(\sfS_\varphi)
    &=\cfrac{1}{K^2}\sum_{j,k=1}^{K}\left|\langle\varphi_j|\varphi_k\rangle\right|^{2t}\\
    &=\cfrac{1}{K^2}\sum_{j,k=1}^K\cos^{2t}{\left(\cfrac{k-j}{K}\pi\right)}\\
    &=\cfrac{1}{K^2}\sum_{j=1-K}^{K-1}(K-|j|)\cos^{2t}{\left(\cfrac{j}{K}\pi\right)}\\
    &=\cfrac{2}{K^2}\sum_{j=0}^{K-1}(K-j)\cos^{2t}{\left(\cfrac{j}{K}\pi\right)}-\cfrac{1}{K}\\
    &=\cfrac{1}{K}\sum_{j=0}^{K-1}\cos^{2t}{\left(\cfrac{j}{K}\pi\right)}, 
    \label{FGK2.1}
\end{align}
where, in the last line, we have used the following.
\begin{align}
    \nonumber
    &2\sum_{j=0}^{K-1}(K-j)\cos^{2t}{\left(\cfrac{j}{K}\pi\right)}\\
    &=\sum_{j=0}^{K-1}(K-j)\cos^{2t}{\left(\cfrac{j}{K}\pi\right)}+\sum_{j=0}^{K}j\cos^{2t}{\left(\cfrac{K-j}{K}\pi\right)} \\
    &=\sum_{j=0}^{K-1}(K-j)\cos^{2t}{\left(\cfrac{j}{K}\pi\right)}+\sum_{j=0}^{K-1}j\cos^{2t}{\left(\cfrac{j}{K}\pi\right)}+K\\
    &=K\left[1+\sum_{j=0}^{K-1}\cos^{2t}{\left(\cfrac{j}{K}\pi\right)}\right].
\end{align}
Moreover, we can further simplify~\cref{FGK2.1} by using Theorem 2.1 in Ref.~\cite{Raman}, which results in
\begin{align}
    F_t(\sfS_\varphi)    &=2^{1-2t}\binom{2t-1}{t-1}. \label{Eq:FGK}
\end{align}
For any constant $t$,~\cref{Eq:FGK} can be computed in classical constant time.
Note that, by using this set, we can vary the cardinality $K_n$ without changing the frame potential $F_t(\sfS_\varphi)$.

We may also generalize~\cref{set1} to 
\begin{equation}
    |\varphi_j\rangle=\left(\cfrac{|0\rangle+e^{2ij\pi/(K_n-2)}|1\rangle}{\sqrt{2}}\right)^{\otimes m}\otimes |0^{n-m-1}\rangle
\end{equation}
for any natural number $1\le m\le n-1$. By a similar calculation, the frame potential can be shown to be
\begin{equation}
    2^{1-2mt}\binom{2mt-1}{mt-1}.
\end{equation}
This value can be dependent on the number $n$ of qubits if we choose $m$ as a function of $n$.\\

From~\cref{Thm:compSFPin,Thm:compSFPhard}, we conclude that $\compSFP \in \FP^{\sharpP}$ and is $\sharpP$-hard. 
Although there is a gap between the result about the inclusion and the hardness, this is merely due to the fact that the answer to $\compSFP$ is a real number, while the answer to $\sharpP$ problems is a non-negative integer.
One way to close the gap is to restrict a set of state vectors and consider a ``rescaled'' frame potential.
A possible example is a set $\sfS_{\cX}$ of state vectors on $n$ qubits in the form of 
\begin{equation}
    \ket{\psi_j} = \chi^{-1/2} \sum_{x \in \cX} (-1)^{a_{jx}} \ket{x},
\end{equation}    
where $\cX \subseteq \{ 0,1\}^n$ with cardinality $\chi:=|\cX|$, and $a_{jx} \in \{0, 1\}$.
We call this type of state a \emph{subset phase state}. The subset phase states and their relations to a state design have been studied in a general context from the viewpoint of entanglement and thermalization phenomena~\cite{NTM2012}. 
In recent years, special types of subset phase states have also been of particular interest in quantum cryptography because the state reduces to a phase-random state~\cite{10.1007/978-3-319-96878-0_5,BS2019,AGQY2023} when $\chi=2^n$ and each $a_{jx}$ is randomly chosen, and to a random subset state~\cite{G-TB2023,JMW2024} when $\cX$ is randomly chosen and $a_{jx}=0$. Both are known to be useful as quantum cryptographic primitives.

For the subset phase states, we can show that computing a rescaled state frame potential is in $\sharpP$.

\begin{theorem} \label{Thm:compSFPin2}
    Let $\sfS_{\cX}$ be a set of $n$-qubit subset phase states with $a_{jx}$ being efficiently computable in classical polynomial time, and $K_n := |\sfS_{\cX}|$ be its cardinality. For any constant $t \in \ZZ^+$, computing 
    \begin{equation}
        \frac{\chi^{2t} K_n^2}{2}\Bigl( 1 - F_t\bigl(\sfS_{\cX}\bigr) \Bigr) \label{Eq:rescaled}
    \end{equation}
    is in $\sharpP$. This is the case even if $K_n \geq K_{\rm st}(2^n, t, 0)$.
\end{theorem}

\begin{Proof}[\cref{Thm:compSFPin2}]
    The frame potential of degree $t$ for $\sfS_{\cX}$ is given by
    \begin{align}
        F_t\bigl(\sfS_{\cX}\bigr)
        &=
        \frac{1}{\chi^{2t} K_n^2} \sum_{j,k=1}^{K_n} \Bigl| \sum_{x \in \cX} (-1)^{a_{jx} + a_{kx}} \Bigr|^{2t}\\
        &=
        \frac{1}{\chi^{2t} K_n^2} \sum_{j,k=1}^{K_n} \sum_{\vec{x} \in \cX^{2t}} (-1)^{|a_{j\vec{x}} + a_{k\vec{x}}|},
    \end{align}
    where $a_{j\vec{x}} = (a_{jx_1}, \dots, a_{jx_{2t}})$, and $|a_{j\vec{x}} + a_{k\vec{x}}| := \sum_{m=1}^{2t} (a_{j x_m} + a_{kx_m})$ with all the addition taken as XOR.

    For simplicity, we introduce a $(K_n \times \chi)$ matrix $A = (a_{jx})_{jx}$ and define a Boolean function $f_A(j,k,\vec{x})$ from $\{0, 1 \}^{2 (\log K_n + t \log \chi)}$ to $\{0, 1\}$ such as
    \begin{equation}
        f_A(j,k, \vec{x}) = |a_{j\vec{x}} + a_{k\vec{x}}|.
    \end{equation}
    As we have assumed that $a_{jx}$ are efficiently computable in classical polynomial time, $f_A(j,k,\vec{x})$ is also computable in classical polynomial time. Using $s(f_A) = \sum_{j,k,\vec{x}} f_A(j,k,\vec{x})$, which is the number of inputs satisfying $f_A(j,k,\vec{x}) = 1$, it follows that
    \begin{align}
        F_t\bigl(\sfS_{\cX}\bigr)
        &=
        \frac{1}{\chi^{2t} K_n^2} \sum_{j,k=1}^{K_n} \sum_{\vec{x} \in \cX^{2t}} (-1)^{f_A(j,k,\vec{x})}\\
        &=
        \frac{\chi^{2t}K_n^2 - 2s(f_A)}{\chi^{2t} K_n^2},
    \end{align}
    which further implies that 
    \begin{equation}
        s(f_A) = \frac{\chi^{2t} K_n^2 }{2}\Bigl( 1 - F_t\bigl(\sfS_{\cX}\bigr) \Bigr).
    \end{equation}
    Thus, computing~\cref{Eq:rescaled} is in $\sharpP$. $\hfill \qed$
\end{Proof}

\subsection{Complexity of \texorpdfstring{$\compUFP$}{compUFP}} \label{SS:CompUFP}

We now turn to $\compUFP$. Both the inclusion and the hardness of $\compUFP$ can be shown in a similar way to $\compSFP$. 

\compUFPin*

The proof of~\cref{Thm:compUFPin} is almost the same as that of~\cref{Thm:compSFPin}, except for the point that, instead of using $\ket{\varrho_{\rm st}}$, we simply use $\ket{\varrho_{\rm uni}}$, which is defined in~\cref{Eq:StateUFPEncoded}.

We next prove~\cref{Thm:compUFPhard}.

\compUFPhard*

\begin{Proof}[\cref{Thm:compUFPhard}]
    Let $f:\{0,1\}^{n-2} \rightarrow \{0,1\}$ be a  Boolean function that is computable in classical polynomial time. Consider a set $\sfU_n= \{U_f, U_p \} \cup \{U_j\}_{j=1}^{K_n-2}$ of unitaries, where
    \begin{align}
        &U_f = (I_{n-1} - 2 \ketbra{f}{f})\otimes Z, \label{Eq:U1}\\
        &U_p = (I_{n-1} - 2 \ketbra{p}{p})\otimes Z,\label{Eq:U2}\\
        &U_j = V_{j} \otimes I, \label{Eq:U3}
    \end{align}
    where $I_{n-1}$ is the identity operator on $(n-1)$ qubits, $\ket{f} = 2^{-(n-2)/2} \sum_{x \in \{0,1\}^{n-2}} \ket{x} \otimes \ket{f(x)}$ and $\ket{p} = \ket{+}^{\otimes (n-2)} \otimes \ket{1}$ are the states on $(n-1)$ qubits, $Z$ is the $1$-qubit Pauli-$Z$ operator and $\sfV_{n-1} = \{V_j\}_{j=1}^{K_n-2}$ is any set of unitaries on $(n-1)$ qubits with a computationally efficient description, the frame potential of which, of degree $t$, can be computed in classical polynomial time.
    As $\sfV_{n-1}$ has a computationally efficient description, so does the set $\sfU_n$.

    A straightforward calculation leads to
    \begin{multline}
        F_t(\sfU_n) 
        =
        2^{2t} \Bigl(1- \frac{2}{K_n} \Bigr)^2 F_t(\sfV_{n-1}) \\
        +
        \frac{2}{K_n^2}
        \Bigl[
        2^{2nt} + g_t\bigl( s(f)\bigr) \Bigr], \label{Eq:FtUnLaterUse}
    \end{multline}
    where $g_t(x) = \bigl( 2^n - 8 + 2^{-2n+7} x^2\bigr)^{2t}$, and $s(f) = \sum_{x \{0,1\}^{n-2}} f(x)$.

    Since we have assumed that $F_t\bigl(\sfV_{n-1} \bigr)$ is computable in classical polynomial time, this equation yields a polynomial-time reduction from a $\sharpP$-complete problem, $\lang{\# SAT}$, to $\compUFP$.
    $\hfill \qed$
\end{Proof}

Similarly to the comment after the proof of~\cref{Thm:compSFPhard}, we can choose various sets $\sfV_{n-1}$ with a computationally efficient description as far as its unitary frame potential can be computed in classical polynomial time.

%%% --------------------------------
\section{Analysis of complexity of \texorpdfstring{$\SFP$ and $\UFP$}{SFP and UFP}} \label{S:promiseFP}
%%% --------------------------------

In this section, we explain our results about the promise problems $\SFP$ and $\UFP$ in detail.
In~\cref{SS:SFPpoly,SS:SFPexp}, we consider $\SFP$ with the promise gap $\Omega(1/\poly(n))$ and $\cO(1/\expf(n))$, respectively. The unitary case, $\UFP$, is investigated in~\cref{SS:UFP}.

\subsection{Complexity of \texorpdfstring{$\SFP$}{SFP} with a promise gap $\Omega(1/\poly(n))$} \label{SS:SFPpoly}

\SFPBQPhard*

This result is obtained by providing a reduction from a $\BQP$-complete problem to $\SFP$.

\begin{Proof}[\cref{Thm:SFPBQPhard}]
    We denote the length of a bit string $x$ by $|x|$.
    
    Let $L=(L_{\rm yes},L_{\rm no})\subseteq\{0,1\}^\ast$ be a $\BQP$-complete problem.
    By definition, for $x \in L$, there exists a unitary $U_x$  corresponding to a polynomial-sized quantum circuit on $m\in\poly(|x|)$ many qubits such that the probability 
    \begin{equation}
        q_1(x)\coloneqq \langle 0^m|U_x^\dag(|1\rangle\langle 1|\otimes I^{\otimes m-1})U_x|0^m\rangle
    \end{equation}
    of measuring the first qubit in state $\ket 1$ satisfies
    \begin{enumerate}
        \item if $x \in L_{\rm yes}$, then $q_1(x)
        \ge\cfrac{2}{3}$,
        \item if $x \in L_{\rm no}$, then $q_1(x)\le\cfrac{1}{3}$. 
    \end{enumerate}
    We assume $m\ge 2$ without loss of generality.
    We define $V_x\coloneqq(U_x^\dag\otimes H)CZ_{1,m+1}(U_x\otimes H)$ with $CZ_{1,m+1}$ and $H$ being the controlled-$Z$ gate applied on the first and $(m+1)$th qubits and the Hadamard gate, respectively.
    As shown in Ref.~\cite{AG2018}, $q_1(x)=|\langle 0^m1|V_x|0^{m+1}\rangle|$ holds.
    
    We consider the set $\{|v_j\rangle,|a_j\rangle\}_{j=1}^K$ of state vectors, where
    \begin{eqnarray}
    |v_j\rangle&\coloneqq& V_x|0^{m+1}\rangle\otimes\cfrac{|0\rangle+e^{2ij\pi/K}|1\rangle}{\sqrt{2}},\\
    |a_j\rangle&\coloneqq&\ket{0^m} \otimes \ket{1} \otimes\cfrac{|0\rangle+e^{2ij\pi/K}|1\rangle}{\sqrt{2}},
    \end{eqnarray}
    and $K$ is set to be larger than $t$.
    Since $\{U_x\}_x$ is a uniform family of polynomial-sized quantum circuits, this is a multiset of state vectors with a computationally efficient description.
    For simplicity, let $|\varphi_j\rangle\coloneqq(|0\rangle+e^{2ij\pi/K}|1\rangle)/\sqrt{2}$ for all $1\le j\le K$.
    
    The state frame potential of degree $t$ for the above set is
    \begin{align}
    &\cfrac{1}{4K^2}\sum_{j,k=1}^K\left(|\langle v_j|v_k\rangle|^{2t}+|\langle a_j|a_k\rangle|^{2t}+2|\langle a_j|v_k\rangle|^{2t}\right)\ \ \ \ \\
    &=\cfrac{1+{q_1(x)}^{2t}}{2}\left(\cfrac{1}{K^2}\sum_{j,k=1}^K|\langle\varphi_j|\varphi_k\rangle|^{2t}\right)\\
    &=\cfrac{1+{q_1(x)}^{2t}}{4^t}\binom{2t-1}{t-1},   \label{Eq:BQPcompFP}
    \end{align}
      where we have used~\cref{Eq:FGK} to derive the last equation.
    From~\cref{Eq:BQPcompFP}, we can solve the $\BQP$-complete problem by deciding whether the frame potential is at least
    \begin{eqnarray}
    \cfrac{\binom{2t-1}{t-1}}{4^t}+\cfrac{\binom{2t-1}{t-1}}{9^t}
    \end{eqnarray}
    or at most
    \begin{eqnarray}
    \cfrac{\binom{2t-1}{t-1}}{4^t}+\cfrac{\binom{2t-1}{t-1}}{36^t}.
    \end{eqnarray}
    Therefore, the promise gap $\epsilon$ is
    \begin{eqnarray}
    \left(\cfrac{1}{9^t}-\cfrac{1}{36^t}\right)\binom{2t-1}{t-1},
    \end{eqnarray}
    and hence, it is constant.
    $\hfill \qed$
\end{Proof}

\subsection{Complexity of \texorpdfstring{$\SFP$}{SFP} with a promise gap $\cO(1/\expf(n))$} \label{SS:SFPexp}

\approxSFPexpin*

Our proof of~\cref{Thm:approxSFPexpin} is based on the complexity class $\PQP$, i.e., is a quantum analogue of $\PP$.

\begin{Proof}[\cref{Thm:approxSFPexpin}]
    Consider the following algorithm:
    \begin{enumerate}
        \item With probability $p=2/(2+\alpha+\beta)$, proceed to the next step.
        With the remaining probability, output ``reject."
        \item Choose $i$ and $j$ from $\{1,2,\ldots,K_n\}$ uniformly at random. 
        \item Perform the swap test for $|\psi_i\rangle^{\otimes t}$ and $|\psi_j\rangle^{\otimes t}$, where $\ket{\psi_i}, \ket{\psi_j} \in \sfS_n$.
        If the outcome is $0$ ($1$), output ``accept" (``reject").
    \end{enumerate}
    Note that the second step, choosing $i$ and $j$, can be done in $\poly(n)$ time since we consider a set with a computationally efficient description, the cardinality of which is $K_n \in \cO(2^{\poly(n)})$.
    
    The acceptance probability of this algorithm is
    \begin{equation}
        \cfrac{p}{K_n^2}\sum_{i,j=1}^{K_n}\frac{1+|\langle\psi_i|\psi_j\rangle|^{2t}}{2}=p\cfrac{1+F_t(\sfS_n)}{2}.
    \end{equation}
    Therefore, the acceptance probability is either
    \begin{equation}
        \ge p\cfrac{1+\alpha}{2}=\cfrac{1}{2}+\cfrac{\alpha-\beta}{2(2+\alpha+\beta)}
    \end{equation}
    or
    \begin{equation}
        \le p\cfrac{1+\beta}{2}=\cfrac{1}{2}-\cfrac{\alpha-\beta}{2(2+\alpha+\beta)}.
    \end{equation}
    This means that $\SFP(t,\epsilon)$ is in $\PQP$ if $\alpha - \beta > \epsilon \in \Omega\bigl(2^{-\poly(n)}\bigr)$. The proof is completed by the fact that $\PQP = \PP$. $\hfill \qed$
\end{Proof}

This proof also provides an alternative proof of~\cref{Thm:SFPBQPin}, stating that $\SFP(t,\epsilon) \in \BQP$ when $\epsilon \in \Omega(1/{\rm poly}(n))$.\\

We next provide the proof of the hardness of $\SFP$ for sufficiently small promise gaps.

\approxSFPexphard*

To show~\cref{Thm:approxSFPexphard}, we provide a polynomial-time reduction from a $\PP$-complete problem, \textsf{MAJ-SAT}, to $\SFP$.
Below, we prove~\cref{Thm:approxSFPexphard} by explicitly embedding the quantity $s(f)$ in \textsf{MAJ-SAT} to the state frame potential $F_t(\sfS_n)$.

\begin{Proof}[\cref{Thm:approxSFPexphard}]
    Let $f: \{0,1\}^{n-2} \rightarrow \{0,1\}$ be a Boolean function that is computable in classical polynomial time. Consider the same set $\sfS_n$ defined by~\cref{Eq:S1,Eq:S2,Eq:S3}, which has a computationally efficient description.
    As in~\cref{Eq:FtSnLaterUse}, the state frame potential $F_t$ of degree $t$ for $\sfS_{n}$ is given by
    \begin{multline}
        F_t(\sfS_n)
        =
        \biggl( 1 - \frac{2}{K_n} \biggr)^2 F_t\bigl(\sfS_{\varphi} \bigr) +
        \frac{2}{K_n^2} \biggl[ 1 + \biggl(\frac{s(f)}{2^{n-2}}\biggr)^{2t}\biggr].
    \end{multline}
    
    We set $\alpha$ and $\beta$ as    
    \begin{align}
        \alpha &\coloneqq 
        \biggl( 1 - \frac{2}{K_n} \biggr)^2 F_t\bigl(\sfS_{\varphi} \bigr)        +
        \frac{2}{K_n^2} \biggl[ 1 + 2^{-2t}\biggr],\\
        \beta &\coloneqq 
        \biggl( 1 - \frac{2}{K_n} \biggr)^2 F_t\bigl(\sfS_{\varphi} \bigr)        +
        \frac{2}{K_n^2} \biggl[ 1 + \Bigl(\frac{2^{n-3} - 1}{2^{n-2}} \Bigr)^{2t}\biggr].
    \end{align}
    It suffices for solving \textsf{MAJ-SAT} to determine either $F_t(\sfS_n) \geq \alpha$ or $\leq \beta$. As \textsf{MAJ-SAT} is $\PP$-complete (\cref{Thm:MAJSAT}), $\SFP(t, \alpha- \beta)$ is $\PP$-hard.

    The promise gap, i.e., $\alpha - \beta$, is evaluated as
    \begin{align}
        \alpha-\beta
        &=\frac{1}{2^{2t-1}K_n^2}\left[1-\left(1-\frac{1}{2^{n-3}}\right)^{2t}\right]\\
        &\ge \frac{t}{2^{2t-1}K_n^2(2^{n-4} + t)},
    \end{align}
    where we have used $(1-\epsilon)^m \leq (1+m\epsilon)^{-1}$.$\hfill \qed$
\end{Proof}

\subsection{Complexity of \texorpdfstring{$\UFP$}{UFP}} \label{SS:UFP}

In a similar way to $\SFP$, we can clarify the computational complexity class of $\UFP$.

\approxUFPin*

\begin{Proof} [\cref{Thm:approxUFPin}]
    Let $\alpha'\coloneqq\alpha/2^{2tn}$ and $\beta'\coloneqq\beta/2^{2tn}$.
    We run the following algorithm:
    \begin{enumerate}
    \item With probability $p=2/(2+\alpha'+\beta')$, proceed to the next step.
    With the remaining probability, output ``reject."
    \item Select $i$ and $j$ from $\{1,2,\ldots,K_n\}$ uniformly at random.
    \item Let $\ket{\Phi} = \frac{1}{\sqrt{2^n}}\sum_{k=0}^{2^n-1}|kk\rangle$ be a maximally entangled state. Perform the swap test for
    \begin{equation}
        \ket{\Psi_i} \coloneqq \bigl(\left(I^{\otimes n}\otimes U_i\right) \ket{\Phi} \bigr)^{\otimes t}
    \end{equation}
    and
    \begin{equation}
        \ket{\Psi_j} \coloneqq \bigl(\left(I^{\otimes n}\otimes U_j\right) \ket{\Phi} \bigr)^{\otimes t},
    \end{equation}
    where $U_i, U_j \in \sfU_n$.
    If the outcome is $0$ ($1$), output ``accept" (``reject").
    \end{enumerate}
    
    The acceptance probability of this algorithm is
    \begin{equation}
        \frac{p}{K_n^2}\sum_{i,j=1}^{K_n}\cfrac{1+|\langle\Psi_i|\Psi_j\rangle|^2}{2}=p\cfrac{1+F_t(\sfU_n)/2^{2tn}}{2}, \label{Eq:UFPSFPdifference}
    \end{equation}
    which is either
    \begin{equation}
        \ge p\cfrac{1+\alpha'}{2}=\cfrac{1}{2}+\cfrac{\alpha'-\beta'}{2(2+\alpha'+\beta')}
    \end{equation}
    or
    \begin{equation}
        \le p\cfrac{1+\beta'}{2}=\cfrac{1}{2}-\cfrac{\alpha'-\beta'}{2(2+\alpha'+\beta')}.
    \end{equation}
    This implies that $\UFP(t,\epsilon)$ is in $\PQP$. The fact that $\PQP=\PP$ completes the proof.
$\hfill \qed$
\end{Proof}

Unlike the case of $\SFP(t,\epsilon)$, this proof does not imply that $\UFP(t,\epsilon)$ is in $\BQP$ even if $\epsilon \in \Omega(1/{\rm poly}(n))$. This is due to the factor $2^{-2tn}$ in $\alpha'$ and $\beta'$.

\approxUFPhard*

\begin{Proof}[\cref{Thm:approxUFPhard}]
    We provide a polynomial-time reduction from a $\PP$-complete problem, \textsf{MAJ-SAT}, to $\UFP(t,\epsilon)$.
    Let $f:\{0,1\}^{n-2} \rightarrow \{0,1\}$ be a Boolean function that is computable in classical polynomial time. Consider the same set $\sfU_n$ defined by~\cref{Eq:U1,Eq:U2,Eq:U3}, of unitaries on $n$ qubits, which has a computationally efficient description.
    As in~\cref{Eq:FtUnLaterUse}, the unitary frame potential for $\sfU_n$ is 
    \begin{multline}
        F_t(\sfU_n) 
        =
        2^{2t}\biggl( 1 - \frac{2}{K_n}\biggr)^2 F_t(\sfV_{n-1}) \\
        +
        \frac{2}{K_n^2}\bigl[2^{2nt}+ g_t\bigl(s(f)\bigr)  \bigr],
    \end{multline}
    where $g_t(x) = \bigl( 2^n - 8 + 2^{-2n+7} x^2\bigr)^{2t}$.

    We set $\alpha$ and $\beta$ such as
    \begin{align}
        &\alpha \coloneqq
         2^{2t}\biggl( 1 - \frac{2}{K_n}\biggr)^2 F_t(\sfV_{n-1}) 
        +
        \frac{2}{K_n^2}\bigl[2^{2nt}+ g_t\bigl(2^{n-3}\bigr)  \bigr],\\
        &\beta \coloneqq
         2^{2t}\biggl( 1 - \frac{2}{K_n}\biggr)^2 F_t(\sfV_{n-1}) 
        +
        \frac{2}{K_n^2}\bigl[2^{2nt}+ g_t\bigl(2^{n-3} - 1\bigr)  \bigr].
    \end{align}
    If one can determine whether $F_t(\sfU_n) \geq \alpha$ or $\leq \beta$, one can also determine if $M < 2^{n-3}$ or $M\geq 2^{n-3}$. Since the latter is \textsf{MAJ-SAT} and $\PP$-complete due to~\cref{Thm:MAJSAT}, the former is \PP-hard.

    The promise gap $\epsilon = \alpha - \beta$ is computed as
    \begin{equation}
        \epsilon \geq \frac{2t(2^n-6)^{2t} \bigl(1 + \cO(2^{-n})\bigr)}{K_n^2[2^{n-6}(2^n-6)+t]} \in \Theta(2^{2(t-1)n}/K_n^2).
    \end{equation}
    This concludes the proof.    $\hfill \qed$
\end{Proof}

%%% --------------------------------
\section{Analysis of complexity of \texorpdfstring{$\StDes$ and $\UniDes$}{StDes and UniDes}} \label{S:promiseDES}
%%% --------------------------------

We investigate the complexity class of $\StDes$ in~\cref{SS:StDes} and that of $\UniDes$ in~\cref{SS:UniDes}.

\subsection{Complexity of \texorpdfstring{$\StDes$}{StDes}} \label{SS:StDes}

The fact that $\StDes$ is in $\PP$ if $\delta'$ and $\delta$ satisfy a certain condition directly follows from~\cref{Thm:approxSFPexpin}, i.e., from the statement that  $\SFP(t, \epsilon) \in \PP$ if $\epsilon$ is not too small.

\StDESin*

\begin{Proof}[\cref{Thm:StDESin}]
    Due to the promise of the problem, it is guaranteed that a given set $\sfS_n$ is either a $\delta$-approximate or not a $\delta'$-approximate state $t$-design. If the former is the case, $F_t(\sfS_n) \leq 1/d_{t} + \delta^2/d_t$, and, if the latter is the case, $F_t(\sfS_n) > 1/d_{t} + \delta'^2/d_t^2$. Since $(1/d_{t} + \delta'^2/d_t^2) - (1/d_{t} + \delta^2/d_t) \in \Omega\bigl(2^{-\poly(n)}\bigr)$ by assumption, checking which is the case is in $\PP$ due to~\cref{Thm:approxSFPexpin}. Hence, $\StDes(t, t, \delta, \delta') \in \PP$. $\hfill \qed$
\end{Proof}

We next show the hardness of $\StDes$. The following is the restatement of~\cref{Thm:StDEShard}.

\StDEShard*

\begin{Proof}[\cref{Thm:StDEShard}]
    In the proof, we use the notation $d=2^n$ for simplicity.
    Let $f: \{0,1\}^{n-1} \rightarrow \{0,1\}$ be a Boolean function that is computable in classical polynomial time, and let $\ket{f}$ and $\ket{p}$ be $n$-qubit state vectors given by, respectively,
    \begin{align}
        &\ket{f} = \sqrt{\frac{2}{d}} \sum_{x \in \{0,1\}^{n-1}} \ket{x}  \ket{f(x)},\\
        &\ket{p} =\ket{+}^{\otimes n-1}  \ket{1}.
    \end{align}
    Let  $\delta_0$ be such that $\delta_0\in o(\delta/d)$, and $\sfD_{t'}$ be a set of $n$-qubit state vectors with a computationally efficient description that forms an exact state $1$-design and a $\delta_0$-approximate state $t'$-design. Such designs exist as $\delta \in \Omega(2^{-\poly(n)})$. For instance, it can be generated by first applying Pauli operators uniformly at random to a fixed state and then applying a local random circuit with $\poly(n)$ depth. The first application of Pauli operators guarantees that the generated states form an exact state $1$-design.
    We denote by $D$ the cardinality of $\sfD_{t'}$.

    We define a multiset $\sfS_n(f)$ of state vectors by
    \begin{equation}
        \sfS_n(f) = \{\ket{f}, \ket{p}\}^{\oplus m} \cup \sfD_{t'},
    \end{equation}
    where $\oplus m$ implies the multiplicity $m \in \mathbb{Z}^+$. 
    A key quantity in the proof is $D/m$, which we will later choose appropriately depending on $\delta$. Here, we note that $D/m$ can be an arbitrary positive rational number because we can choose arbitrarily large positive integers $m$ and $D$. Note that this is the case for $D$ since, given a $\delta_0$-approximate state $t'$-design with some cardinality, one can increase its cardinality by changing the multiplicity of each state vector. The resulting multiset is a $\delta_0$-approximate state $t'$-design with cardinality that is an integer multiple of the original one.    
    
    For $\tau \in \{1, t, t'\}$, the set $\sfS_n(f)$ is a $\delta_{\tau}(f)$-approximate state $\tau$-design, where 
    \begin{equation}
        \delta_{\tau}(f) \coloneqq d_{\tau} \left\| \Delta^{(\tau)}_{\sfS_n(f)} \right\|_{\infty}    
    \end{equation}
    with
    \begin{equation}
        \Delta^{(\tau)}_{\sfS_n(f)} \coloneqq\frac{\Pi_{\textrm{sym}}^{(\tau)}}{d_\tau} - \mathbb{E}_{\ket{\varphi}\sim \sfS_n(f)}\bigl[ \ketbra{\varphi}{\varphi}^{\otimes \tau} \bigr].
    \end{equation}
    Here, $\Pi_{\textrm{sym}}^{(\tau)}$ is the projection onto the symmetric subspace $\cH^{(\tau)}_{\rm sym}$, $d_{\tau}=\tr{\Pi_{\textrm{sym}}^{(\tau)}}$, and
    \begin{multline}
        \mathbb{E}_{\ket{\varphi}\sim \sfS_n(f)}\bigl[ \ketbra{\varphi}{\varphi}^{\otimes \tau} \bigr] \coloneqq \\         
        \frac{1}{2 m + D} \Bigl[m \bigl(\ketbra{f}{f}^{\otimes \tau} + \ketbra{p}{p}^{\otimes \tau} \bigr) + D \,\mathbb{E}_{\ket{\varphi}\sim \sfD_{t'}}\bigl[ \ketbra{\varphi}{\varphi}^{\otimes \tau} \bigr] \Bigr].
    \end{multline}
    Below, we compute $\delta_{\tau}(f)$.

    We begin by checking the eigenvalues of $\ketbra{f}{f}^{\otimes \tau} + \ketbra{p}{p}^{\otimes \tau}$. By expanding $\ket{f}^{\otimes \tau}$ as $\alpha_{\tau} \ket{p}^{\otimes \tau} + \sqrt{1 - \alpha_{\tau}^2} \ket{q}$, where 
    \begin{equation}
        \alpha_{\tau} = \braket{p}{f}^{\tau} = \left(\frac{2s(f)}{d}\right)^{\tau},   \label{Eq:alphadefSt}
    \end{equation}
    with $s(f) = \sum_{x \in \{0,1\}^{n-1}} f(x)$, and $\ket{q}$ orthogonal to $\ket{p}^{\otimes \tau}$, we have
    \begin{equation}
    \ketbra{f}{f}^{\otimes \tau} + \ketbra{p}{p}^{\otimes \tau}
    =
    \begin{pmatrix}
        1 + \alpha_{\tau}^2 & \alpha_{\tau} \sqrt{1 - \alpha_{\tau}^2} \\
        \alpha_{\tau} \sqrt{1 - \alpha_{\tau}^2} & 1 - \alpha_{\tau}^2
    \end{pmatrix},
    \end{equation}
    where the matrix representation is in terms of $\{ \ket{p}^{\otimes \tau}, \ket{q} \}$. 
    Hence, the eigenvalues are $1 \pm \alpha_{\tau}$.  Below, we denote the corresponding eigenvectors by $\ket{\alpha_{\tau,\pm}}$.

    When $\tau =1$, we can use the fact that $\sfD_{t'}$ is an exact state $1$-design. That is,
    \begin{equation}
        \mathbb{E}_{\ket{\varphi}\sim \sfD_{t'}}\bigl[ \ketbra{\varphi}{\varphi} \bigr]   = \frac{I}{d}.
    \end{equation}
    Also, $\Pi_{\textrm{sym}}^{(1)} = I$ and $d_1 = d$. Hence, it follows that 
    \begin{align}
        \delta_1(f) &= \frac{m}{2m+D} \left\| 2 I - d\left(\ketbra{f}{f} + \ketbra{p}{p}\right)\right\|_{\infty}\\
        &= \frac{1}{D/m+2}\left( d + 2s(f) - 2 \right),
    \end{align}
    where we have used that the eigenvalues of $\ketbra{f}{f} + \ketbra{p}{p}$ are $1 \pm \alpha_1$, which is $1 \pm 2s(f)/d$.

    When $\tau > 1$, we derive an upper and a lower bound on $\delta_{\tau}(f)$.
    Using the eigenvectors $\ket{\alpha_{\tau, \pm}}$, a lower bound on $\delta_{\tau}(f)$ is given as 
    \begin{multline}
        d_{\tau} \max\Bigl\{ \bigl| \bra{\alpha_{\tau, +}} \Delta^{(\tau)}_{\sfS_n(f)} \ket{\alpha_{\tau, +}} \bigr|, \bigl| \bra{\alpha_{\tau, -}} \Delta^{(\tau)}_{\sfS_n(f)} \ket{\alpha_{\tau, -}}\bigr| \Bigr\}\\
        \leq \delta_{\tau}(f), \label{Eq:maxforlowerbound}
    \end{multline}
    which turns out to be
    \begin{equation}
        d_{\tau} | \bra{\alpha_{\tau, +}} \Delta^{(\tau)}_{\sfS_n(f)} \ket{\alpha_{\tau, +}} | \leq \delta_{\tau}(f). \label{Eq:aaaaa}
    \end{equation}

    As $\ket{\alpha_{\tau, +}} \in \cH^{(\tau)}_{\rm sym}$, we have
    \begin{multline}
        \bra{\alpha_{\tau, +}} \Delta^{(\tau)}_{\sfS_n(f)} \ket{\alpha_{\tau, +}}
        =
        \frac{1}{d_{\tau}} -
        \frac{1}{2m+D}\biggl( m(1+\alpha_{\tau}) + \\ 
        D \bra{\alpha_{\tau, +}} \mathbb{E}_{\ket{\varphi}\sim \sfD_{t'}}\bigl[ \ketbra{\varphi}{\varphi}^{\otimes \tau} \bigr] \ket{\alpha_{\tau, +}} \biggr).\label{Eq:bbbb}
    \end{multline}
    Hence,
    \begin{multline}
        \delta_{\tau}(f)
        \geq
        \biggl|
        1 -
        \frac{d_{\tau}}{2m+D}\biggl( m(1+\alpha_{\tau}) +  \\
        D \bra{\alpha_{\tau, +}} \mathbb{E}_{\ket{\varphi}\sim \sfD_{t'}}\bigl[ \ketbra{\varphi}{\varphi}^{\otimes \tau} \bigr] \ket{\alpha_{\tau, +}} \biggr)
        \biggr|. \label{Eq:aegr;oaerg}
    \end{multline}
    Moreover, as $\tau \in \{1,t, t'\}$ and $t\leq t'$, $\sfD_{t'}$ is a $\delta_0$-approximate state $\tau$-design. This implies that
    \begin{equation}
        \frac{1-\delta_0}{d_{\tau}} 
        \leq
        \bra{\alpha_{\tau, +}} \mathbb{E}_{\ket{\varphi}\sim \sfD_{t'}}\bigl[ \ketbra{\varphi}{\varphi}^{\otimes \tau} \bigr] \ket{\alpha_{\tau, +}}
        \leq
        \frac{1+\delta_0}{d_{\tau}}. \label{Eq:ccccc}
    \end{equation}
    From~\cref{{Eq:aegr;oaerg},Eq:ccccc}, we obtain a lower bound on $\delta_{\tau}(f)$, which reads as
    \begin{equation}
        \frac{1}{D/m+2} \left[(1+\alpha_{\tau})d_{\tau} - 2 - \frac{D}{m} \delta_0 \right] \leq \delta_{\tau}(f).
    \end{equation}
    Remember that $\alpha_{\tau} = (2s(f)/d)^{\tau}$.
    For the later convenience, we introduce a real-valued function $\textrm{LB}_{\tau} (q, x)$ as
    \begin{equation}
        \textrm{LB}_{\tau} (q, x) \coloneqq \frac{1}{q+2} \left\{ \left[ 1 + \Bigl(\frac{2x}{d} \Bigr)^{\tau} \right] d_{\tau} - 2  - q \delta_0\right\}. \label{Eq:DefLBSt}
    \end{equation}
    Using this notation, we can rephrase the lower bound as $\textrm{LB}_{\tau}\bigl(D/m, s(f) \bigr) \leq \delta_{\tau}(f)$.

    It is easy to obtain an upper bound on $\delta_{\tau}(f)$. From the triangle inequality, we obtain
    \begin{multline}
        \delta_{\tau}(f)
        \leq 
        d_{\tau}\Bigl\| \frac{\Pi_{\textrm{sym}}^{(\tau)}}{d_\tau} -  \frac{D}{2m + D} \mathbb{E}_{\ket{\varphi}\sim \sfD_{t'}}\bigl[ \ketbra{\varphi}{\varphi}^{\otimes \tau} \bigr]\Bigr\|_{\infty} \\
        +
        \frac{m d_{\tau}}{2m + D}  \Bigl\| \ketbra{f}{f}^{\otimes \tau} + \ketbra{p}{p}^{\otimes \tau} \Bigr\|_{\infty}. 
    \end{multline}
    As $\sfD_{t'}$ is a $\delta_0$-approximate state $\tau$-design, it holds that
    \begin{multline}
        \Bigl\| \frac{\Pi_{\textrm{sym}}^{(\tau)}}{d_\tau} -  \frac{D}{2m + D} \mathbb{E}_{\ket{\varphi}\sim \sfD_{t'}}\bigl[ \ketbra{\varphi}{\varphi}^{\otimes \tau} \bigr]\Bigr\|_{\infty}\\
        \leq
        \Bigl( 1 - \frac{D}{2m + D} \Bigr)\frac{1}{d_{\tau}} + \frac{D}{2m + D} \frac{\delta_0}{d_{\tau}}.
    \end{multline}
    We also have
    \begin{equation}
        \Bigl\| \ketbra{f}{f}^{\otimes \tau} + \ketbra{p}{p}^{\otimes \tau} \Bigr\|_{\infty} = 1+\alpha_{\tau},
    \end{equation}
    from the diagonalization of $\ketbra{f}{f}^{\otimes \tau} + \ketbra{p}{p}^{\otimes \tau}$. Hence, we obtain
    \begin{equation}
        \delta_{\tau}(f) \leq \frac{1}{D/m+2} \left[ (1+\alpha_{\tau})d_{\tau} + 2 +\frac{D}{m} \delta_0\right].
    \end{equation}
    Similarly to the lower bound, we introduce a real-valued function $\textrm{UB}_{\tau}(q, x)$ as
    \begin{equation}
        \textrm{UB}_{\tau} (q, x) \coloneqq \frac{1}{q+2} \left\{ \left[ 1 + \Bigl(\frac{2x}{d} \Bigr)^{\tau} \right] d_{\tau} +2 +  q \delta_0\right\}, \label{Eq:DefUBSt}
    \end{equation}
    and write the upper bound as $\delta_{\tau}(f) \leq \textrm{UB}_{\tau}\bigl(D/m, s(f) \bigr)$.\\

    To summarize so far, we have obtained
    \begin{equation}
        \delta_{1}(f) =  \frac{d + 2s(f) - 2}{D/m + 2}, \label{Eq:delta1State}
    \end{equation}
    and for $\tau >1$, a lower and an upper bound on $\delta_{\tau}(f)$:
    \begin{equation}
        \textrm{LB}_{\tau}\left(\frac{D}{m}, s(f) \right) \leq \delta_{\tau}(f) \leq \textrm{UB}_{\tau}\left(\frac{D}{m}, s(f) \right). \label{Eq:ULboundsSt}
    \end{equation}
    Importantly, for any fixed $q \geq 0$, both $\textrm{LB}_{\tau}(q, x)$ and $\textrm{LB}_{\tau}(q, x)$ are monotonically increasing in $x$ when $x\geq 0$.
    
    We now choose $D/m$. 
    For our purpose, we specifically use $D/m$ that approximately satisfies the following: if $t = 1$
    \begin{equation}    
        \frac{d + 2s_0 - 2}{D/m + 2} = \delta, \label{Eq:delta11State}
    \end{equation}
    with $s_0 = d/4-2/3$, and if $t>1$,
    \begin{equation}
        \textrm{LB}_{t}\left(\frac{D}{m}, \frac{d}{4} - \frac{2}{3} \right) = \delta. \label{Eq:lowerdelta}
    \end{equation}
    As $\textrm{LB}_{t}$ is defined as~\cref{Eq:DefLBSt}, the latter equation is explicitly given by
    \begin{equation}
        \frac{1}{D/m+2} \biggl[ \Bigl\{ 1 + \Bigl(\frac{1}{2} - \frac{4}{3d} \Bigr)^{t} \Bigr\} d_{t} - 2 \biggr] - \frac{D/m}{D/m+2} \delta_0 =\delta.
    \end{equation}
    As $D/m$ can be an arbitrary positive rational number by choosing the appropriate cardinality $D$ of $\sfD_{t'}$ and multiplicity $m$ of $\{\ket{f}, \ket{p}\}$ in $\sfS_n(f)$, the solutions of these equations can be approximated by $D/m$ with arbitrary precision. 
    
    With this choice of $D/m$, we set $\delta'$ as follows. If $t'=1$, 
    \begin{equation}    
        \delta' = \frac{d + 2s_1 - 2}{D/m + 2},\label{Eq:delta111State}
    \end{equation}
    with $s_1 = d/4-1/3$, and if $t'>1$,
    \begin{align}
        \delta' &\coloneqq \textrm{UB}_{t'}\left(\frac{D}{m}, \frac{d}{4} - \frac{1}{3} \right) \label{Eq:Defdelta'}\\
        &= \frac{1}{D/m+2} \biggl[ \Bigl\{ 1 + \Bigl(\frac{1}{2} - \frac{2}{3d} \Bigr)^{t'} \Bigr\} d_{t'} + 2 \biggr] + \frac{D/m}{D/m+2} \delta_0.
    \end{align}
    From simple calculations and using $\delta_0 \in o(\delta/d)$, it follows that 
    \begin{equation}
        \delta' \in \Theta(d^{t'-t}\delta).
    \end{equation}
    Clearly, $\delta' - \delta \in \Theta(\delta') = \Theta(d^{t'-t}\delta)$ for $t<t'$. When $t=t'$, it is also straightforward to see that $\delta' - \delta \in \Theta(\delta/d)$. 
    
    Below, we show that \textsf{MAJ-SAT} can be solved by a single use of the algorithm $\cA$ for solving $\StDes(t,t',\delta,\delta')$. By definition,  
    the algorithm $\cA$ can determine whether $\sfS_n(f)$ satisfies
    \begin{equation}
        \delta_{t'}(f) >\delta', \ \ \textrm{or} \ \ \delta_t(f) \leq \delta, \label{Eq:ppqsrtstate}
    \end{equation}
    under the exclusive assumption that only one of the two holds.

    We begin by checking that, for any Boolean function $f: \{0,1\}^{n-1} \rightarrow \{0,1\}$, the corresponding set $\sfS_n(f)$ must exclusively satisfy one of the two conditions in~\cref{Eq:ppqsrtstate}. 
    We check this condition separately for three cases. The first case is $t=t' =1$, the second case is $t=1<t'$, and the last case is $1< t\leq t'$.

    In the first case, in which $t=t'=1$, we observe from~\cref{Eq:delta1State,Eq:delta11State,Eq:delta111State} that 
    \begin{align}
        \delta_1(f) \leq \delta &\Longleftrightarrow s(f) \leq s_0 = \frac{d}{4}-\frac{2}{3},\\  
        \delta_{1}(f) >\delta' &\Longleftrightarrow \frac{d}{4}-\frac{1}{3} = s_1 < s(f).
    \end{align}
    As $s(f)$ is a non-negative integer, these cases are equivalent to
    \begin{align}
        \delta_1(f) \leq \delta &\Longleftrightarrow s(f) \leq  \frac{d}{4}-1, \label{Eq:case1a}\\  
        \delta_{1}(f) >\delta' &\Longleftrightarrow \frac{d}{4} \leq s(f).\label{Eq:case1b}
    \end{align}
    It is clear that all Boolean functions must exclusively satisfy one of the two.

    In the second case, in which $t=1<t'$, as shown above, $\delta_1(f) \leq \delta$ is equivalent to $s(f) \leq d/4-1$. 
    On the other hand, if $\delta_{t'}(f) >\delta'$, it follows from~\cref{Eq:Defdelta',Eq:ULboundsSt} that 
    \begin{equation}
         \textrm{UB}_{t'}\left(\frac{D}{m}, \frac{d}{4} - \frac{1}{3} \right) = \delta' < \delta_{t'}(f)\le \textrm{UB}_{t'}\left(\frac{D}{m}, s(f) \right).
    \end{equation}
    As $\textrm{UB}_{t'}(q,x)$ is monotonically increasing in $x$ for any fixed $q \geq 0$, this implies that $s(f)>d/4-1/3$. As $s(f)$ is a non-negative integer, we conclude that $\delta_{t'}(f) >\delta'$ implies that $s(f) \geq d/4$.
    We also show that, if $s(f) \geq d/4$, then $\delta_{t'}(f) >\delta'$. 
    That is, $\delta_{t'}(f) >\delta'$ is equivalent to $s(f) \geq d/4$.
    We show the contraposition of this statement. To this end, it is crucial to note that 
    \begin{equation}
        \textrm{UB}_{t'}\left(\frac{D}{m}, \frac{d}{4}-\frac{1}{3}\right)< \textrm{LB}_{t'}\left(\frac{D}{m}, \frac{d}{4}\right), \label{Eq:a;oierwfgno;q4wjikg'3q4'a}
    \end{equation}
    which is obtained by the direct calculation using $1<t'$, and the fact that $\sfD_{t'}$ is a $\delta_0$-approximate state $t'$-design with $\delta_0 \in o(\delta/d)$. As $\delta'$ for $t'>1$ is defined as~\cref{Eq:Defdelta'}, this implies that
    \begin{equation}
        \delta'  < \textrm{LB}_{t'}\left(\frac{D}{m}, \frac{d}{4} \right). \label{Eq:La;okwnomovmear}
    \end{equation}
    Thus, if $\delta_{t'}(f) \leq \delta'$, we obtain, by using~\cref{Eq:ULboundsSt},
    \begin{equation}
        \textrm{LB}_{t'}\left(\frac{D}{m}, s(f) \right) \leq \delta_{t'}(f) \leq \delta' < \textrm{LB}_{t'}\left(\frac{D}{m},\frac{d}{4}\right).
    \end{equation}
    Using the monotonicity of $\textrm{LB}_{t'}(q,x)$, this reduces to $s(f) < d/4$. That is, $\delta_{t'}(f) \leq \delta'$ implies that $s(f) < d/4$. Taking the contraposition of this, we obtain that $s(f) \geq d/4$ implies that $\delta_{t'}(f) > \delta'$.

    All together, in the case of $t=1<t'$, we have
    \begin{align}
        \delta_1(f) \leq \delta &\Longleftrightarrow s(f) \leq \frac{d}{4} - 1, \label{Eq:case2a}\\
        \delta_{t'}(f) > \delta' &\Longleftrightarrow \frac{d}{4} \leq s(f). \label{Eq:case2b}
    \end{align}
    Clearly, all Boolean functions must exclusively satisfy one of the two.

    Finally, when $1<t\leq t'$, we begin by rephrasing $\delta_{t}(f) \leq \delta$ in terms of $s(f)$.
    If $\delta_{t}(f) \leq \delta$, we have from~\cref{Eq:ULboundsSt,Eq:lowerdelta}
    \begin{equation}
        \textrm{LB}_{t}\left(\frac{D}{m}, s(f) \right) \leq \delta_t(f) \leq \delta = \textrm{LB}_{t}\left(\frac{D}{m}, \frac{d}{4} - \frac{2}{3} \right).
    \end{equation}
    As $\textrm{LB}_{t}(q,x)$ is monotonically increasing in $x$ for any fixed $q \geq 0$, together with the fact that $s(f)$ is a non-negative integer, this implies that $s(f) \leq d/4 - 1$.
    We can also show that $d/4 - 1 \geq s(f)$ implies that $\delta_{t}(f) \leq \delta$. To this end, we consider the contraposition. The crucial observation is
    \begin{align}
        &\textrm{UB}_{t}\left(\frac{D}{m}, \frac{d}{4}-1 \right) \leq  \textrm{LB}_{t}\left(\frac{D}{m}, \frac{d}{4}-\frac{2}{3}\right),
    \end{align}
    which can be shown using $1<t$ and the fact that $\sfD_{t'}$ is a $\delta_0$-approximate state $t$-design with $\delta_0 \in o(\delta/d)$. Note that $t\leq t'$. 
    From~\cref{Eq:lowerdelta}, this implies that
    \begin{align}
        &\textrm{UB}_{t}\left(\frac{D}{m}, \frac{d}{4}-1\right) \leq \delta. \label{Eq:aargoopq34'pgaek}
    \end{align}
    Hence, if $\delta_t(f) > \delta$, then 
    \begin{align}
        &\textrm{UB}_{t}\left(\frac{D}{m}, \frac{d}{4}-1 \right) \leq \delta < \delta_t(f)  \leq \textrm{UB}_{t}\left(\frac{D}{m},s(f) \right),
    \end{align}
    where we have used~\cref{Eq:ULboundsSt}. Using the monotonicity of $\textrm{UB}_t(q,x)$, this further implies that $d/4-1 < s(f)$. By taking the contraposition, we obtain that $d/4-1 \geq s(f)$ implies that $\delta_t(f) \leq \delta$.

    Therefore, when $1<t\leq t'$, we have
    \begin{align}
        \delta_{t}(f) \leq \delta & \Longleftrightarrow s(f) \leq \frac{d}{4} -1,\label{Eq:case3a}\\
        \delta_{t'}(f) > \delta' &\Longleftrightarrow \frac{d}{4} \leq s(f). \label{Eq:case3b}
    \end{align}
    Note that the latter follows from~\cref{Eq:case2b}.
    Thus, also in the case of $1<t\le t'$, all Boolean functions must exclusively satisfy one of the two conditions in~\cref{Eq:ppqsrtstate}.\\

    We finally show that the algorithm $\cA$ for solving $\StDes(t,t',\delta,\delta')$ can be used to solve \textsf{MAJ-SAT}. 
    We simply apply the algorithm $\cA$ to the set $\sfS_n(f)$ constructed from a Boolean function $f$. As we have shown, $\sfS_n(f)$ satisfies the exclusive promise of $\StDes(t,t',\delta,\delta')$ and, hence, the algorithm $\cA$ can determine which of the following two is the case:
    \begin{equation}
        \delta_{t'}(f) >\delta', \ \ \textrm{or} \ \ \delta_t(f) \leq \delta.
    \end{equation}
    From~\cref{Eq:case1a,Eq:case1b,Eq:case2a,Eq:case2b,Eq:case3a,Eq:case3b}, it is clear that, for any $t\leq t'$, the algorithm $\cA$ can decide
    \begin{equation}
        s(f) \geq \frac{d}{4} = 2^{n-2}, \ \ \textrm{or} \ \ s(f) \leq \frac{d}{4} - 1 = 2^{n-2} -1.
    \end{equation}
    Hence, one can efficiently solve \textsf{MAJ-SAT} with one query to $\cA$. 
    $\hfill \qed$
\end{Proof}

\subsection{Complexity of \texorpdfstring{$\UniDes$}{UniDes}} \label{SS:UniDes}

The same proof technique for~\cref{Thm:StDESin,Thm:StDEShard} can be applied to proving~\cref{Thm:UniDESin,Thm:UniDEShard}, respectively.

\UniDESin*

\begin{Proof}[\cref{Thm:UniDESin}]
    Due to the promise of the problem, it is guaranteed that a given set $\sfU_n$ is either a $\delta$-approximate or not a $\delta'$-approximate unitary $t$-design. If the former is the case, $F_t(\sfU_n) \leq t! + \delta^2$, and, if the latter is the case, $F_t(\sfU_n) > t! + \delta'^2/2^{2tn}$. Since $(t! + \delta'^2/2^{2tn}) - (t! + \delta^2) \in \Omega\bigl(2^{-\poly(n)}\bigr)$ by assumption, checking which is the case is in $\PP$ due to~\cref{Thm:approxUFPin}. Hence, $\UniDes(t, t, \delta, \delta') \in \PP$.    $\hfill \qed$    
\end{Proof}

Below is a restatement of~\cref{Thm:UniDEShard} for clarity.

\UniDEShard*

\begin{Proof}[\cref{Thm:UniDEShard}]
    The proof is almost in parallel with that of $\StDes$, though we make use of unitaries instead of state vectors.
    Let $f: \{0,1\}^{n-1} \rightarrow \{0,1\}$ be a Boolean function that is computable in classical polynomial time, and let $\ket{f}$ and $\ket{p}$ be $n$-qubit state vectors given by, respectively,
    \begin{align}
        &\ket{f} = \sqrt{\frac{2}{d}} \sum_{x \in \{0,1\}^{n-1}} \ket{x}  \ket{f(x)}, \label{Eq:fforuni}\\
        &\ket{p} =\ket{+}^{\otimes n-1}  \ket{1}. \label{Eq:pforuni}
    \end{align}
    Using these states, we define unitaries as
    \begin{equation}
        V_f = I - 2 \ketbra{f}{f},\ \ \text{and}\ \ V_p = I - 2 \ketbra{p}{p}.
    \end{equation}
    Let $\delta_0$ be such that $\delta_0\in O(\delta^2/d^{2t})$, and let $\sfD_{t'}$ be a $\delta_0$-approximate unitary $t'$-design of $n$ qubits with a computationally efficient description, which can be generated by, e.g., local random circuits with $\poly(n)$ depth.
    We denote by $D$ the cardinality of $\sfD_{t'}$.

    We define a multiset $\sfU_n(f)$ of unitaries by
    \begin{equation}
        \sfU_n(f) = \{V_f, V_p\}^{\oplus m} \cup \sfD_{t'},
    \end{equation}
    where $\oplus m$ implies the multiplicity $m \in \mathbb{Z}^+$.  
    Later, we determine $D$ and $m$ depending on $\delta$ such that $D/m$ is appropriately set for our purpose. Note that $D/m$ can be an arbitrary positive rational number. 
    
    Let $\tau \in \{t, t'\}$. The set $\sfU_n(f)$ is a $\delta_{\tau}(f)$-approximate unitary $\tau$-design, where
    \begin{equation}
        \delta_{\tau}(f) \coloneqq \left\| M^{(\tau)}_{\sfU_n(f)}-M^{(\tau)}_\Haar \right\|_{1}.
    \end{equation}
    Here, $M^{(\tau)}_{\mu} = \mathbb{E}_{U \sim \mu}[U^{\otimes \tau} \otimes \bar{U}^{\otimes \tau}]$ is the moment operator.
    Below, we compute a lower and an upper bound on $\delta_{\tau}(f)$. 

     The moment operator of the Haar measure is a projector onto the irreducible representation of $\pi_{\tau}: U \mapsto U^{\otimes \tau}\otimes \bar{U}^{\otimes \tau}$. Denoting the projector by $\Pi_{\textrm{irrep}}^{(\tau)}$, the moment operator of the uniform measure on $\sfU_n(f)$ is in the form of 
     \begin{align}
         M^{(\tau)}_{\sfU_n(f)}
         &=
         \Pi_{\textrm{irrep}}^{(\tau)} + \Pi_{\textrm{irrep}}^{(\tau)\perp} M^{(\tau)}_{\sfU_n(f)} \Pi_{\textrm{irrep}}^{(\tau)\perp},
     \end{align}
     where $\Pi_{\textrm{irrep}}^{(\tau)\perp} = I - \Pi_{\textrm{irrep}}^{(\tau)}$. Note that we have used the irreducibility. 
     Hence, 
     \begin{equation}
        \delta_{\tau}(f)  =  \left\|  \Pi_{\textrm{irrep}}^{(\tau)\perp} M^{(\tau)}_{\sfU_n(f)} \Pi_{\textrm{irrep}}^{(\tau)\perp} \right\|_{1},
    \end{equation}
    with
    \begin{align}
        M^{(\tau)}_{\sfU_n(f)}
        &=\mathbb{E}_{U \sim \sfU_n(f)}\left[ \pi_{\tau}(U) \right]\\
        &=
        \frac{1}{D+2m}\left[ m \left(\pi_{\tau}(V_f) + \pi_{\tau}(V_p)\right) + D M_{\sfD_{t'}}^{(\tau)} \right].
    \end{align}
    From the triangle inequality, we have that
    \begin{multline}
        \frac{1}{D+2m} \left| m A  - D B \right| \\
        \leq
        \delta_{\tau}(f)
        \leq 
        \frac{1}{D+2m} \left(m A  + D B \right), \label{Eq:unitarydesLandUB}
    \end{multline}
    where
    \begin{align}
        A &= \left\|  \Pi_{\textrm{irrep}}^{(\tau)\perp}   \left(\pi_{\tau}(V_f) + \pi_{\tau}(V_p)\right) \Pi_{\textrm{irrep}}^{(\tau)\perp} \right\|_{1}\\
        B &=  \left\|  \Pi_{\textrm{irrep}}^{(\tau)\perp} M_{\sfD_{t'}}^{(\tau)} \Pi_{\textrm{irrep}}^{(\tau)\perp} \right\|_{1}\leq \delta_0.        
    \end{align}
    Note that the upper bound on $B$ follows from the fact that $\sfD_{t'}$ is a $\delta_0$-approximate $\tau$-design as $\tau \in \{t, t'\}$ and $t\leq t'$.

    To compute $A$, we use the expansion that
    \begin{multline}
        \pi_{\tau}(V_f) + \pi_{\tau}(V_p) = \\
        2\Pi_{\textrm{irrep}}^{(\tau)} +
        \Pi_{\textrm{irrep}}^{(\tau)\perp}   \left(\pi_{\tau}(V_f) + \pi_{\tau}(V_p)\right) \Pi_{\textrm{irrep}}^{(\tau)\perp},
    \end{multline}
    and the fact that the first and the remaining term in the right-hand side have orthogonal support. Hence, using $\tau! = \tr{\Pi_{\textrm{irrep}}^{(\tau)}}$, we obtain
    \begin{multline}
         \left\|  \pi_{\tau}(V_f) + \pi_{\tau}(V_p) \right\|_{1}
         =\\
         \left\|  \Pi_{\textrm{irrep}}^{(\tau)\perp}   \left(\pi_{\tau}(V_f) + \pi_{\tau}(V_p)\right) \Pi_{\textrm{irrep}}^{(\tau)\perp} \right\|_{1} + 2 \tau!.
    \end{multline}
    In~\cref{App:Lambda}, we show that 
    \begin{equation}
        \left\|  \pi_{\tau}(V_f) + \pi_{\tau}(V_p) \right\|_{1}
        =
        \Lambda_{\tau}\left(s(f) \right). \label{Eq:Lambda1}
    \end{equation}
    Here, $\Lambda_{\tau}(x)$ is defined for $x \in [0,d/2]$ as
    \begin{multline}
        \Lambda_{\tau}(x)
        \coloneqq\\
        2\sum_{s=0}^{2\tau} \binom{2 \tau}{s}(d-2)^{2\tau-s} \sum_{r=0}^s \binom{s}{r} \left| T_{|2r-s|}\left(\frac{2x}{d}\right) \right|,
    \end{multline}
    and $T_m(z) = \cos m\theta$, where $z = \cos \theta$, are the Chebyshev polynomials of the first kind.
    Hence,
    \begin{equation}
        A = \Lambda_{\tau}\left(s(f) \right) - 2\tau!.
    \end{equation}

   Substituting $A$ and $B$ into~\cref{Eq:unitarydesLandUB}, we obtain
    \begin{multline}
        \frac{1}{D/m+2}\left( \Lambda_{\tau}\left(s(f) \right) - 2\tau!    -
        \frac{D}{m} \delta_0\right)
        \leq  \\
        \delta_{\tau}(f)   
        \leq 
        \frac{1}{D/m+2}\left( \Lambda_{\tau}\left(s(f) \right) - 2\tau!    +
        \frac{D}{m} \delta_0\right).
    \label{Eq:ap'rgkmqaper}
    \end{multline}
    In the lower bound, we have assumed that $\delta_0$ is sufficiently small. We will check this condition later.
    
    For later convenience, we introduce functions $\textrm{UB}_{\tau}(q, x)$ and $\textrm{LB}_{\tau}(q, x)$ by
    \begin{align}
        &\textrm{UB}_{\tau}(q, x)
        =
        \frac{1}{q+2} \left(  \Lambda_{\tau}\left(x \right) - 2\tau!  + q\delta_0\right) \label{Eq:UBtauUni}\\
        &\textrm{LB}_{\tau}(q, x)
        =
        \frac{1}{q+2} \left(  \Lambda_{\tau}\left(x \right) - 2\tau!  - q\delta_0\right) \label{Eq:LBtauUni}
    \end{align}
    using which we have
    \begin{equation}
         \textrm{LB}_{\tau}\left(\frac{D}{m}, s(f)\right)\leq \delta_{\tau}(f)  \leq \textrm{UB}_{\tau}\left(\frac{D}{m}, s(f)\right). \label{Eq:UBLBunitary}
    \end{equation}
    
    Importantly, for any fixed $q \geq 0$, these functions are monotonically increasing as $x$ increases for sufficiently large $d$. Indeed, the leading order expansion of $\Lambda_{\tau}(x)$ is given by
    \begin{equation}
        \Lambda_{\tau}(x)
         =
         2d^{2 \tau} - 8 \tau d^{2 \tau - 1} \left(1 - \frac{2x}{d}\right)   + \cO\left( d^{2\tau - 2}\right), \label{Eq:LeadingOrderLambda}
    \end{equation}
    as shown in~\cref{App:Lambda}.
    Thus, if $d$ is sufficiently large, $\Lambda_{\tau}(x)$ is monotonically increasing in $x$, and so are $\textrm{UB}_{\tau}(q, x)$ and $\textrm{LB}_{\tau}(q, x)$ for a fixed $q\geq 0$.

    We now choose $D$ and $m$ such that $D/m$ satisfies
    \begin{equation}
        \textrm{LB}_{t}\left(\frac{D}{m}, \frac{d}{4} - \frac{2}{3}\right) = \delta. \label{Eq:defdeltauni}
    \end{equation}
    From the order estimation of $\Lambda_t(x)$ and the facts that $\delta_0 \in \cO(\delta^2/d^{2t})$ and $\delta \in o(d^{2(t-1)})$, which implies that $\delta \in o(d^{2t})$, we have $D/m \in \Theta(d^{2t}/\delta)$. We also check the assumption on $\delta_0$ in~\cref{Eq:ap'rgkmqaper}, which is that $\delta_0$ is sufficiently small such that $\textrm{LB}_{\tau}\left(D/m, s(f)\right) \geq 0$. As $D/m\in \Theta(d^{2t}/\delta)$, this condition is satisfied when $\delta_0 \in \cO(\delta^2/d^{2t})$, which we have assumed.
    We also set $\delta'$ as
    \begin{equation}
        \delta' = \textrm{UB}_{t'}\left(\frac{D}{m}, \frac{d}{4} - \frac{1}{3}\right) \in \Theta(d^{2(t'-t)}\delta). \label{Eq:delta'defuni}
    \end{equation}
    Clearly, $\delta' - \delta \in \Theta(d^{2(t'-t)}\delta)$ for $t<t'$.
    When $t=t'$, we need more careful analysis based on~\cref{Eq:LeadingOrderLambda}, from which it turns out that $\delta' - \delta \in \Theta(d^{-2}\delta)$ for $t=t'$.

    With this choice of $D/m$, we below show that an algorithm $\cA$ to solve $\UniDes(t,t',\delta,\delta')$ can be used to solve \textsf{MAJ-SAT}. 
    By definition, the algorithm can determine
    \begin{equation}
        \delta_{t'}(f) >\delta', \ \ \textrm{or} \ \ \delta_{t}(f) \leq \delta, \label{Eq:ppqsrtuni}
    \end{equation}
    under the exclusive promise that only one of the two holds. 

    Let us first check if $\sfU_n(f)$ constructed from any Boolean function $f: \{0,1\}^{n-1} \rightarrow \{0,1\}$ satisfies the exclusive promise. 
    To this end, we show that
    \begin{align}
        \delta_{t'}(f) > \delta' &\Longleftrightarrow s(f) \geq \frac{d}{4}, \label{Eq:Unicasea}\\
        \delta_{t}(f) \leq \delta &\Longleftrightarrow s(f) \leq \frac{d}{4}-1. \label{Eq:Unicaseb}
    \end{align}
    From these two, it is clear that all Boolean functions must exclusively satisfy one of the two conditions $\delta_t(f)\leq \delta$ or $\delta_{t'}(f) > \delta'$.

    To show~\cref{Eq:Unicasea}, we begin by assuming that $\delta_{t'}(f) > \delta'$. If this is satisfied, it follows from~\cref{Eq:UBLBunitary,Eq:delta'defuni} that
    \begin{equation}
       \textrm{UB}_{t'}\left(\frac{D}{m}, \frac{d}{4} - \frac{1}{3}\right) = \delta' < \delta_{t'}(f) \leq \textrm{UB}_{t'}\left(\frac{D}{m}, s(f)\right).
    \end{equation}
    Using the monotonicity of $\textrm{UB}_{t'}(q,x)$ in $x$, we have $d/4-1/3 < s(f)$. Moreover, $s(f)$ is a non-negative integer, leading to $d/4 \leq s(f)$.

    On the other hand, if $d/4 \leq s(f)$, $\delta_{t'}(f) > \delta'$. To show this, we use the fact that, for sufficiently large $d$,
    \begin{equation}
        \textrm{UB}_{t'}\left(\frac{D}{m}, \frac{d}{4} - \frac{1}{3}\right) < \textrm{LB}_{t'}\left(\frac{D}{m}, \frac{d}{4}\right), \label{Eq:UBLB}
    \end{equation}
     which is shown in~\cref{App:UBLBcomparison} using the assumption that $\delta \in o (d^{2(t'-1)})$. Hence, if $\delta_{t'}(f) \leq \delta'$, it follows from~\cref{Eq:UBLBunitary,Eq:delta'defuni} that
    \begin{multline}
       \textrm{LB}_{t'}\left(\frac{D}{m}, s(f)\right) \leq \delta_{t'}(f) \leq \delta' \\
       = \textrm{UB}_{t'}\left(\frac{D}{m}, \frac{d}{4}-\frac{1}{3}\right)  < \textrm{LB}_{t'}\left(\frac{D}{m}, \frac{d}{4}\right).
    \end{multline}
    From the monotonicity of $\textrm{LB}_{t'}(q,x)$ in $x$, we conclude that $\delta_{t'}(f) \leq \delta'$ implies that $s(f) < d/4$. Taking the contraposition, we obtain that, if $s(f) \geq d/4$, then $\delta_{t'}(f) > \delta'$.

    To show~\cref{Eq:Unicaseb}, if $\delta_{t}(f) \leq \delta$, we have from~\cref{Eq:UBLBunitary,Eq:defdeltauni} that
    \begin{equation}
       \textrm{LB}_{t}\left(\frac{D}{m}, s(f)\right) \leq \delta_t(f) \leq \delta = \textrm{LB}_{t}\left(\frac{D}{m}, \frac{d}{4}- \frac{2}{3}\right).
    \end{equation}
    Using the monotonicity of $\textrm{LB}_t(q,x)$ in $x$ and the fact that $s(f)$ is a non-negative integer, this implies that $s(f) \leq d/4-1$.
    To show that $s(f) \leq d/4-1$ implies that $\delta_{t}(f) \leq \delta$, we consider its contraposition.
    A key observation is 
    \begin{equation}
        \textrm{UB}_{t}\left(\frac{D}{m}, \frac{d}{4} - 1\right) < \textrm{LB}_{t}\left(\frac{D}{m}, \frac{d}{4}-\frac{2}{3}\right), \label{Eq:LBUB}
    \end{equation}
    shown in~\cref{App:UBLBcomparison} using the assumption that $\delta \in o (d^{2(t'-1)})$. 
    Hence, if $\delta_{t}(f) > \delta$, we have from~\cref{Eq:UBLBunitary,Eq:defdeltauni} that
    \begin{multline}
       \textrm{UB}_{t}\left(\frac{D}{m}, \frac{d}{4}-1\right) <    \textrm{LB}_{t}\left(\frac{D}{m}, \frac{d}{4}-\frac{2}{3}\right)=   \delta \\ 
       < \delta_t(f)
       \leq \textrm{UB}_{t}\left(\frac{D}{m}, s(f)\right).
    \end{multline}
    From the monotonicity, we have $d/4-1<s(f)$. By taking the contraposition, we obtain that $s(f) \leq d/4-1$ implies that $\delta_{t}(f) \leq \delta$.
    \\

    We finally show how the algorithm $\cA$ for solving $\UniDes(t,t',\delta,\delta')$ can be used to solve \textsf{MAJ-SAT}. As $\sfU_n(f)$ satisfies the exclusive promise of $\UniDes(t,t',\delta,\delta')$ for any Boolean function $f$, the algorithm is capable of determining whether $\sfU_n(f)$ satisfies
    \begin{equation}
        \delta_{t'}(f) >\delta', \ \ \textrm{or} \ \ \delta_t(f) \leq \delta.
    \end{equation}
    From~\cref{Eq:Unicasea,Eq:Unicaseb}, this implies that $\mathcal{A}$ can determine
    \begin{equation}
        s(f) \geq \frac{d}{4} = 2^{n-2}, \ \ \textrm{or} \ \ s(f) \leq \frac{d}{4}-1=2^{n-2}-1.
    \end{equation}
    Hence, one can efficiently solve \textsf{MAJ-SAT} with one query to $\cA$.
    $\hfill \qed$
\end{Proof}

%%% --------------------------------
\section{Conclusions and discussion} \label{S:conclusion}
%%% --------------------------------

In this paper, we have initiated computational complexity-theoretic analyses about the problems related to unitary and state designs and have paved the way to understanding randomness in quantum systems from the complexity perspective.
We have begun with an explicit quantum algorithm for computing the state and unitary frame potentials and then clarified the complexity classes of exactly computing the frame potentials, $\compSFP(t)$ and $\compUFP(t)$, which have been shown to be $\sharpP$-hard and to be achievable by a single query to a $\sharpP$-oracle.

We have then introduced promise problems, $\SFP(t, \epsilon)$ and $\UFP(t, \epsilon)$, to decide whether the frame potential is larger than or smaller than certain values with a promise gap $\epsilon$. These problems are important to address the complexity of approximately computing the frame potentials.
For $\SFP(t, \epsilon)$, we have proven that the problem is $\BQP$-complete for $\epsilon \in \Theta(1/\poly(n))$, but is $\PP$-complete for $\epsilon \in \Theta(2^{-\poly(n)})$. The latter has been shown to be the case for $\UFP(t, \epsilon)$.

We have finally addressed promise problems that are directly related to state and unitary designs: whether a given set is a $\delta$-approximate $t$-design or not a $\delta'$-approximate $t'$-design, under the promise that only one of the two is the case. We have shown that these problems, $\StDes(t, t', \delta, \delta')$ and $\UniDes(t, t',\delta, \delta')$, are in $\PP$ if $t=t'$ and the gap between $\delta$ and $\delta'$ is not too small. We have also shown the hardness result. Specifically, these problems are $\PP$-hard if the gap 
is $\delta' - \delta \in \Theta(2^{c(t'-t)n}\delta)$ for $t < t'$ and $\delta' - \delta \in \Theta(2^{-cn}\delta)$ for $t=t'$, where $c=1$ for $\StDes(t, t', \delta, \delta')$ and $c=2$ for $\UniDes(t, t', \delta, \delta')$. 
We would like to emphasize that these results imply the computational hardness of deciding if a given set is an exponentially-accurate approximation to a $t$-design or a constant-away approximation to a $(t+1)$-design that is not an accurate $t$-design, highlighting the computationally hard nature of checking design properties.

These results have far-reaching implications, owing to the broad applicability of complexity theory and the universal significance of designs. Specifically, we have demonstrated their relevance to variational constructions of designs, the characterization of quantum chaos via the average OTOC, and the emergence of designs in Hamiltonian systems. Notably, our results imply that, for algorithms addressing these problems to scale efficiently to large systems, it is crucial to tailor them to leverage the specific structure of the underlying problems.

While we have thoroughly investigated the computational complexity classes of the introduced problems, the complexity in some parameter regions, such as $\UFP(t, \epsilon)$ for $\epsilon \in \Omega(1/\poly(n))$, remains open. Investigating these cases is an important future work.
While it is natural to expect that $\UFP(t, \epsilon)$ with inverse-polynomial or constant $\epsilon$ is computationally easy, showing this is not straightforward, at least in our approach. 
The difficulty is intrinsic in our approach, where we start with a quantum algorithm to compute the purity of a quantum state that encodes all the information of the unitary frame potential. 
This approach inevitably introduces the prefactor $2^{-2nt}$ between the unitary frame potential and the purity. Therefore, for estimating the frame potential with constant approximation in our approach, it is necessary to evaluate the purity with exponential accuracy, which is, in general, computationally intractable. This is, however, the difficulty specific to our approach, and it is open if $\UFP(t, \epsilon)$ with inverse-polynomial or constant $\epsilon$ is computationally feasible. Developing a new approach toward this is an interesting open problem.

Regarding $\StDes(t, t', \delta, \delta')$ and $\UniDes(t, t', \delta, \delta')$, it is of great interest to further investigate the problems when $t=t'$. In this case, we have succeeded in proving the hardness only with small gap $\delta' - \delta$. To this end, however, our methods, i.e., to hide a hard instance in a set, may not work.
It may also be interesting to consider the computational problems related to other definitions of unitary designs, such as the one based on the diamond norm. As different definitions do not lead to significant differences in applications, this is not practically important, but may still be of theoretical interest. In this case, the key is to tightly compute the approximation accuracy as we have done for proving the $\PP$-hardness of $\UniDes$ in terms of the moment operators.

Our results, showing the computationally hard nature of checking design properties, indicate that there may exist \emph{pseudo} $t$-designs, namely, a set of state vectors or unitaries that is not a $t$-design but indistinguishable from $t$-designs in polynomial time. This direction may be of interest in regard to the recent progress in computationally-secure quantum cryptography based on pseudorandom states or unitaries.
We expect that our results, that good $t$-design cannot be efficiently distinguished from bad $(t+1)$-design that fails to be a good $t$-design, could offer an interesting starting point toward this direction.

From a more practical viewpoint, it will also be important to consider restricted sets of state vectors or unitaries, for which the problems that we have introduced in this work become solvable in classical or quantum polynomial time.
This direction is of significant importance for developing the practical variational ansatz for constructing designs or algorithms for studying emergent designs in Hamiltonian systems.

%%% --------------------------------
\section{Acknowledgments}
%%% --------------------------------

We would like to thank Özgün Kum for discussions on computational complexity theory and Masaki Owari for discussions about the algorithms for computing the frame potential. 
Y.~N.\ is supported by JSPS KAKENHI Grant Number JP22K03464, MEXT-KAKENHI Grant-in-Aid for Transformative Research Areas (A) ``Extreme Universe'' Grant Numbers JP21H05182 and JP21H05183, JST CREST Grant Number JPMJCR23I3, and JST PRESTO Grant Number JPMJPR2456.
Y.~T.\ is partially supported by JST [Moonshot R\&D -- MILLENNIA Program] Grant Number JPMJMS2061.
M.~K.\ is supported by 
the Deutsche Forschungsgemeinschaft (DFG, German Research Foundation) within the Emmy Noether program (grant number 441423094) 
and the Fujitsu Germany GmbH as part of the endowed professorship ``Quantum Inspired and Quantum Optimization.''
A.~D.\ is supported by Grant-in Aid for Transformative Research Areas (A) ``Extreme Universe'' Grant Number 21H05183.

\bibliography{nakata,mk}

%%% ==============================================
\appendix
\addcontentsline{toc}{section}{Appendix}
%%% ==============================================

%%% --------------------------------
\section{Proofs of~\cref{Prop:StateFPBasicProperties,Prop:UnitaryFPBasicProperties}} \label{App:FramePotential}
%%% --------------------------------

\subsection{Proof of~\cref{Prop:StateFPBasicProperties}}

We here provide a proof of~\cref{Prop:StateFPBasicProperties} for completeness.
Note that we assume that $t \leq d$, but similar statements hold even for $t>d$.\\

\StateFPBasicProperties*

\begin{Proof}[\cref{Prop:StateFPBasicProperties}]
    The state frame potential for a uniformly distributed state can be calculated directly using~\cref{Eq:symproj,Eq:StFrame2norm}.

    To show the second statement, we begin with the fact that the support of the state $\sum_{j=1}^K \ketbra{\psi_j}{\psi_j}^{\otimes t}/K$ is the symmetric subspace, as it is a probabilistic mixture of symmetric states.
    The frame potential is defined by the squared Hilbert-Schmidt norm of the state, and its minimum $1/d_t$ is attained by the completely mixed state on the subspace, which corresponds to a state averaged over an exact state $t$-design.
    The upper bound follows from the simple facts that $|\braket{\phi}{\psi}| \leq 1$ for any state vectors $\ket{\phi}$ and $\ket{\psi}$.
    For the uniform probability measure over a finite set $\sfS$ with cardinality $K$, the frame potential further satisfies
    \begin{align}
        F_t(\sfS) &= \frac{1}{K^2}\sum_{i,j=1}^K \bigl| \braket{\psi_i}{\psi_j} \bigr|^{2t}\\
        & \geq \frac{1}{K^2 }\sum_{j=1}^K \bigl| \braket{\psi_j}{\psi_j} \bigr|^{2t}\\
        &=\frac{1}{K}. \label{Eq:FtStateLowerfinite}
    \end{align}

    To show the third statement, we start with the definition of a state design. If a set $\sfS$ is a $\delta$-approximate state $t$-design,~\cref{Eq:DefStateDesign} holds.
    Since $\| \cdot \|_2 \leq \sqrt{ \operatorname{rank}(\argdot)} \| \argdot\|_{\infty}$,
    we have  
    \begin{equation}
        \Bigl\| \frac{1}{K}\sum_{j=1}^K \ketbra{\psi_j}{\psi_j}^{\otimes t} - \frac{\Pi_{\rm sym}(n,t)}{d_t} \Bigr\|_2 \leq \frac{\delta}{\sqrt{d_t}},
    \end{equation}
    where we used~\cref{Eq:symproj}. 
    By direct calculation and~\cref{Eq:StFrame2norm}, the left-hand side satisfies
    \begin{multline}
        \Bigl\| \frac{1}{K}\sum_{j=1}^K \ketbra{\psi_j}{\psi_j}^{\otimes t} - \frac{\Pi_{\rm sym}(n,t)}{d_t} \Bigr\|_2^2
        =
        F_t(\sfS) - \frac{1}{d_t}.
    \end{multline}
    Hence, the third statement holds.

    The fourth statement is shown as follows.
    Suppose that a set $\sfS =\{ \ket{\psi_j} \}_{j=1, \dots, K}$ satisfies $F_t(\sfS) \leq 1/d_t + \delta^2/d_t^2$. A direct calculation leads to
    \begin{equation}
        \Bigl\| \frac{1}{K}\sum_{j=1}^K \ketbra{\psi_j}{\psi_j}^{\otimes t} - \frac{\Pi_{\rm sym}(n,t)}{d_t}\Bigr\|_2
        \leq
        \delta/d_t.
    \end{equation}
    As $\| \cdot \|_{\infty}\leq \| \cdot \|_2$, we obtain
    \begin{equation}
        \Bigl\| \frac{1}{K}\sum_{j=1}^K \ketbra{\psi_j}{\psi_j}^{\otimes t} - \frac{\Pi_{\rm sym}(n,t)}{d_t} \Bigr\|_{\infty}
        \leq
        \delta/d_t,
    \end{equation}
    which implies that $\sfS$ is a $\delta$-approximate state $t$-design.

    Finally, suppose that the cardinality $K$ of a set $\sfS =\{ \ket{\psi_j}\}_{j=1, \dots, K}$ is less than $d_t/(1+\delta^2)$. Together with~\cref{Eq:FtStateLowerfinite}, we have 
    \begin{equation}
        F_t(\sfS) \geq \frac{1}{K} > \frac{1}{d_t} +\frac{\delta^2}{d_t}.
    \end{equation}
    Due to the contraposition of the third statement, this implies that $\sfS$ is not a $\delta$-approximate state $t$-design.
    $\hfill \qed$
\end{Proof}

\subsection{Proof of~\cref{Prop:UnitaryFPBasicProperties}}

We next show~\cref{Prop:UnitaryFPBasicProperties}. Similarly to the state frame potential, we assume that $t \leq d$ just for simplicity. Similar statements hold even for $t>d$.

\UnitaryFPBasicProperties*

\begin{Proof}[\cref{Prop:UnitaryFPBasicProperties}]
    The unitary frame potential for the Haar measure is given by the number of trivial irreducible representations of $U \mapsto U^{\otimes t} \otimes \bar{U}^{\otimes t}$, which is $t!$ when $t \leq 2^{n}$. This is the first statement.

    For the second statement, since the Haar measure is the left- and right-unitary invariant measure, 
    \begin{equation}
        \bigl\| M^{(t)}_\Haar - M^{(t)}_\mu\bigr\|_2^2 = F_t(\mu) - F_t(\Haar), \label{Eq:Aaaaa}
    \end{equation}
    for any probability measure $\mu$ on $\mathbb{U}(d)$.
    As the left-hand side is non-negative, $F_t(\Haar) \leq F_t(\mu)$. Also using the first property, we have $t! \leq F_t(\mu)$. The upper bound is obtained from the fact that $\tr{U^{\dagger}V} \leq d$ for any unitary $U, V \in \mathbb{U}(d)$.
    For the uniform probability measure on a finite set $\sfU$ with cardinality $K$, the frame potential satisfies
    \begin{align}
        F_t(\sfU) &= \frac{1}{K^2}\sum_{i,j=1}^K \bigl| {\rm Tr}[U_i^{\dagger}U_j] \bigr|^{2t}\\
        & \geq \frac{1}{K^2 }\sum_{j=1}^K \bigl| {\rm Tr}[U_j^{\dagger}U_j] \bigr|^{2t}\\
        &=\frac{d^{2t}}{K}. \label{Eq:FtUlowerbound}
    \end{align}
    
    If a set $\sfU$ is a $\delta$-approximate unitary $t$-design, it satisfies
    \begin{equation}
        \bigl\| M^{(t)}_\Haar - M^{(t)}_\sfU \bigr\|_1 \leq \delta.
    \end{equation}
    Since $\| \cdot \|_2 \leq \| \cdot \|_1$, the first statement and~\cref{Eq:Aaaaa} lead to $F_t(\sfU) \leq t! +\delta^2$.

    Suppose that a set $\sfU$ satisfies $F_t(\sfU) \leq t! + \delta^2/d^{2t}$. From the first statement and~\cref{Eq:Aaaaa}, it follows that
    \begin{equation}
        \bigl\| M^{(t)}_\Haar - M^{(t)}_\sfU \bigr\|_2 \leq \delta/d^t.
    \end{equation}
    As $\| \argdot \|_1 \leq \sqrt{\operatorname{rank}(\argdot)} \| \argdot \|_2$, we obtain $\bigl\| M^{(t)}_\Haar - M^{(t)}_\sfU \bigr\|_1 \leq \delta$. That is, $\sfU$ is a $\delta$-approximate unitary $t$-design.

    Finally, suppose that the cardinality $K$ of a set $\sfU =\{ U_j\}_{j=1, \dots, K}$ of unitaries is less than $d^{2t}/(t!+\delta^2)$. Together with~\cref{Eq:FtUlowerbound}, the frame potential satisfies
    \begin{align}
        F_t(\sfU) \geq \frac{d^{2t}}{K} > t!+\delta^2.
    \end{align}
    Due to the contraposition of the third statement, $\sfS$ is not a $\delta$-approximate unitary $t$-design.

    To show the last statement, we use the Choi-Jamio\l kowski representation of a superoperator. For $\cT$, it is given by $\mathfrak{J}(\cT) := \cT \otimes {\rm id}(\ketbra{\Phi}{\Phi})$, where $\ket{\Phi}$ is a maximally entangled state. It is known that $\| \mathfrak{J}(\cT) \|_1 \leq \| \cT \|_{\diamond} \leq d \| \mathfrak{J}(\cT) \|_1$, where $d$ is the dimension of the input Hilbert space. Using this relation, we have
    \begin{multline}
        \bigl\| \mathfrak{J}\bigl(\cG^{(t)}_{\sfU}  -  \cG^{(t)}_{\Haar} \bigr) \bigr\|_1 \leq \bigl\| \cG^{(t)}_{\sfU} - \cG^{(t)}_{\Haar} \bigr\|_{\diamond} \leq d^t \bigl\| \mathfrak{J}\bigl(\cG^{(t)}_{\sfU}  -  \cG^{(t)}_{\Haar}\bigr) \bigr\|_1. \label{Eq:CJDesign}
    \end{multline}

    We further use the vectorization of an operator: $\cO = \sum_{j,k} c_{jk} \ketbra{j}{k} \mapsto \ket{o} = \sum_{j,k} c_{jk} \ket{j} \otimes \ket{\bar{k}}$, where $\bar{\argdot}$ represents the complex conjugate in a fixed orthonormal basis. This map preserves the Hilbert-Schmidt norm, i.e., $\| \cO \|_2 = \| \ket{o} \|_2$.
    We consider the vectorization of $\mathfrak{J}\bigl(\cG^{(t)}_{\sfU}  -  \cG^{(t)}_{\Haar} \bigr)$. For simplicity, let $A$ be denoted by the $d^t$-dimensional input Hilbert space of $\cG^{(t)}_{\mu}$ ($\mu \in \{\sfU, \Haar\})$, and $B,C$, and $D$ by the systems isomorphic to $A$. Denoting the systems, on which operators act, by superscript, the vectorization is given by
    \begin{equation}
        \bigl( M_{\sfU}^{(t) AB} - M_{\Haar}^{(t) AB} \bigr) \ket{\Phi}^{AC} \otimes \ket{\Phi}^{BD}.
    \end{equation}
    A direct calculation shows that the Hilbert-Schmidt norm of this vector is $d^{-t} \sqrt{F_t(\sfU) - F_t(\Haar)}$. Thus, we have
    \begin{equation}
        \bigl\| \mathfrak{J}\bigl(\cG^{(t)}_{\sfU}  -  \cG^{(t)}_{\Haar} \bigr) \bigr\|_2
        =
        \frac{1}{d^t} \sqrt{F_t(\sfU) - F_t(\Haar)}.
    \end{equation}
    The desired statement is obtained by combining~\cref{Eq:CJDesign}, this equality, and the relations between norms given in~\cref{Eq:Schatten}.
    $\hfill \qed$
\end{Proof}

%%% ==============================================
\section{Analysis of quantum algorithm for computing frame potentials} \label{S:QAforFP}
%%% ==============================================

Here, we provide proofs of~\cref{Lemma:FPencodedState} and~\cref{Thm:QA_FP}.

\FPencodedState*

Note that the states $\ket{\varrho_{\rm st}}^{AB}$ and $\ket{\varrho_{\rm uni}}^{ABR}$ are defined by
\begin{align}
    &\ket{\varrho_{\rm st}}^{AB} 
    = \mathrm{ctrl}^A\mathchar`-\sfU_n^{B} \left(\ket{+_{\kappa_n}}^{A} \bigotimes_{m=1}^t \ket{0_n}^{B_m} \right), \\
    &\ket{\varrho_{\rm uni}}^{ABR} = 
    \mathrm{ctrl}^A\mathchar`-\sfU^{B}_n \left(\ket{+_{\kappa_n}}^{A} \bigotimes_{m=1}^t \ket{\Phi_n}^{B_mR_m} \right),
\end{align}
respectively. The controlled unitary is given by~\cref{Eq:ctrlU}.

\begin{Proof}[\cref{Lemma:FPencodedState}]
    By a direct calculation, it can be shown that
    \begin{align}
        &\varrho_{\rm st}^A = \frac{1}{K_n} \sum_{i,j=1}^{K_n} \bigl( \bra{0_n}U_{j}^{\dagger} U_{i}\ket{0_n} \bigr)^t \ketbra{i}{j}^A,\\
        &\varrho_{\rm uni}^A = \frac{1}{2^{nt} K_n} \sum_{i,j=1}^{K_n} \bigl(\tr{U_{j}^{\dagger} U_{i}} \bigr)^t \ketbra{i}{j}^A.
    \end{align}
    This implies~\cref{Eq:ar;ogknaerge3ar}. $\hfill \qed$
\end{Proof}

We next show~\cref{Thm:QA_FP}.

\QAforFP*

To show this, we refer to a quantum algorithmic result for computing the Hilbert-Schmidt distance between two states~\cite{GL2019}.

\begin{theorem}[Quantum algorithm for $\ell^2$-closeness with purified query-access~\cite{GL2019}] \label{Thm:l2Estimation}
    Given $\epsilon, \nu \in (0,1)$ and mixed states $\rho$ and $\sigma$ on a $D$-dimensional system, the purified states of which can be generated by $U_{\rho}$ and $U_{\sigma}$, respectively, it takes 
    \begin{equation}
        \cO\biggl(\min \biggl\{ \frac{\sqrt{D}}{\epsilon},\  \frac{1}{\epsilon^2} \biggr\} \frac{1}{\nu} \biggr)
    \end{equation}
    queries to $U_{\rho}, U_{\rho}^{\dagger}, U_{\sigma}, U_{\sigma}^{\dagger}$ to decide whether $\| \rho - \sigma\|_2 \geq \epsilon$ or $\| \rho - \sigma\|_2 \leq (1-\nu) \epsilon$ with success probability at least $2/3$.
\end{theorem}

The former query complexity in~\cref{Thm:l2Estimation}, i.e., $\cO\bigl( \sqrt{D}/(\epsilon\nu)\bigr)$, is achieved by using block encoding, and the latter by combining the SWAP test and amplitude estimation.

\begin{Proof}[\cref{Thm:QA_FP}]
    For our purpose, we may estimate $\| \varrho_{\rm st/uni}^A - \pi^A \|_2$ instead of $\tr{(\varrho_{\rm st/uni}^A)^2}$, where $\pi^{A}$ is the completely mixed state on $A$, since
    \begin{align}
        &\| \varrho_{\rm st/uni}^A - \pi^A \|_2^2\\
        &=
        {\rm Tr}\bigl[ (\varrho_{\rm st/uni}^A - \pi^A)^2 \bigr]\\
        &=
        {\rm Tr}\bigl[ (\varrho_{\rm st/uni}^A)^2 \bigr] - 2{\rm Tr}\bigl[ \varrho_{\rm st/uni}^A \pi^A \bigr]
        +{\rm Tr}\bigl[ (\pi^A)^2 \bigr]\\
        &=
        {\rm Tr}\bigl[ (\varrho_{\rm st/uni}^A)^2 \bigr] - \frac{1}{K_n}.
    \end{align}
    Using~\cref{Eq:ar;ogknaerge3ar}, we obtain
    \begin{align}
        &F_t(\sfS_n) = \| \varrho_{\rm st}^A - \pi^A \|_2^2 + \frac{1}{K_n},\\
        &F_t(\sfU_n) = 2^{2nt} \left(\| \varrho_{\rm uni}^A - \pi^A \|_2^2 + \frac{1}{K_n}\right).
    \end{align}
    From this fact, it is easy to obtain~\cref{Thm:QA_FP}. $\hfill \qed$    
\end{Proof}

\section{Derivation of~\cref{Eq:Lambda1}} \label{App:Lambda}

We show $\| \pi_{\tau}(V_f) + \pi_{\tau}(V_p) \|_1 = \Lambda_{\tau}(s(f))$, where $\pi_{\tau}$ is a representation given by $U \mapsto U^{\otimes \tau} \otimes \bar{U}^{\otimes \tau}$, $V_w = I - 2\ketbra{w}{w}$ for $w = f, p$ with $\ket{f}$ and $\ket{p}$ defined by~\cref{Eq:fforuni,Eq:pforuni}, respectively, and 
\begin{multline}
    \Lambda_{\tau}(x)
    =\\
    2\sum_{s=0}^{2\tau} \binom{2 \tau}{s}(d-2)^{2\tau-s} \sum_{r=0}^s \binom{s}{r} \left| T_{|2r-s|}\left(\frac{2x}{d}\right) \right|.
\end{multline}
Here, $T_m(z) = \cos m\theta$ with $z = \cos \theta$ are the Chebyshev polynomials of the first kind.

We first decompose the Hilbert space $\cH_d = (\mathbb{C}^2)^{\otimes n}$ to a 2-dimensional subspace $\cH_2$ spanned by $\ket{f}$ and $\ket{p}$, and the remaining one $\cH_{d-2}$. We also fix the orthonormal basis of $\cH_2$ as $\{ \ket{p}, \ket{p^{\perp}}\}$, where $\ket{p^{\perp}}$ is orthogonal to $\ket{p}$ such that $\braket{f}{p^{\perp}} \in \mathbb{R}$. 
Under this decomposition, we can write $V_w$ for $w=f, p$ as $\tilde{w} \oplus I_{d-2}$. The $2 \times 2$ matrix $\tilde{w}$ is given by $I_2 - 2\ketbra{w}{w}$ or, more explicitly, 
\begin{equation}
    \tilde{f} = \begin{pmatrix} 1- 2 \alpha^2 & -2 \alpha \beta \\ -2 \alpha \beta & 1-2\beta^2 \end{pmatrix}, \ \ \text{and} \ \ 
    \tilde{p} = \begin{pmatrix} -1 & 0 \\ 0 & 1 \end{pmatrix}, \label{Eq:tildefp}
\end{equation}
in our fixed orthonormal basis, where $\alpha = \braket{f}{p} = 2s(f)/d$, and $\beta = \sqrt{1- \alpha^2}$.

Using this notation, we have
\begin{align}
    \pi_{\tau}(V_w) &= V_w^{\otimes 2 \tau}\\
    &=\left(\tilde{w} \oplus I_{d-2} \right)^{\otimes 2 \tau}\\
    &=\bigoplus_{s=0}^{2 \tau} \bigoplus_{\textrm{sym}} \left(\tilde{w}^{\otimes s} \otimes  I_{d-2}^{\otimes (2\tau - s)}\right),
\end{align}
where
\begin{widetext}
\begin{equation}
    \bigoplus_{\textrm{sym}} \left(\tilde{w}^{\otimes s} \otimes  I_{d-2}^{\otimes (2\tau - s)}\right)
    \coloneqq 
    \overbrace{
    \left(\tilde{w}^{\otimes s} \otimes  I_{d-2}^{\otimes (2\tau - s)}\right)
    \oplus
    \left(\tilde{w}^{\otimes s-1} \otimes I_{d-2} \otimes \tilde{w} \otimes  I_{d-2}^{\otimes (2\tau - s-1)}\right)
    \oplus \dots \oplus
    \left(I_{d-2}^{\otimes (2\tau - s)} \otimes \tilde{w}^{\otimes s}\right)}^{\textrm{$\binom{2\tau}{s}$ terms}}
\end{equation}
\end{widetext}
is a full symmetrization of the tensor product. As this holds both for $w=f, p$, we have
\begin{equation}
    \pi_{\tau}(V_f) + \pi_{\tau}(V_p) =\bigoplus_{s=0}^{2 \tau} \bigoplus_{\textrm{sym}} \left( \left(\tilde{f}^{\otimes s} + \tilde{p}^{\otimes s} \right) \otimes  I_{d-2}^{\otimes (2\tau - s)}\right).
\end{equation}
Hence, we have
\begin{equation}
    \left\| \pi_{\tau}(V_f) + \pi_{\tau}(V_p) \right\|_1 =
    \sum_{s=0}^{2 \tau} \binom{2 \tau}{s} (d-2)^{2 \tau - s} \left\| \tilde{f}^{\otimes s} + \tilde{p}^{\otimes s}\right\|_1. \label{Eq:laststeptotheend}
\end{equation}

To compute $\left\| \tilde{f}^{\otimes s} + \tilde{p}^{\otimes s}\right\|_1$, we note that both $\tilde{f}$ and $\tilde{p}$ are unitary, and use the unitary invariance of the trace norm. 
This leads to
\begin{equation}
    \left\| \tilde{f}^{\otimes s} + \tilde{p}^{\otimes s}\right\|_1
    =
    \left\| \left(\tilde{f} \tilde{p} \right)^{\otimes s} + I_2^{\otimes s}\right\|_1,
\end{equation}
where $I_2$ is the $2 \times 2$ identity matrix.
From the matrix representation of $\tilde{f}$ and $\tilde{p}$ given in~\cref{Eq:tildefp}, we have
\begin{equation}
    \tilde{f} \tilde{p} = \begin{pmatrix} 2 \alpha^2-1 & - 2 \alpha \beta \\ 2 \alpha\beta & 1- 2 \beta^2 \end{pmatrix}.
\end{equation}
The eigenvalues of this unitary are $e^{\pm i \theta}$ with $\theta = \arccos (2 \alpha^2 -1)$.
Hence,
\begin{align}
    \left\| \tilde{f}^{\otimes s} + \tilde{p}^{\otimes s}\right\|_1
    &=
    \left\| \left(\tilde{f} \tilde{p} \right)^{\otimes s} + I_2^{\otimes s}\right\|_1\\
    &=
    \sum_{r=0}^s \binom{s}{r} \left| e^{i r\theta} e^{-i (s-r)\theta} +1\right|\\
    &=
    \sum_{r=0}^s \binom{s}{r} \left| e^{i (2r-s)\theta} +1 \right|\\
    &=
    2\sum_{r=0}^s \binom{s}{r} \left| \cos \left(2r-s\right) \frac{\theta}{2} \right|\\
    &=
    2\sum_{r=0}^s \binom{s}{r} \left| \cos \left|2r-s\right| \frac{\theta}{2} \right|.
\end{align}
Using the Chebyshev polynomials of the first kind, we have $\cos \left|2r-s\right| \frac{\theta}{2} = T_{|2r-s|}\left( \cos \theta/2 \right)$. It further turns out that $\cos \theta/2 = \pm \alpha$. Using the fact that $|T_m(\pm x)| = |T_m(x)|$, we have
\begin{align}
    \left\| \tilde{f}^{\otimes s} + \tilde{p}^{\otimes s}\right\|_1
    &=
    2\sum_{r=0}^s \binom{s}{r} \left| T_{|2r-s|}(\alpha) \right|.
\end{align}
Substituting this into~\cref{Eq:laststeptotheend} completes the derivation.\\

We next show that, for $x \in [0, d/2]$, $\Lambda_{\tau}(x)$ is given by
\begin{equation}
    \Lambda_{\tau}(x)
    =
    2d^{2 \tau} - 8 \tau d^{2 \tau - 1} \left(1 - \frac{2x}{d}\right)   + \cO\left( d^{2\tau - 2}\right), \label{Eq:LambdaLeadingOrder}
\end{equation}
to the second leading order of $d$.

From the definition of $\Lambda_{\tau}(x)$, an expansion up to the second leading order is obtained by expanding the summation over $s$ up to $s=1$, such as
\begin{multline}
    \Lambda_{\tau}(x)/2
    =
    (d-2)^{2 \tau} + 4 \tau (d-2)^{2\tau-1}\left| T_1\left(\frac{2x}{d}\right) \right|.
\end{multline}
The Chebyshev polynomials of the first kind can be explicitly given by
\begin{align}
    &T_0\left(\frac{2x}{d}\right) = 1,\\
    &T_1\left(\frac{2x}{d}\right) = \cos \left( \arccos \left(\frac{2x}{d}\right) \right) = \frac{2x}{d}.
\end{align}
Substituting these results in~\cref{Eq:LambdaLeadingOrder}.

\section{Derivations of~\cref{Eq:LBUB,Eq:UBLB}} \label{App:UBLBcomparison}
We show that, for sufficiently large $d$,
\begin{align}
    &\textrm{UB}_{t'}\left(\frac{D}{m}, \frac{d}{4} - \frac{1}{3}\right) < \textrm{LB}_{t'}\left(\frac{D}{m}, \frac{d}{4}\right),\\
    &\textrm{UB}_{t}\left(\frac{D}{m}, \frac{d}{4} - 1\right) < \textrm{LB}_{t}\left(\frac{D}{m}, \frac{d}{4}-\frac{2}{3}\right),
\end{align}
which are used in the analysis of $\UniDes(t,t',\delta,\delta')$. These functions are defined by~\cref{Eq:UBtauUni,Eq:LBtauUni}.

For the former, a direct calculation leads to
\begin{multline}
    \textrm{R.H.S.} - \textrm{L.H.S.}\\
    =
    \frac{1}{D/m+2}\left[\Lambda_{t'}\left(\frac{d}{4}\right) - \Lambda_{t'}\left(\frac{d}{4}-\frac{1}{3}\right)  \right]  
    - \frac{2D/m}{D/m + 2} \delta_0.
\end{multline}
We also have 
\begin{widetext}
\begin{multline}
    \Lambda_{t'}\left(\frac{d}{4}\right) - \Lambda_{t'}\left(\frac{d}{4}-\frac{1}{3}\right)
    =
    8t'(d-2)^{2t'-1}\left( \left| T_1\left(\frac{1}{2}\right)\right| - \left| T_1\left(\frac{1}{2} - \frac{2}{3d}\right)\right|\right)\\
    +2 \sum_{s=2}^{2t'} \binom{2 t'}{s}(d-2)^{2t'-s} \sum_{r=0}^s \binom{s}{r}
    \left( \left| T_{|2r-s|}\left(\frac{1}{2}\right)\right| - \left| T_{|2r-s|}\left(\frac{1}{2} - \frac{2}{3d}\right)\right|\right).
\end{multline}
\end{widetext}
As $T_1(x) = \cos (\arccos x) = x$,
\begin{equation}
    \left| T_1\left(\frac{1}{2}\right)\right| - \left| T_1\left(\frac{1}{2} - \frac{2}{3d}\right)\right|
    =
    \frac{2}{3d}. \label{Eq:ave'rpiohj43gea4g43qe}
\end{equation}
From the Taylor expansion of the Chebyshev polynomials $T_n(x)$ around $x=1/2$, we have
\begin{multline}
    T_n \left(\frac{1}{2} - \varepsilon\right) = \cos\left(\frac{\pi n}{3}\right) + \frac{2\sqrt{3}n}{3} \sin\left(\frac{\pi n}{3} \right) \varepsilon \\
    + \cO\left(\varepsilon^{2}\right).
\end{multline}
This leads to
\begin{multline}
    \left|\left( \left| T_{|2r-s|}\left(\frac{1}{2}\right)\right| - \left| T_{|2r-s|}\left(\frac{1}{2} - \frac{2}{3d}\right)\right|\right)  \right|\\
    \leq
    \frac{4\sqrt{3}|2r-s|}{9d} + \cO\left(d^{-2}\right)
    \leq
        \frac{8\sqrt{3}t'}{9d} + \cO\left(d^{-2}\right).\label{Eq:Poaw4er;ogihver}
\end{multline}
Note that we have used the fact that $s, r \in [0,2t']$.
Hence, it holds that
\begin{align}
    &\Lambda_{t'}\left(\frac{d}{4}\right) - \Lambda_{t'}\left(\frac{d}{4}-\frac{1}{3}\right) \notag \\
    &\geq
    \frac{16t'}{3}d^{2t'-2} + \cO\left(d^{2t'-3}\right)\notag \\
    &\hspace{10mm} - \frac{16\sqrt{3}t'}{9d} \sum_{s=2}^{2t'} \binom{2 t'}{s}(d-2)^{2t'-s} \sum_{r=0}^s \binom{s}{r}\\
    &=\frac{16t'}{3}d^{2t'-2} + \cO\left(d^{2t'-3}\right). 
\end{align}
All together, we obtain
\begin{multline}
    \textrm{R.H.S.} - \textrm{L.H.S.}\\
    \geq
    \frac{1}{D/m+2}\left(\frac{16t'}{3}d^{2t'-2} + \cO\left(d^{2t'-3}\right) \right)
    - \frac{2D/m}{D/m + 2} \delta_0. \label{Eq:LHSRHS}
\end{multline}
As $D/m \in \Theta(d^{2t}/\delta)$ and $\delta_0 \in \cO(\delta^2/d^{2t})$, this reduces to
\begin{equation}
    \textrm{R.H.S.} - \textrm{L.H.S.}
    \geq
    \frac{\delta}{d^{2t}}\left(\frac{16t'c}{3}d^{2t'-2} - c' \delta\right),
\end{equation}
with some positive constants $c$ and $c'$, to the leading order.
Hence, as far as $\delta \in o(d^{2(t'-1)})$ that follows from the assumption $\delta \in o(d^{2(t-1)})$ of the theorem, the leading term is positive. That is, for sufficiently large $d$, we have
\begin{equation}
    \textrm{UB}_{t'}\left(\frac{D}{m}, \frac{d}{4} - \frac{1}{3}\right) < \textrm{LB}_{t'}\left(\frac{D}{m}, \frac{d}{4}\right).
\end{equation} 

The latter inequality can be similarly shown. In this case,
\begin{multline}
    \textrm{R.H.S.} - \textrm{L.H.S.}\\
    =
    \frac{1}{D/m+2}\left[\Lambda_{t}\left(\frac{d}{4}-\frac{2}{3}\right) - \Lambda_{t}\left(\frac{d}{4}-1\right)  \right]  
    - \frac{2D/m}{D/m + 2} \delta_0.
\end{multline}
The right-hand side of this equation is exactly the same as that of~\cref{Eq:LHSRHS} since 
\begin{equation}
    \left| T_1\left(\frac{1}{2} - \frac{4}{3d} \right)\right| - \left| T_1\left(\frac{1}{2} - \frac{2}{d}\right)\right|
    =
    \frac{2}{3d},
\end{equation}
which has the same value as~\cref{Eq:ave'rpiohj43gea4g43qe}, and
\begin{multline}
    \left|\left( \left| T_{|2r-s|}\left(\frac{1}{2} - \frac{4}{3d}\right)\right| - \left| T_{|2r-s|}\left(\frac{1}{2} - \frac{2}{d}\right)\right|\right)  \right|\\
    \leq
    \frac{8\sqrt{3}t'}{9d} + \cO\left(d^{-2}\right),
\end{multline}
which is the same as~\cref{Eq:Poaw4er;ogihver}.
Thus, the difference between L.H.S. and R.H.S. in this case is the same as previously, implying that R.H.S.-L.H.S. is positive as well, to leading order.

\end{document}